\documentclass{article}
\usepackage{emulateapjAASTeX4}
\usepackage{onecolfloat}
\usepackage{graphicx} 

\lefthead{Dalcanton \& Bernstein}
\righthead{Edge-on Disks (II): Vertical Color Gradients and Ubiquitous Thick Disks}

\newcommand\degree{{^\circ}}
\newcommand\pc{{\rm\,pc}}

\newcommand\kpc{{\rm\,kpc}}
\newcommand\Mpc{{\rm\,Mpc}}

\newcommand\Gyr{{\rm\,Gyr}}
\newcommand\kmsec{{\rm\,km\,s^{-1}}}
\newcommand\kms{\kmsec}

\newcommand\msun{{\rm\,M_\odot}}

\newcommand\surfb{{\rm\,mag\,arcsec^{-2}}}

\def\aj{{\rm A.~J.}, }  
\def\apj{{\rm Ap.~J.}, }  
\def\apjs{{\rm Ap.~J.~Suppl.}, }  
\def\apjl{{\rm Ap.~J.~(Letters)}, } 
\def\pasp{{\rm Pub.~A.S.P.}, }      
\def\nat{{\rm Nature}, }      
\def\aa{{\rm Astr.~Ap.}, }     
\newcommand\mn{{\rm M.N.R.A.S.}, }      
\newcommand\aasup{{\rm Astr.~Ap.~Suppl.}, }     
\newcommand\ass{{\rm Ast.Sp.Sci.}, }  

\begin{document}

\twocolumn[

\title{A Structural and Dynamical Study of Late-Type, Edge-On Galaxies:
II. Vertical Color Gradients and the Detection of Ubiquitous Thick Disks}

\author{Julianne J. Dalcanton\altaffilmark{1,2}}
\affil{Department of Astronomy, University of Washington, Box 351580,
Seattle WA, 98195}

\author{Rebecca A. Bernstein\altaffilmark{3,4}}
\affil{Department of Astronomy, University of Michigan, Ann Arbor MI, 48109}

\altaffiltext{1}{e-mail address: jd@astro.washington.edu}
\altaffiltext{2}{Alfred P. Sloan Research Fellow}
\altaffiltext{3}{e-mail address: rab@astro.lsa.umich.edu}
\altaffiltext{4}{Hubble Fellow}
  
\begin{abstract}

  We present an analysis of optical ($B-R$) and optical-infrared
  ($R-K_s$) color maps for 47 extremely late-type, edge-on, unwarped,
  bulgeless disk galaxies spanning a wide range of mass.  The color
  maps show that the thin disks of these galaxies are embedded within
  a low surface brightness red envelope.  This component is
  substantially thicker than the thin disk ($a/b\,\sim\,$4:1, vs
  $>\,$8:1), extends to at least 5 vertical disk scale heights above
  the galaxy midplane, and has a radial scale length that appears to
  be uncorrelated with that of the embedded thin disk.  The color of
  the red envelope is similar from galaxy to galaxy, even when the
  thin disk is extremely blue, and is consistent with a relatively old
  ($>\!6\Gyr$) stellar population that is not particularly metal-poor.
  The color difference between the embedded thin disk and the red
  stellar envelope varies systematically with rotation speed,
  reflecting an increasing age difference between the thin and thick
  components in lower mass galaxies, driven primarily by changes in
  the age of the thin disk.

  The red stellar envelopes are similar to the thick disk of the Milky
  Way, having common surface brightnesses, spatial distributions, mean
  ages, and metallicities.  We argue that the ubiquity of the red
  stellar envelopes implies that the formation of the thick disk is a
  nearly universal feature of disk formation and is not necessarily
  connected to the formation of a bulge.  Our data suggest that the
  thick disk forms early ($>6\Gyr$ ago), even within galaxies where
  the bulk of the stars formed very recently ($<2\Gyr$).  
  We argue that several aspects of our data and the observed properties
  of the Milky Way thick disk argue in favor of a merger origin for
  the thick disk population.  If so, then the age of the thick disk
  marks the end of the epoch of major merging, and the age difference
  between the younger thin disk and the older thick disk can become a
  strong constraint on cosmological constants and models of galaxy
  and/or structure formation.

\end{abstract}

]

\setcounter{footnote}{0}

\section{Introduction}

Currently, most observational studies of galaxy formation focus on
two epochs -- extremely high redshift, where one can observe
galaxy formation in progress, and zero redshift, where one can
disentangle past history using individual stars within the Milky
Way.  While each of these approaches has been essential in shaping our
current view of galaxy formation, they each have fundamental
limitations.

At high redshifts, it is extremely difficult to match galaxies to
their low redshift descendents due to morphological transformation,
luminosity evolution, and merging; only changes in the mean galaxy
population can be tracked, revealing few details of the physical
mechanisms which drive evolution.  In contrast, within the Milky Way a
wealth of detail can be extracted from the ages, metallicities, and
kinematics of stars, allowing us to trace the formation of the
faintest individual components of the Galaxy (the stellar halo, the
thick disk, tidal streams, etc).  However, in the end these data
address only the formation of the Milky Way, and give no constraints
on how galaxy formation proceeds in the mean population, or varies
with fundamental parameters (e.g.\ mass, angular momentum, local
density, etc.).

An opportunity to bridge these regimes lies in the realm of nearby
galaxies, just beyond the confines of the Local Group.  At
moderate distances ($cz\lesssim5000\kms$), galaxies of all types are
plentiful and are extremely well resolved spatially
($\Delta\theta\lesssim200h^{-1}\pc$ from the ground, or
$\Delta\theta\lesssim20h^{-1}\pc$ from space), allowing us to trace their
morphology and internal dynamics on small spatial scales.  These
features can be studied at very high signal-to-noise and/or
at low surface brightnesses inaccessible at higher redshifts.
To place observational constraints on the process of disk galaxy
formation using nearby galaxies, we are engaged in a comprehensive
program to study the dynamics, gas content, metallicity, and stellar
populations of a population of late-type, bulgeless disk galaxies.
This population forms a structurally uniform sample, allowing us to
isolate changes in the physical properties of the galaxies (i.e.\ mass,
angular momentum, etc.) independent of changes in morphology, akin to
what is possible with the nearly single parameter sequence spanned by
elliptical galaxies.  By avoiding systems with bulges, we also limit
the degree to which the baryonic component of the galaxy may have been
affected by dissipation or angular momentum transport during
formation.  This yields a sample which represents the purest endpoint
of the disk galaxy formation process.  Details of the sample selection
and the optical and infrared imaging can be found in Dalcanton
\& Bernstein 2000 (hereafter ``Paper I'').

In this paper we use the imaging presented in Paper I to undertake an
analysis of the color maps of the bulgeless, edge-on disks which
comprise our sample.  We focus our attention on the vertical
color gradients within the galaxies, probing the stellar populations
of the galaxies at many scale heights above the thin disk.  Buried
within these low surface brightness components are the remnants of
some of the earliest epochs in the assembly of galaxies, namely the
thick disk and stellar halo.

Because the typical metallicity of a galaxy tends to increase with
time, it has long been recognized that the low metallicity thick disk
and stellar halo are fossil records of the very early
history of the Milky Way.  Detailed studies of their kinematics and
metal abundance have revealed signatures of the processes which led to
their formation over 10 billion years ago, even though these
components contain only a small fraction of the total stellar mass of
the Milky Way.  In general, these two components are thought to be the
leftovers from either the monolithic, dissipative collapse of the
early galaxy (Eggen et al.\ 1962, hereafter ``ELS'') or the buildup of
the galaxy through hierarchical merging (Searle \& Zinn 1978).
Because of the low surface brightness of stellar halos and thick
disks, it is impossible to study their formation directly at high
redshift (due to cosmological $(1+z)^4$ dimming), and we are confined
to deducing their history from very low redshift data alone.

Almost all of the detailed knowledge of the formation of thick disks
and stellar halos comes from evidence within the Milky Way alone (see
van den Bergh 1996 for a review), teaching us little about galaxy
formation {\emph{in general}}.  For this reason, astronomers have
attempted to identify these faint components in other very nearby
galaxies, particularly in the edge-on orientation where the light from
the younger thin disk can be minimized (Burstein 1979, Tsikoudi 1979,
van der Kruit \& Searle 1981).  Previous detections of possible halo
or thick disk stellar light in external galaxies have been made in a
scant handful of nearby edge-on galaxies (e.g.\ recently Neeser et
al.\ 2000, Fry et al.\ 1999, Zheng et al.\ 1999, Morrison et al.\
1994, N\"aslund \& J\"ors\"ater 1997, Morrison et al.\ 1997, Sackett
et al.\ 1994, van Dokkum et al.\ 1994, Shaw \& Gilmore 1990; see
\S\ref{otherworksec} below for further discussion).  Typically, the
presence of a thick disk or stellar halo has been identified by the
need for an additional disk component when attempting to fit models of
the light distribution in a deep image.  Not all galaxies have
required this second component, however.  Instead, thick disks have
only been identified in a handful of relatively massive Sc (or
earlier) galaxies with substantial bulges (see summary by Morrison
1999).

One limitation of the previous searches for thick disks is that almost
all have been based on imaging in a single passband, discriminating
between the thick disk and thin disk components through subtle changes
in the surface brightness profile perpendicular to the plane.  It is
therefore difficult to make a unique decomposition of the thick and
thin disks when neither dominates in the region studied, as noted by
Morrison et al.\ (1997).  In this paper, however, we use multi-color
imaging to identify thick disks via the systematic changes in broad
band colors produced by the variation in the stellar populations of
the thick and thin disks.  As we show below, we find unambiguous
evidence for stellar envelopes surrounding the majority of the nearly
50 disks in our sample, across all galaxy masses.

The structure of the paper is as follows.  We begin by briefly
summarizing the galaxy sample and imaging data in \S\ref{datasec}.  We
present color maps in \S\ref{colormapsec} and discuss the general,
qualitative implications for vertical color gradients, radial color
gradients, and the presence of dust in the sample.  We further
quantify the results in \S\ref{extractionsec} and interpret them
based on comparison with stellar population models in
\S\ref{interpsec}.  We show that the color gradients and color maps
argue for the presence of old, red stellar envelopes around most, if
not all, disk galaxies.  We analyze the shapes of the stellar
envelopes in \S\ref{isophotesec}.  Color gradients and isophotes are
more difficult to interpret in the more massive galaxies for a variety
of reasons which we discuss in \S\ref{massivegalaxysec}.  In
\S\S\ref{interpretationsec} \& \ref{formationsec}, we suggest that
the stellar envelopes in this sample are analogous to the thick disk
of the Milky Way and have properties consistent with those expected by a
stochastic merging scenario for the formation of the thick disk.  To
conclude, we discuss the general constraints which can be placed on
galaxy formation based on the observation of ubiquitous thick disks
(\S\ref{conclusionsec}).

\section{Data}                  \label{datasec}

As described in detail in Paper I (Dalcanton \& Bernstein 2000), our
sample of edge-on bulgeless galaxies was initially selected from the
Flat Galaxy Catalog (FGC) of Karanchentsev et al.\ (1993), a catalog
of 4455 edge-on galaxies with axial ratios greater than 7, and major
axis lengths of $>\!0.6\arcmin$.  The FGC was originally selected by
visual inspection of the O POSS plates in the north ($\delta>-27\deg$)
and the J films of the ESO/SERC survey in the south ($\delta <
-17\deg$); galaxies from the ESO plates are known as the FGCE, and
have slightly different properties due to small differences in the
plate material\footnote{The FGC catalog has been recently revised to
  make the RFGC (Karachentsev et al.\ 1999), which has a different
  numbering scheme.  However, we have chosen to retain the original
  FGC numbers from the original 1993 catalog for
  consistency with Paper I.}.  From the combined FGC/FGCE catalog we
selected galaxies which appeared undisturbed, bulgeless, and have no
signs of inclination (major-to-minor axis ratio $a/b>8$; see Figure 2
in Paper I).  Our final sample contains 49 galaxies.  One of these
galaxies, FGC 1971, appears to be a polar ring galaxy, and we do not
include it in the analysis presented here.  Another galaxy, FGC 2292,
is between two relatively bright stars, producing significant
scattered light in our images, and prohibiting surface brightness
measurements at the faint levels possible in the rest of the survey.
Results based upon this galaxy are included for completeness, but
should be used with caution.  Finally, while the initial sample
appeared to be ``bulge-free'' on digitized POSS-II survey plates,
our deeper imaging revealed the presence of small bulges in a few of
the galaxies.  FGC 227, FGC 395, FGC 1043, FGC 1440, FGC 2217, FGC
E1447, and FGC E1619 all have small bulges, as indicated by visual
inspection of the $K_s$ infrared imaging, and/or by a significantly
increased quality of fit to the radial profile when a double
exponential disk model is used instead of a single exponential model
(judged via the ratio of the $\chi^2$ values).  The ``bulges'' in
FGC 227, FGC 395, FGC 1043, and FGC 1440 are
in general extremely small and may not represent truly kinematically
distinct components in most cases; they may be edge-on manifestations
of ``pseudo-bulges'' (Kormendy 1992).  Only in FGC E1447,
FGC E1619, and possibly FGC 2217 do the bulges appear prominent and
vertically extended in the $K_s$ band images.  Our analysis includes all
these galaxies for completeness, but our results do no change if they
are excluded.  FGC E1447 is excluded from our analysis, however,
because we do not have a measured rotation speed for the galaxy.

We obtained multi-wavelength imaging for the sample at the Las
Campanas du Pont 2.5m telescope as part of an extensive observational
program.  The optical ($B$ \& $R$) and infrared ($K_s$) imaging, and
its reduction and calibration, have all been presented in Paper I, and
the resulting calibrated images form the basis for the work presented
in this paper.  The optical data have photometric calibration
uncertainties of $0.01-02$ mag, sub-arcsecond spatial resolution, and
flat fielding accurate to $29-30\surfb$ in $B$ and $28-29\surfb$ in
$R$ on scales larger than 10$\arcsec$ (see Paper I).  The majority of
the infrared data were taken in photometric conditions ($\Delta m <
0.04$ mag) with seeing of $0.9-1.2\arcsec$.  However, due to
limitations in sky subtraction and flat fielding, the infrared data
can reach only to $22.5\surfb$ in $K_s$ on 10$\arcsec$ scales.  Any
regions with contaminating foreground or background sources or without
complete bandpass coverage are masked from our surface brightness and
color analysis.

In addition to our deep imaging, we have substantial information on
the dynamics of the galaxies in the sample.  More than $3/4$ of the
sample have single-dish HI observations, yielding corrected line
widths at 50\% peak flux ($W_{50,c}$), as compiled in Paper I.  The
majority of these measurements are from a large survey of FGC galaxies
observed at Arecibo by Giovanelli et al.\ (1997).  These are
supplemented with measurements for FGC 164 from Schneider et al.\ 
(1990), for FGC 84 \& 2264 from Matthews \& van Driel (2000), and for
FGC 349 from Haynes et al.\ (1997).  We have also obtained long-slit
H$\alpha$ rotation curves for 34 galaxies in the sample, using the du
Pont 2.5m telescope.  These rotation curves (Dalcanton \& Bernstein
2000, Dalcanton \& Bernstein 2003 in prep.) are used to supplement the
dynamical information available from the HI observations.  Throughout
this paper we will consider galaxies as a function of their rotational
velocity $V_c$, which we take to be $W_{50,c}/2$ if HI data is
available, or the maximum $V_{c,opt}$ measured for the optical
rotation curve if no radio data exists.  For the 23 galaxies where
both are available, we find that $(W_{50,c}/2) / V_{c,opt,max} = 1 \pm
0.1$, suggesting that on average, $W_{50,c}/2$ and $V_{c,opt,max}$ are
equivalent measures of rotation speed.  However, for 5 of these 23
galaxies with both optical and HI kinematics, $(W_{50,c}/2)$ and
$V_{c,opt,max}$ differ by more than 20\%, and thus there may be
substantial offsets in the adopted $V_c$ in individual cases where HI
data was lacking.

\section{Color Maps}                    \label{colormapsec}

In Figure \ref{colormapfig} we present $B-R$ (left column) and $R-K_s$
(right column) color maps of the sample galaxies, sorted in order of
decreasing rotation speed $V_c$.  Sky-subtracted images of the
galaxies in $B$, $R$, and $K_s$ can be found in Paper I, but for
reference the faintest $R$-band contours from Figure 3 in Paper I have
been superimposed on the color maps to show the maximum detected
extent of the galaxies.  All of the galaxies are displayed with the
same minimum and maximum color, and have been corrected for foreground
extinction using the Schlegel et al.\ (1998) dust maps, assuming an
R=3.1 extinction law.  Apparent variations from galaxy to galaxy
therefore reflect true variations in color.  For display purposes, we
have generated the color maps using ``asinh magnitudes'' (Lupton et
al.\ 1999).  The asinh magnitudes are effectively identical to
traditional logarithmic magnitudes at high signal-to-noise, but unlike
normal magnitudes, they are mathematically well-behaved even for
negative fluxes (such as occur at low signal-to-noise).  We found that
the use of asinh magnitudes greatly increased our ability to visually
detect features in the color maps.
        
Before we begin a quantitative analysis, there are a number of qualitative
features to note about the color maps in Figure \ref{colormapfig}.  We discuss
these briefly below.

\subsection{Strong Vertical Color Gradients}

Among of the strongest features recognizable in the color maps in
Figure~\ref{colormapfig} are strong color gradients with increasing
height above the mid-plane.  In the most massive galaxies, these color
gradients are clearly due to the presence of strong dust lanes, which
create a very red midplane, embedded in a bluer, unreddened stellar
envelope.  However, in the less massive galaxies, the situation is
quite different.  These low mass galaxies have a very {\emph{blue}}
midplane, embedded in a much rounder {\emph{red}} stellar envelope.
Looking at the high signal-to-noise $B-R$ maps of the most spatially
well-resolved low mass galaxies in the sample (for example FGC 51, FGC
780, or FGC 1285), a thin blue disk superimposed on a much rounder red
stellar population is evident.  We will quantify these results and
discuss them in detail below.

\subsection{Strong Radial Color Gradients}      \label{radialgradsec}

In Figure~\ref{colormapfig}, strong radial color gradients are also
evident along the midplane, even in those disks which are partially
obscured by dust. These gradients are particularly noticeable in the
higher signal-to-noise $B-R$ color maps.  The gradients are all in the
sense of having a very blue outer disk and red inner disk.  If these
disks are optically thin over much or all of their extent, then this
apparent radial gradient is simply the edge-on manifestation of the
well-studied radial gradients seen in face-on galaxies.  In the
massive galaxies with clear dust lanes, the gradient is likely to be
accentuated by an edge-on viewing geometry, due to larger reddening in
the inner regions.  A systematic analysis of radial color gradients
has recently been presented by Bell \& de Jong (2000) for a large
sample of face-on galaxies, using techniques similar to those adopted
in this paper.  They find that age, not metallicity or dust, is the
primary driver for the bluing in galaxy disks with increasing radius.
As there is no reason to believe that the disks in our sample are not
comparable to the face-on late-type disks in theirs, it is likely that
age is the principal driver of the radial color gradient in our disks
as well.  We will not pursue an analysis of radial color gradients
independently with our sample because dust will always complicate the
interpretation of edge-on colors relative to comparable face-on disks.
However, it should be possible to compare the face-on and edge-on
samples to limit the {\emph{dust}} properties within the disks. We
defer that analysis to a later paper.

\clearpage

\begin{figure*}[t]
\hbox{ 
\includegraphics[width=3.5in]{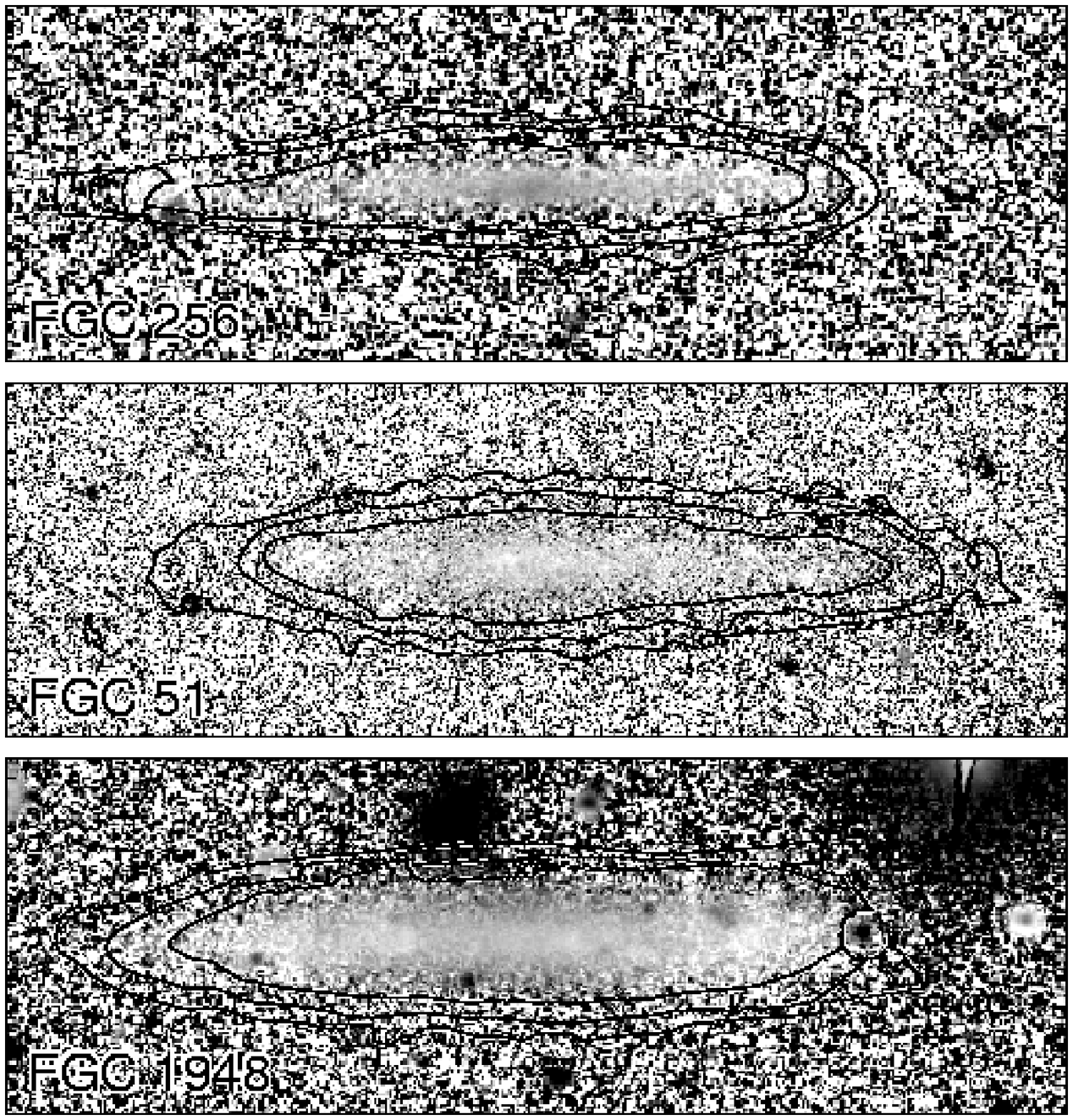}
\includegraphics[width=3.5in]{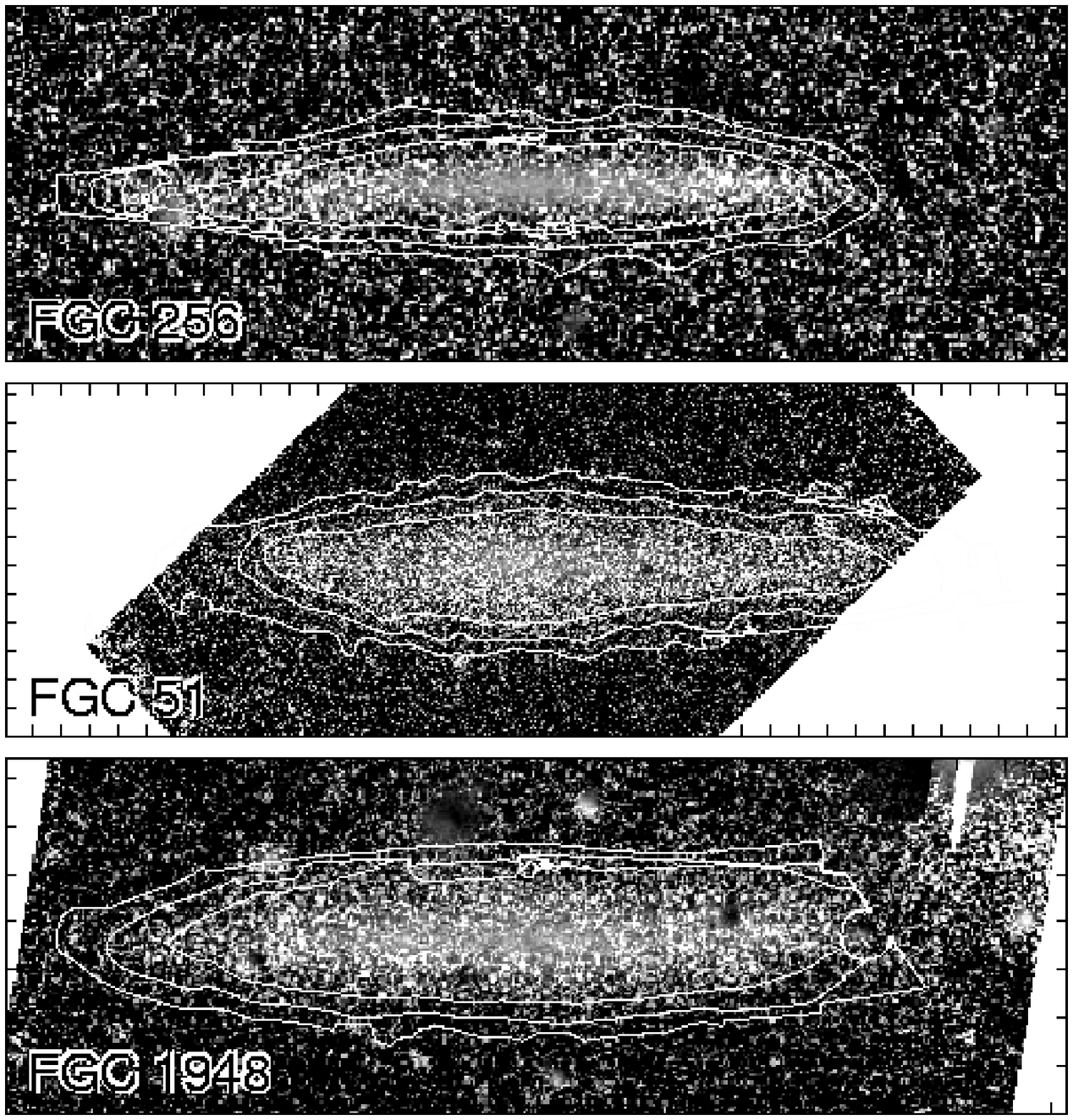}}
\vspace{0.1in}
\hbox{
\includegraphics[width=3.5in]{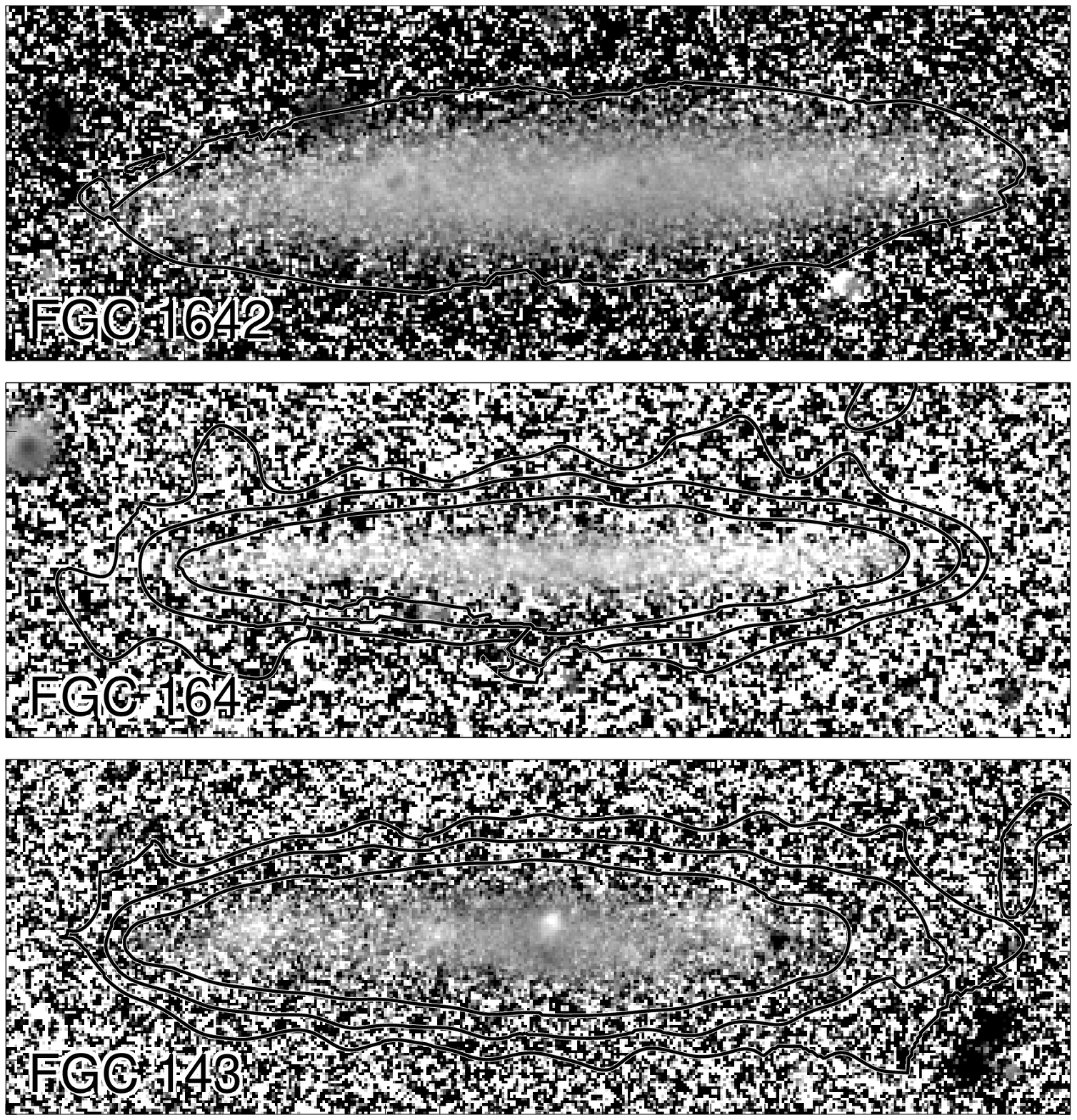}
\includegraphics[width=3.5in]{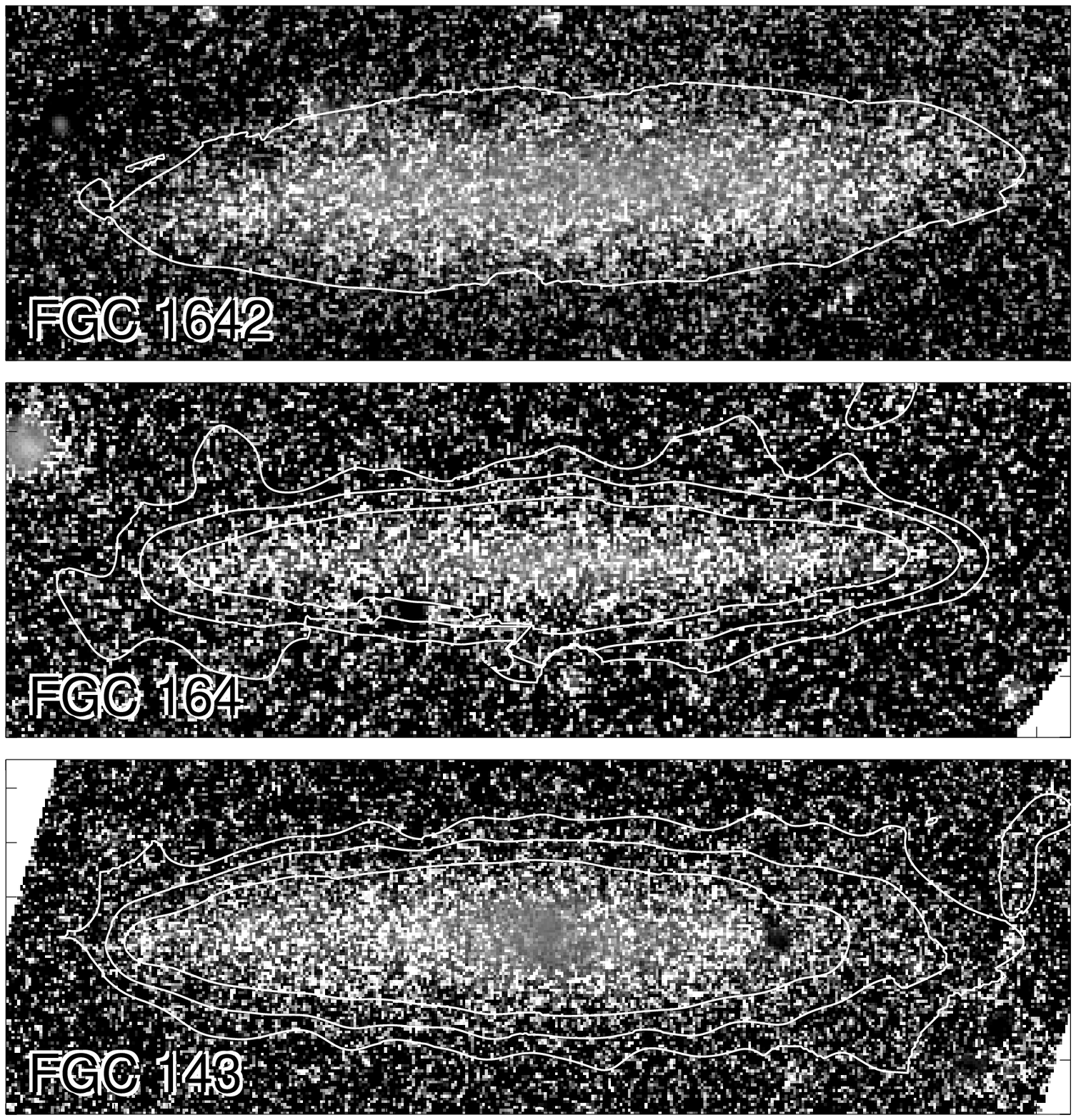}}
\caption{\footnotesize
$B-R$ (left column) and $R-K_s$ (right column) color maps of FGC
galaxies, sorted in order of increasing rotation speed.  Images are
plotted in a grey-scale, linear stretch such that black corresponds to
the red limit of the range ($B-R=1.9$, $R-K_s=4$) and white to the
blue limit ($B-R=0.3$, $R-K_s=0.3$).  Overlayed contours are from the
$R$-band images presented in Paper I, and are separated by $\Delta\mu
= 1\surfb$.  Tick marks are drawn at 5$\arcsec$ intervals.  The color
maps were generated with ``asinh'' magnitudes (Lupton et al. 1999),
which are well behaved for negative fluxes and which are a less biased
estimator of the true color than traditional magnitudes at low
signal-to-noise.
\label{colormapfig}}
\end{figure*}

\setcounter{figure}{0}
\begin{figure*}[t]
\hbox{ 
\includegraphics[width=3.5in]{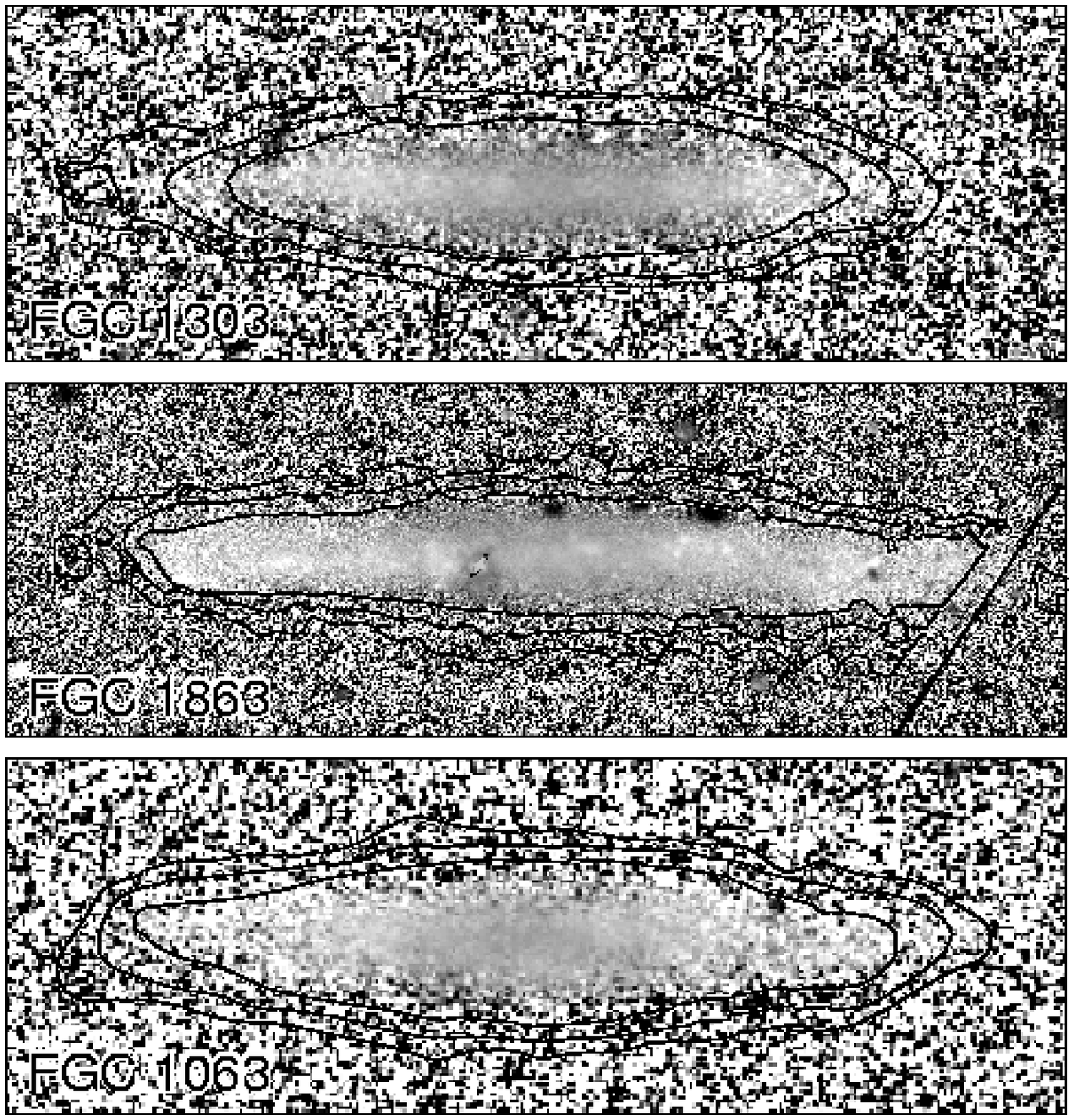}
\includegraphics[width=3.5in]{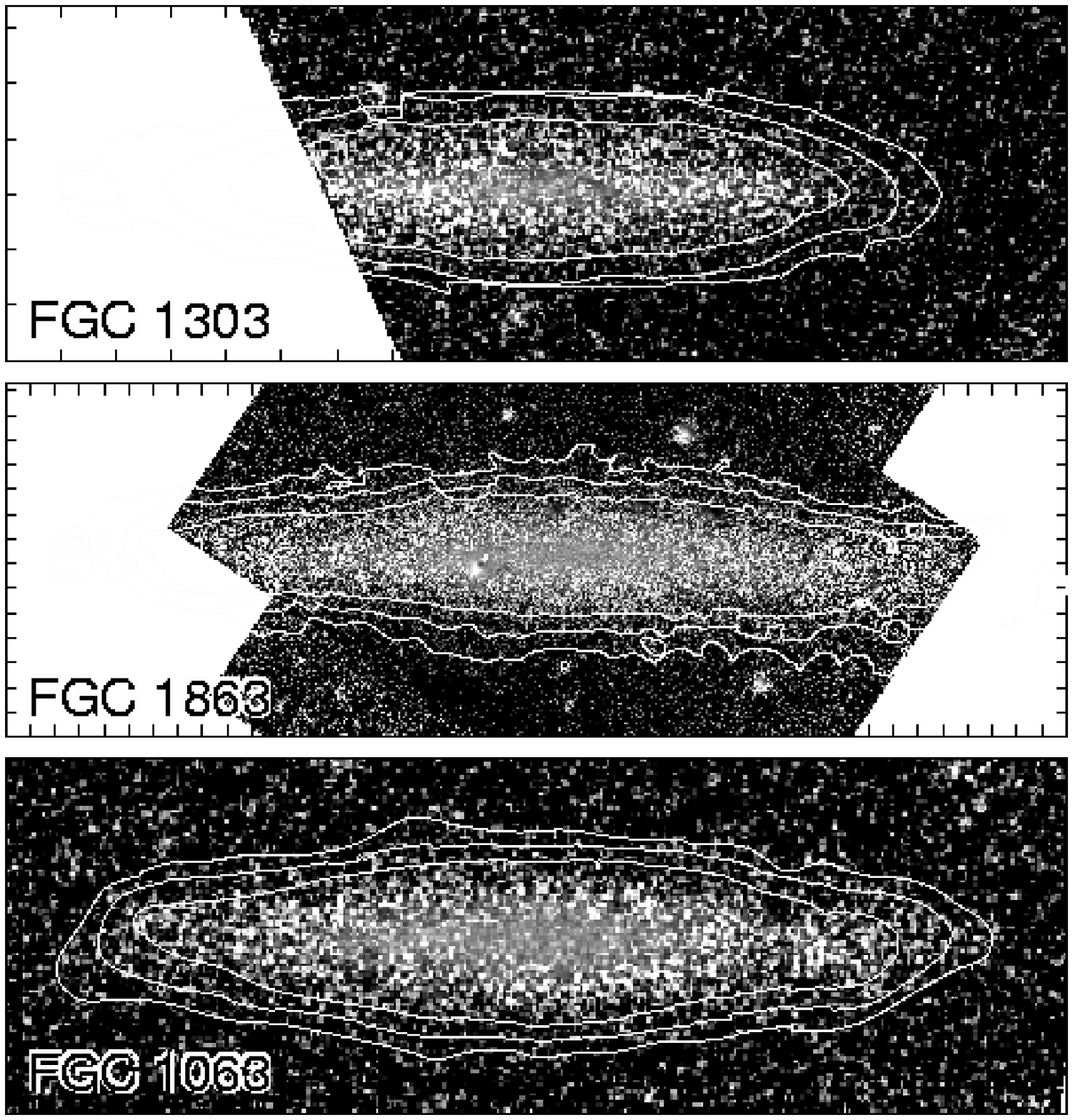}}
\vspace{0.1in}
\hbox{
\includegraphics[width=3.5in]{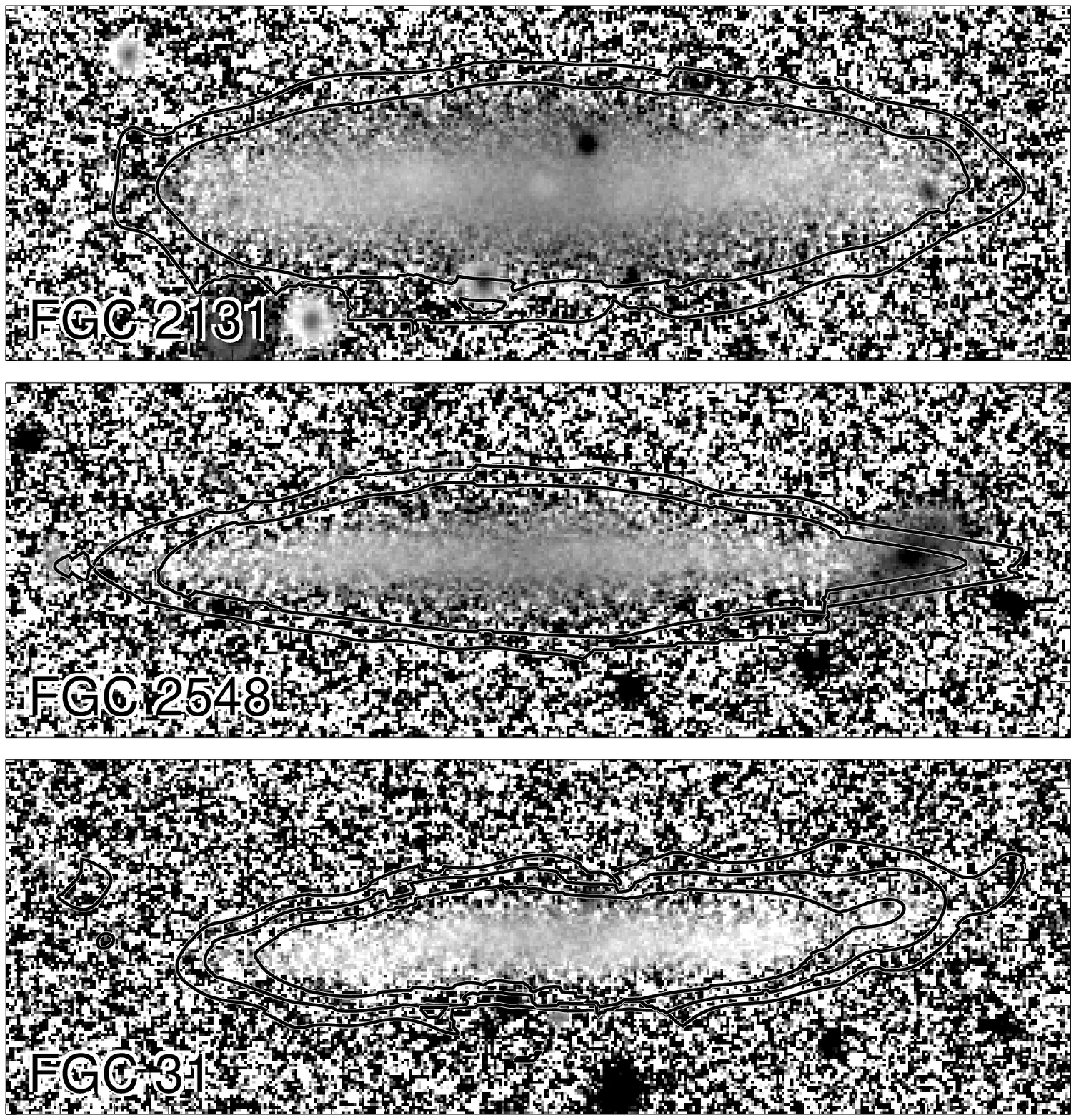}
\includegraphics[width=3.5in]{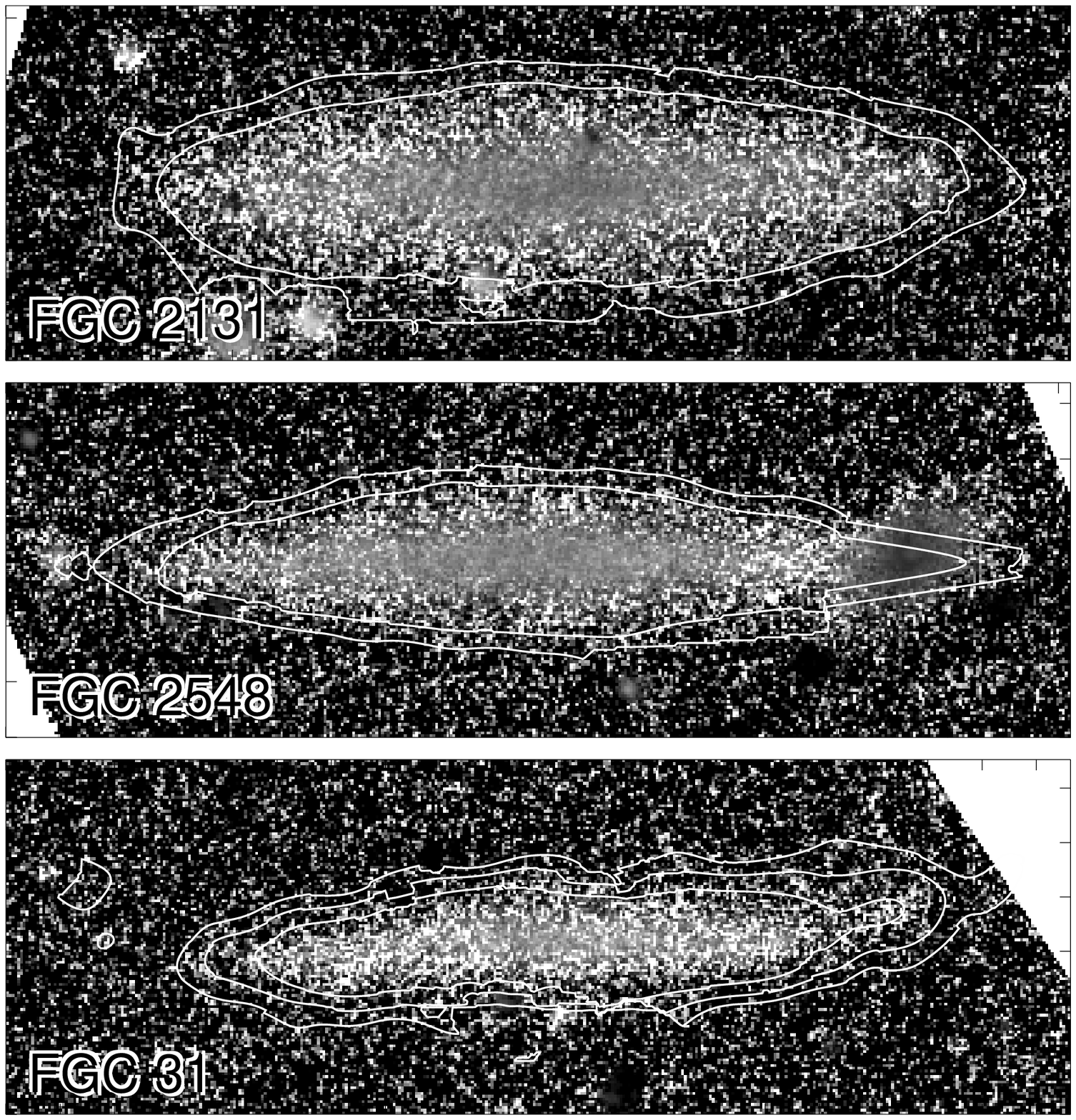}}
\caption{\footnotesize (continued)}
\end{figure*}

\setcounter{figure}{0}
\begin{figure*}[t]
\hbox{ 
\includegraphics[width=3.5in]{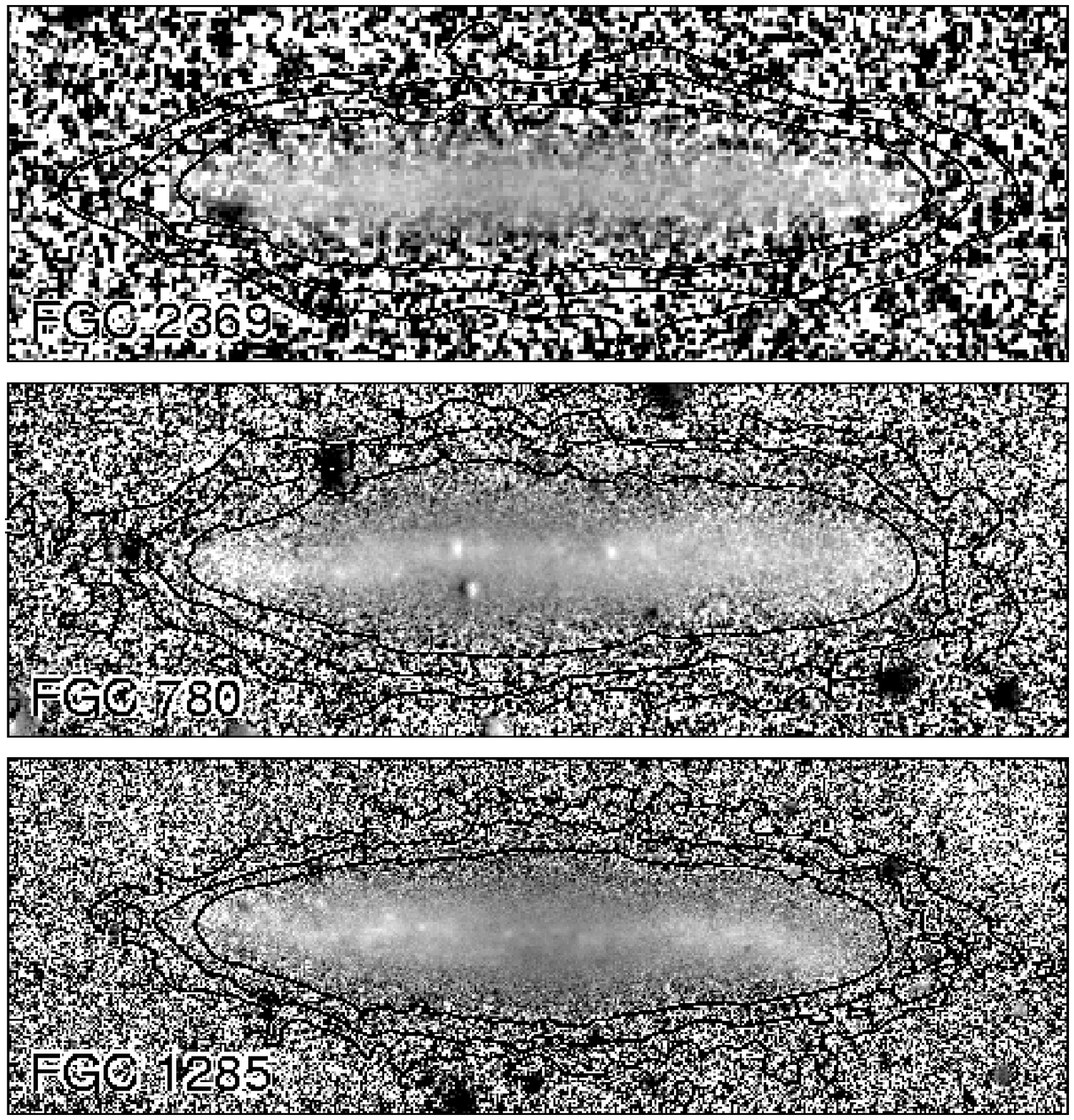}
\includegraphics[width=3.5in]{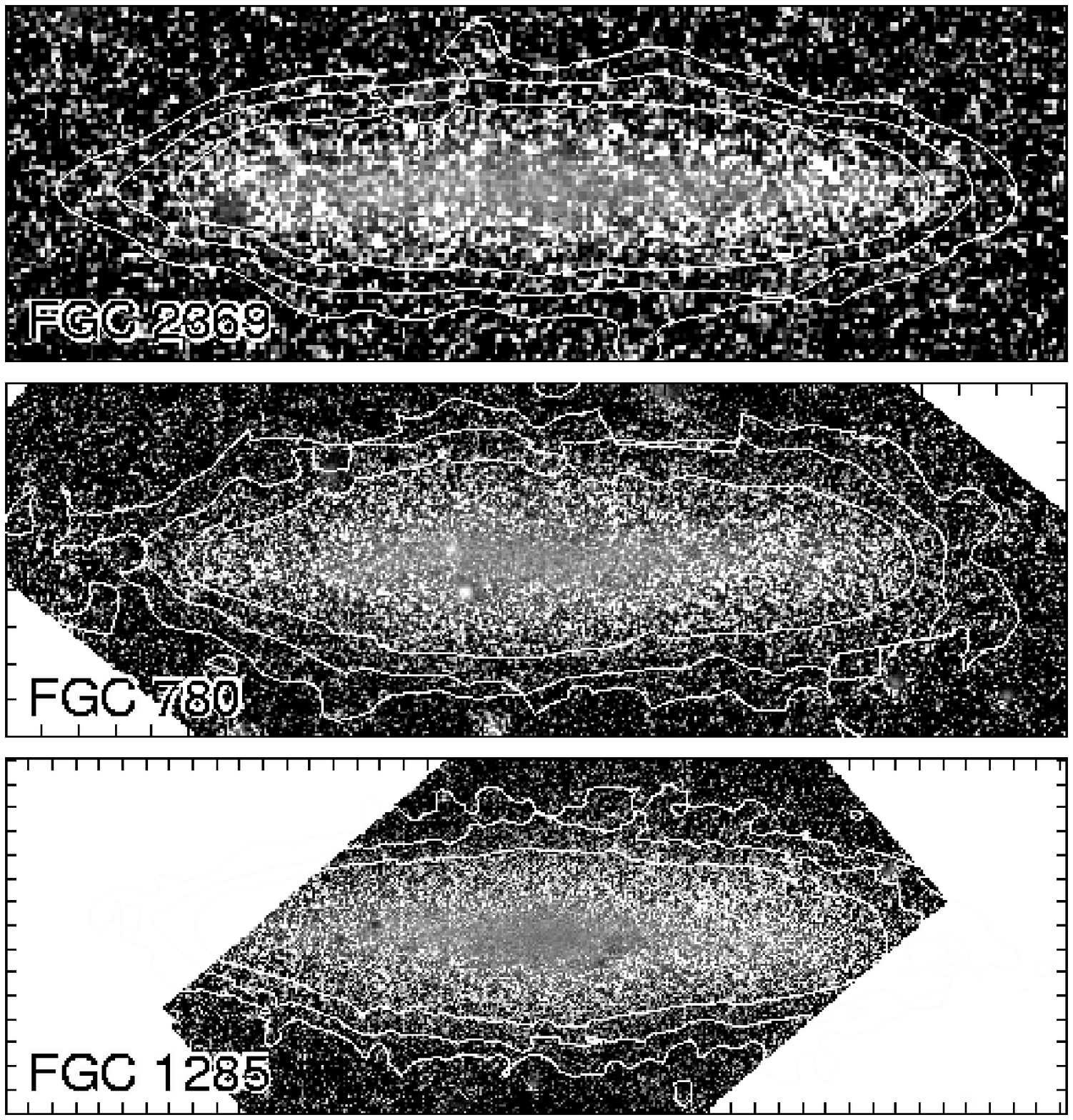}}
\vspace{0.1in}
\hbox{
\includegraphics[width=3.5in]{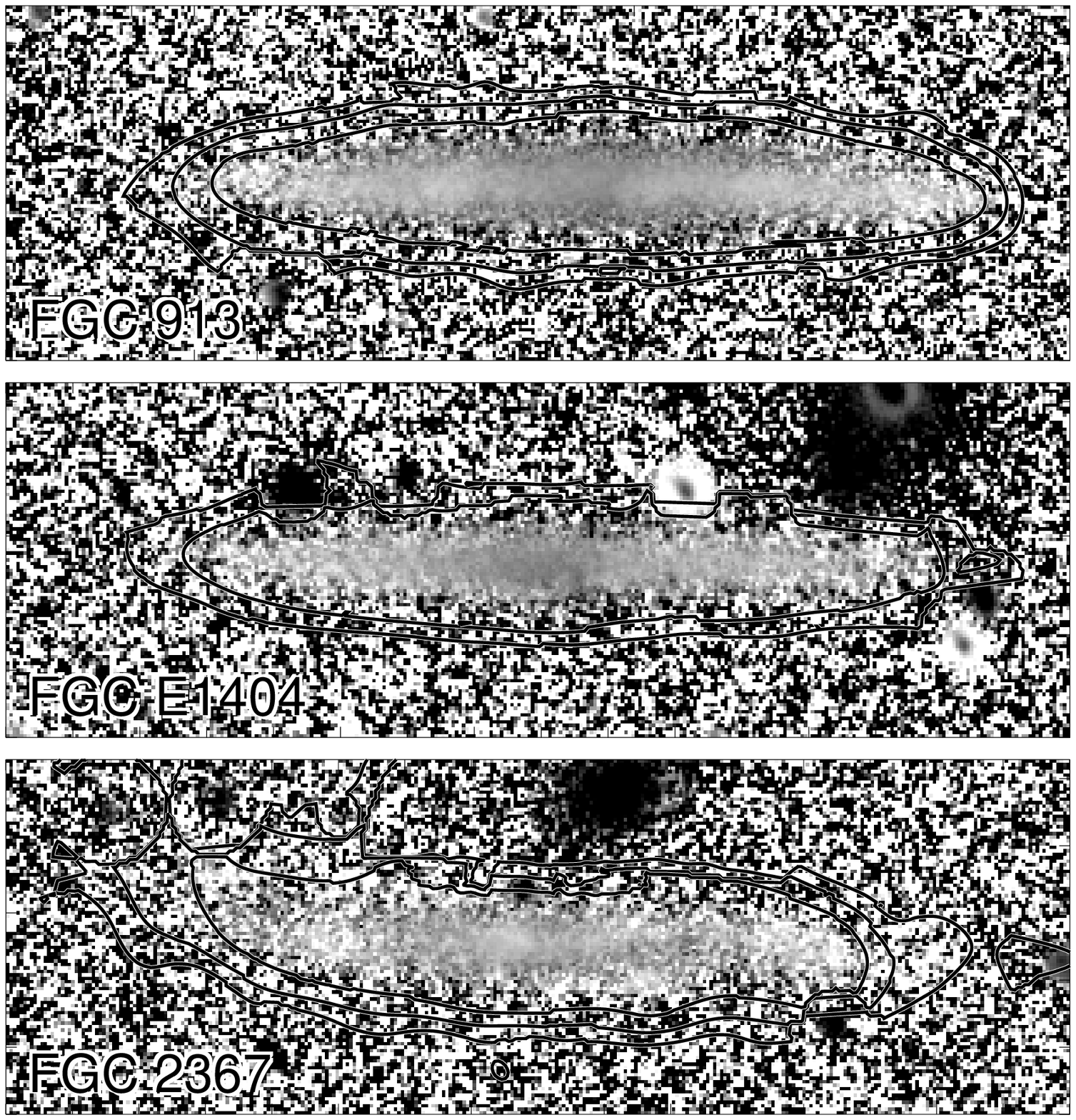}
\includegraphics[width=3.5in]{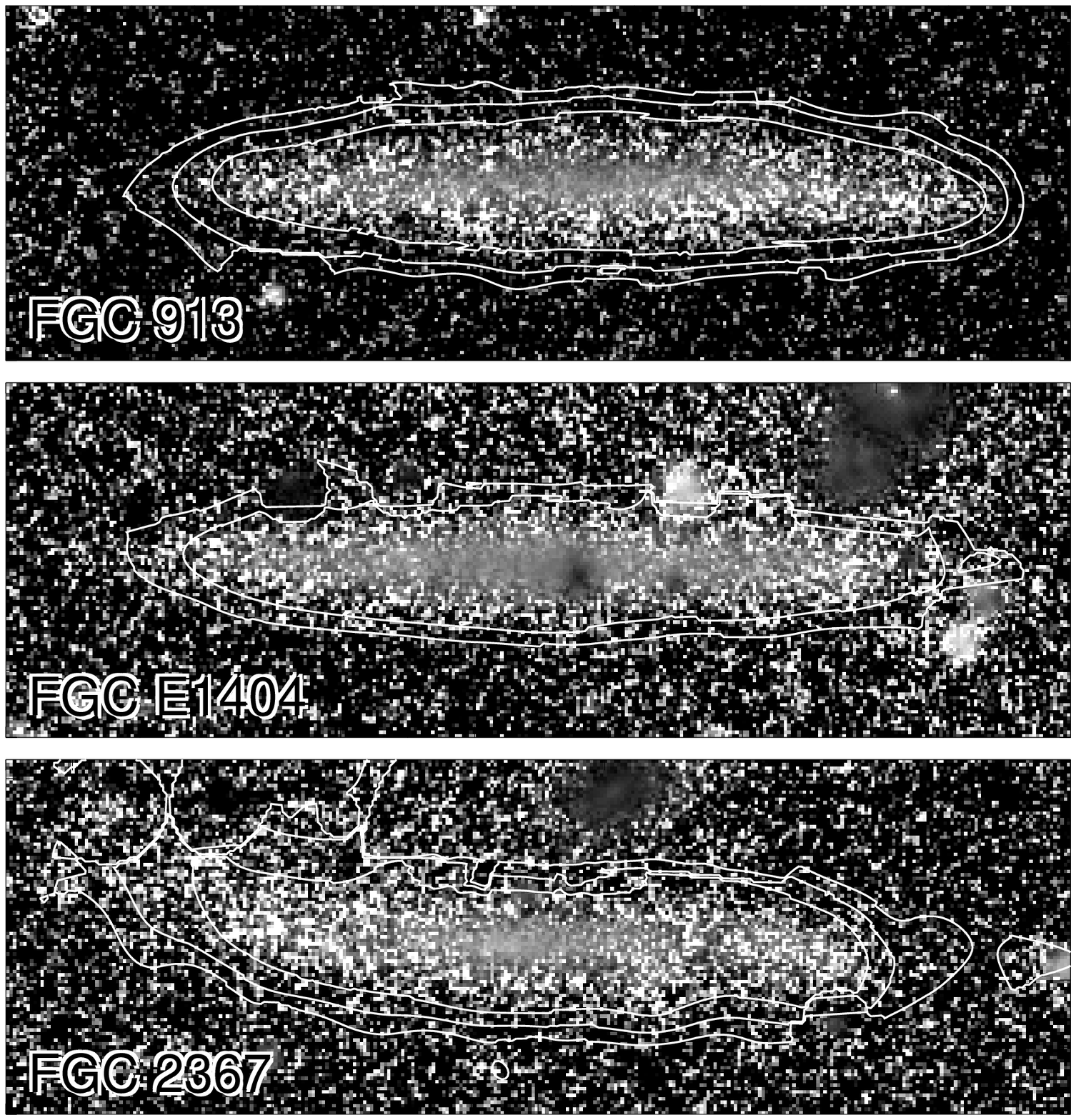}}
\caption{\footnotesize (continued)}
\end{figure*}

\setcounter{figure}{0}
\begin{figure*}[t]
\hbox{ 
\includegraphics[width=3.5in]{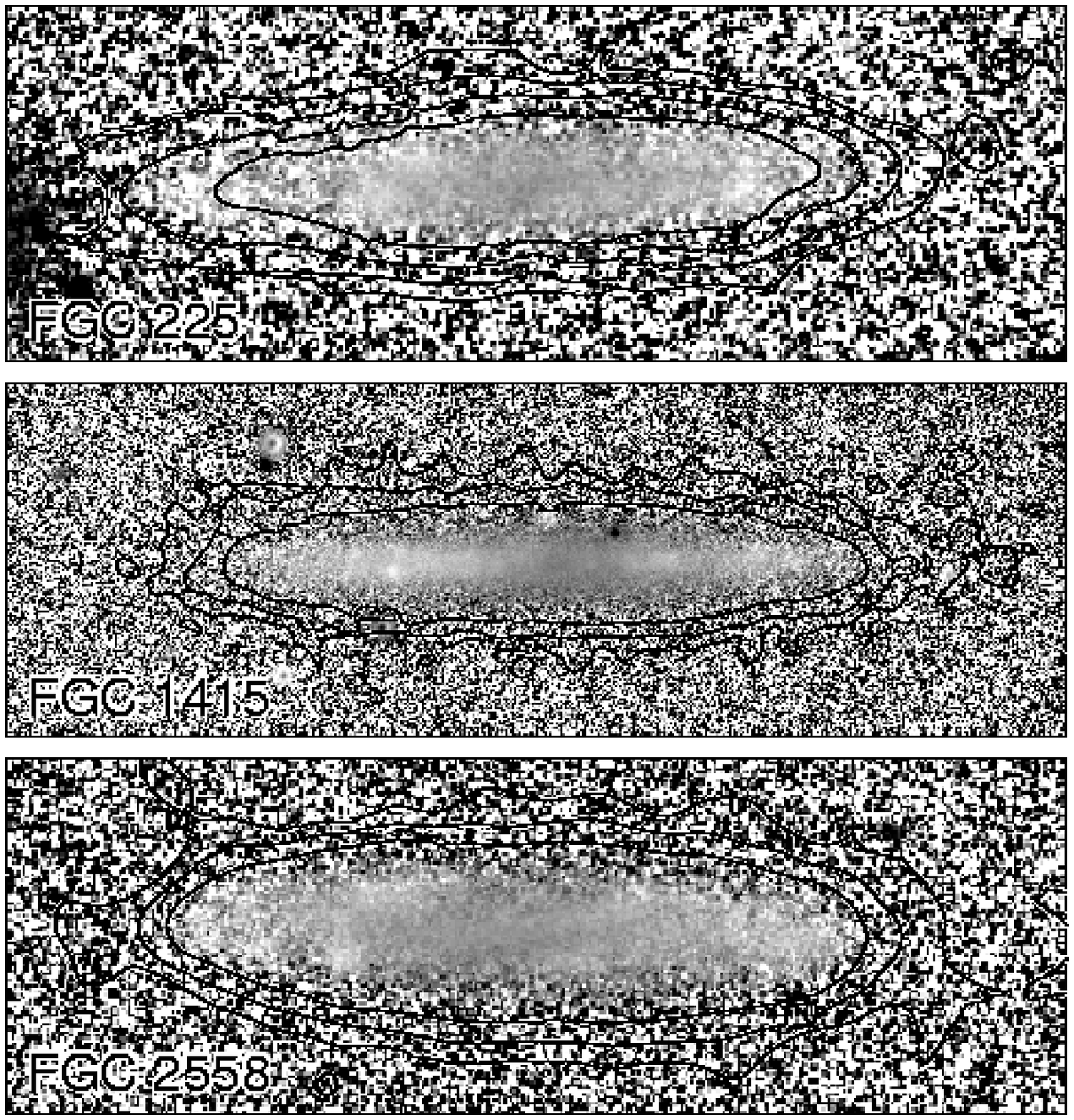}
\includegraphics[width=3.5in]{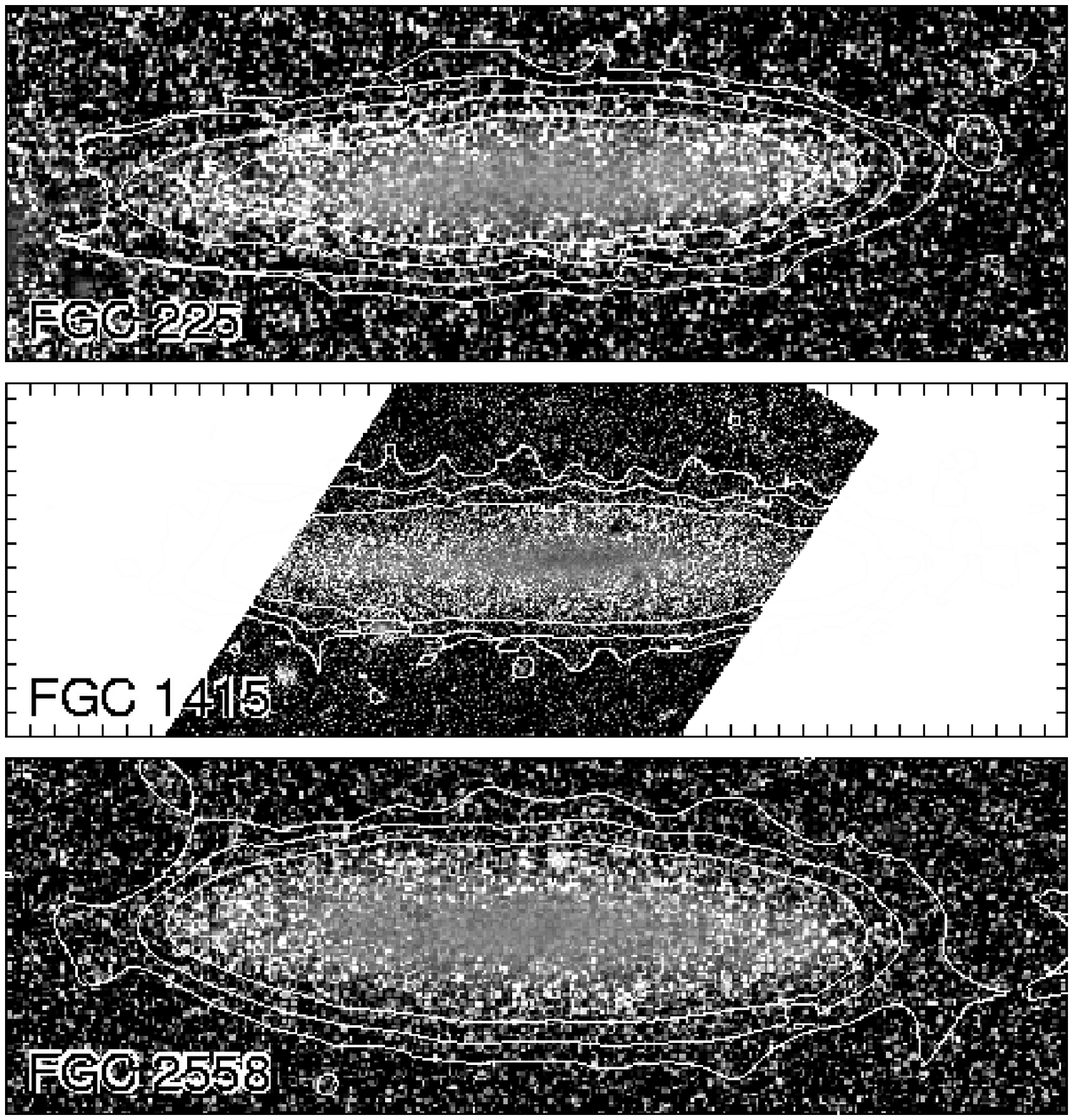}}
\vspace{0.1in}
\hbox{
\includegraphics[width=3.5in]{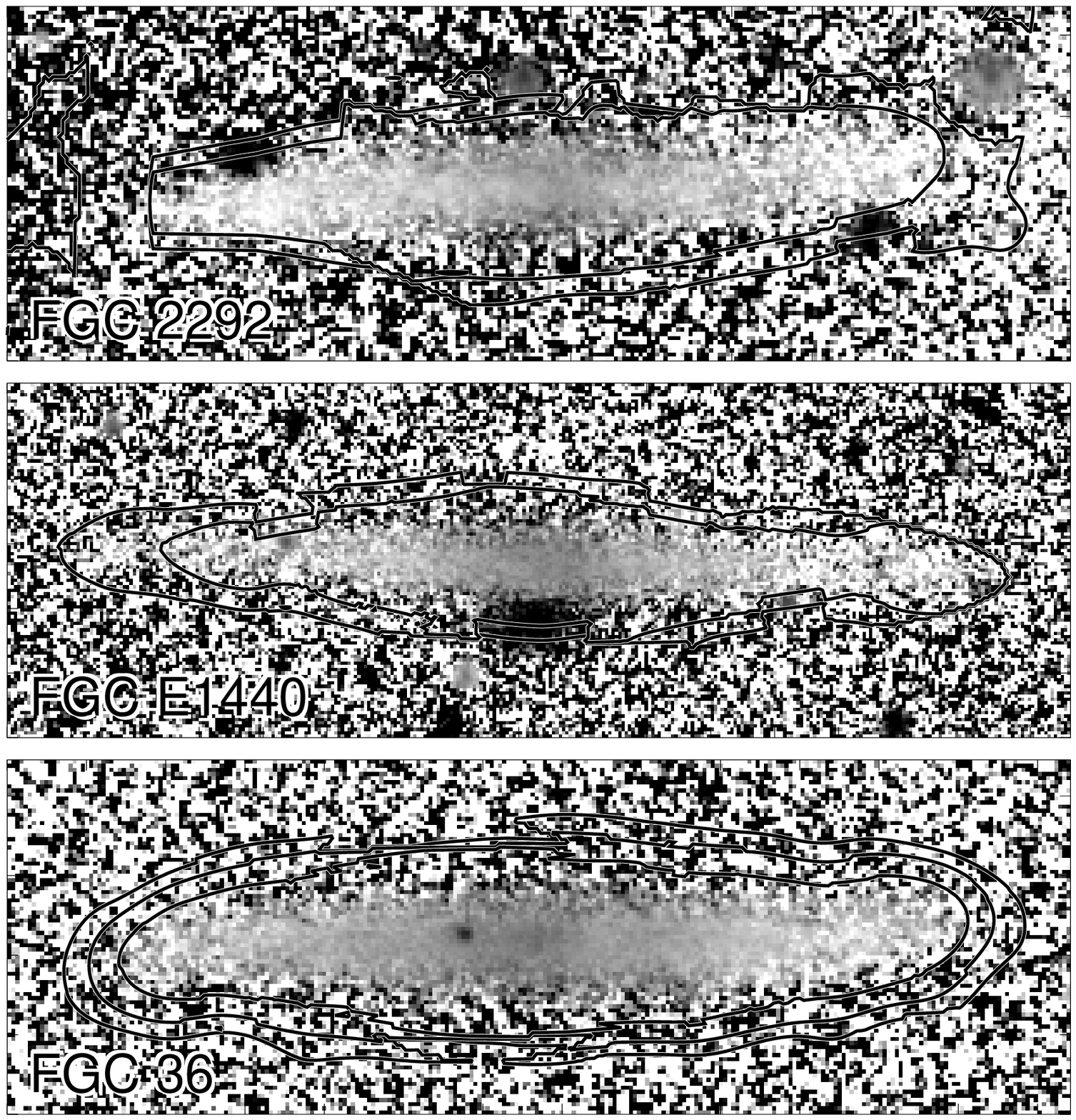}
\includegraphics[width=3.5in]{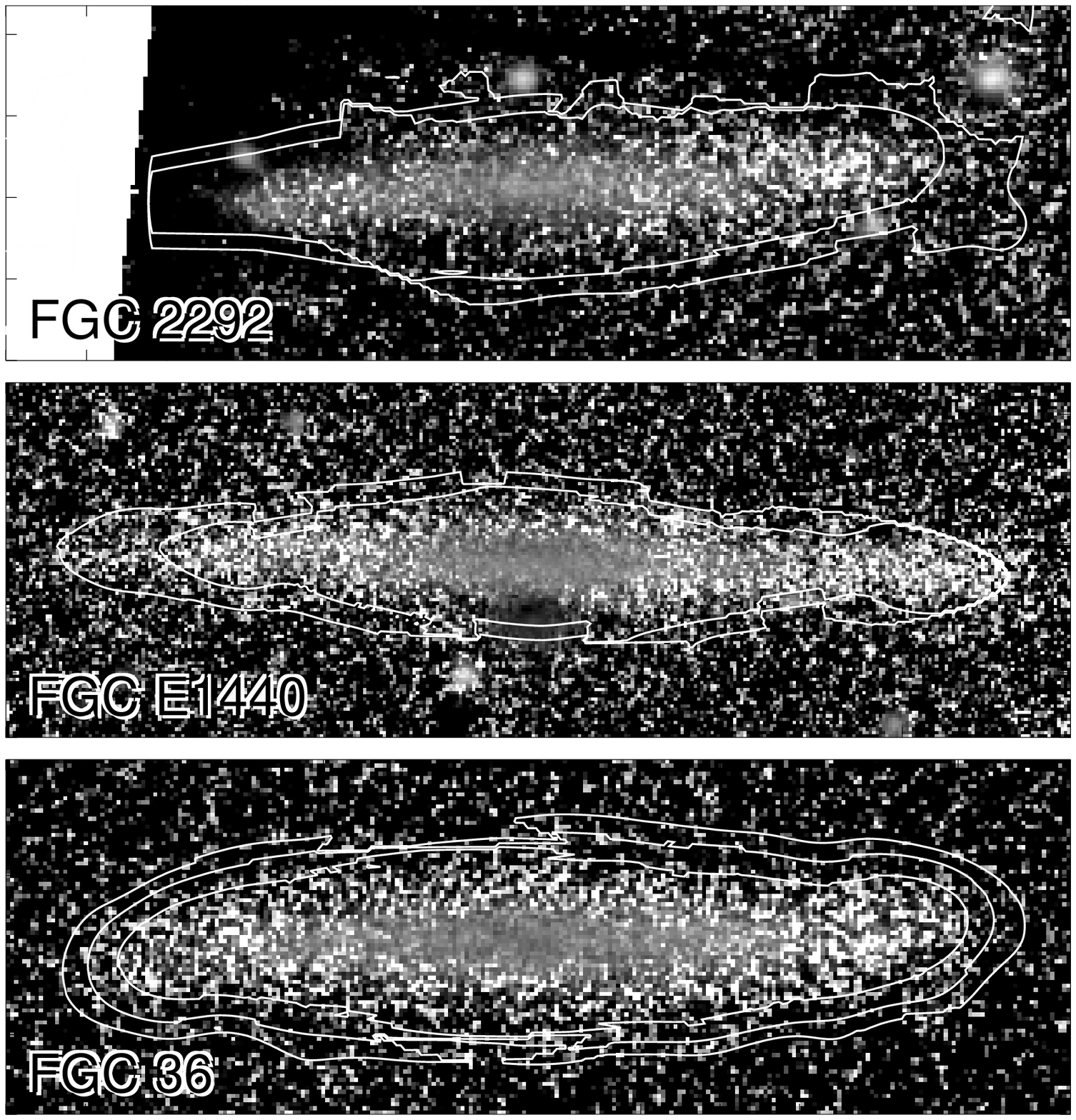}}
\caption{\footnotesize (continued)}
\end{figure*}

\setcounter{figure}{0}
\begin{figure*}[t]
\hbox{ 
\includegraphics[width=3.5in]{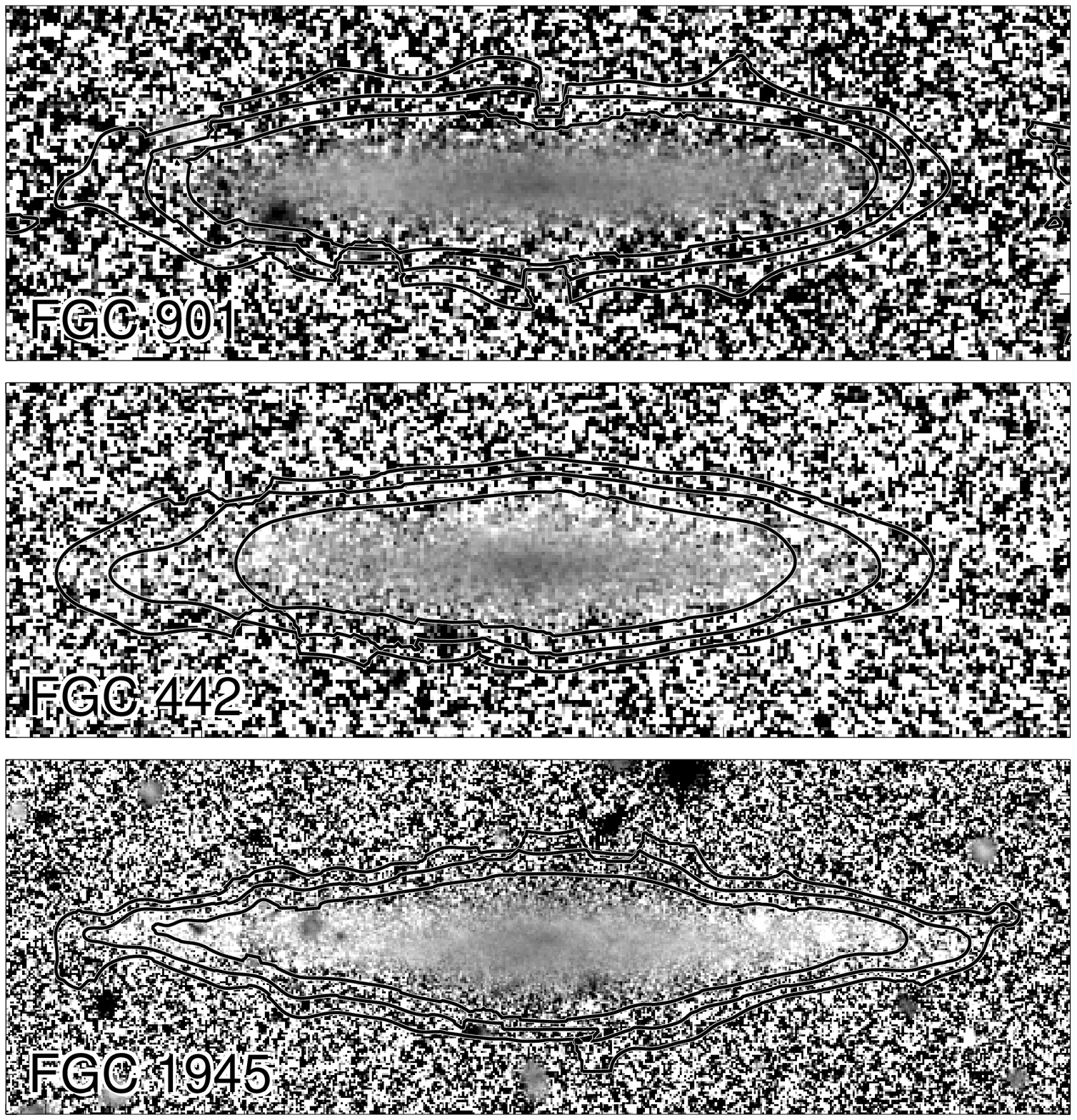}
\includegraphics[width=3.5in]{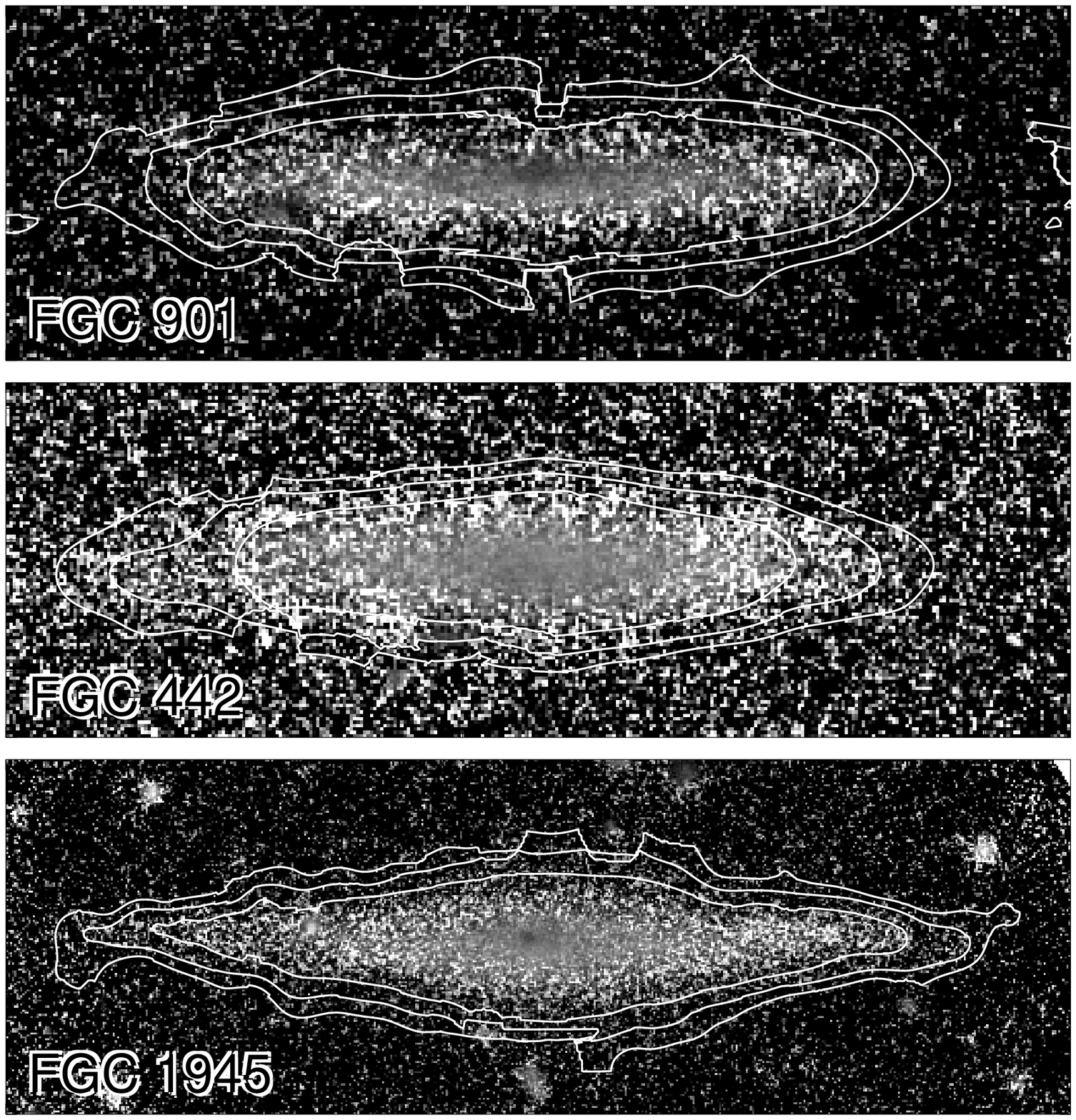}}
\vspace{0.1in}
\hbox{
\includegraphics[width=3.5in]{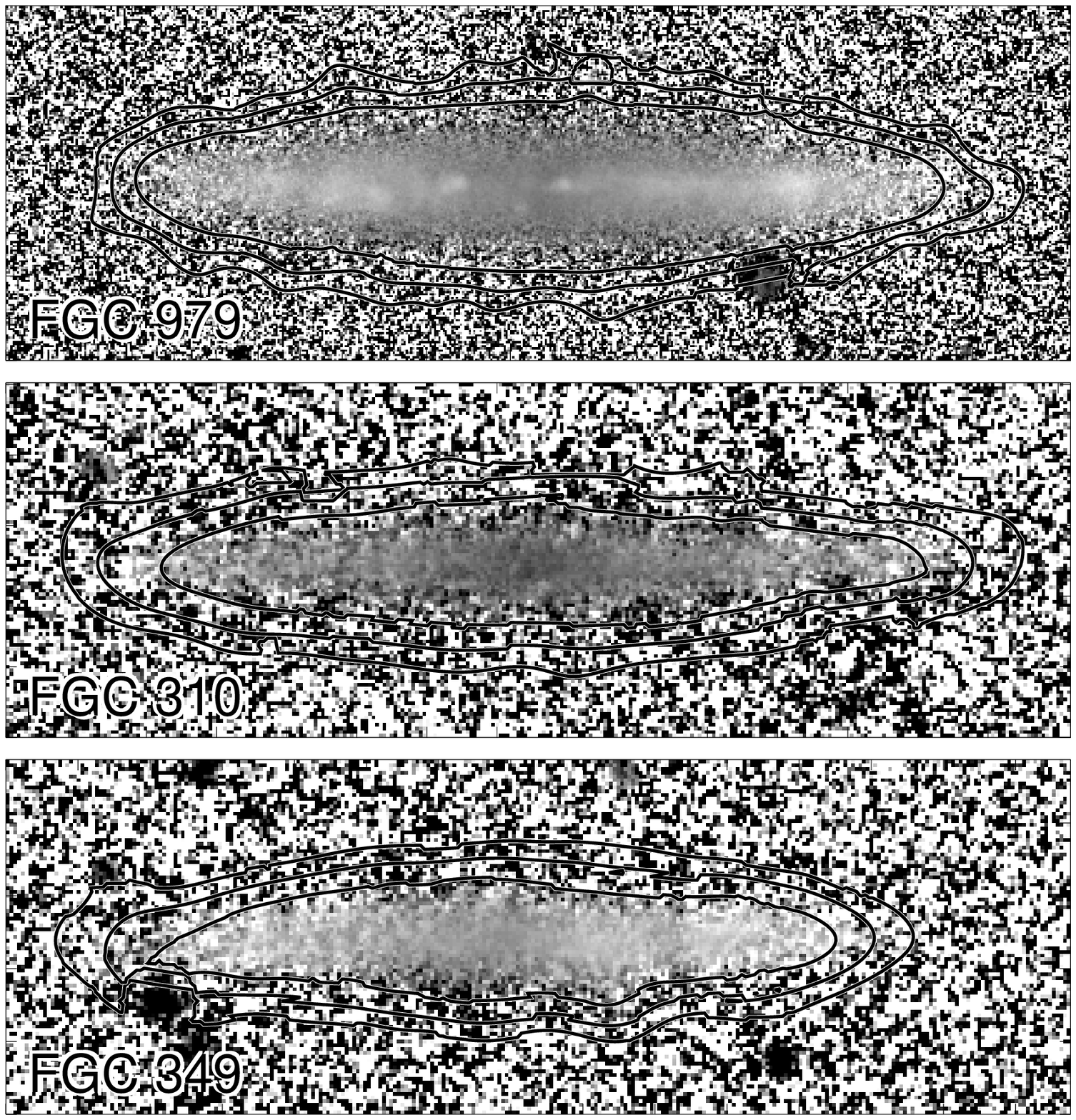}
\includegraphics[width=3.5in]{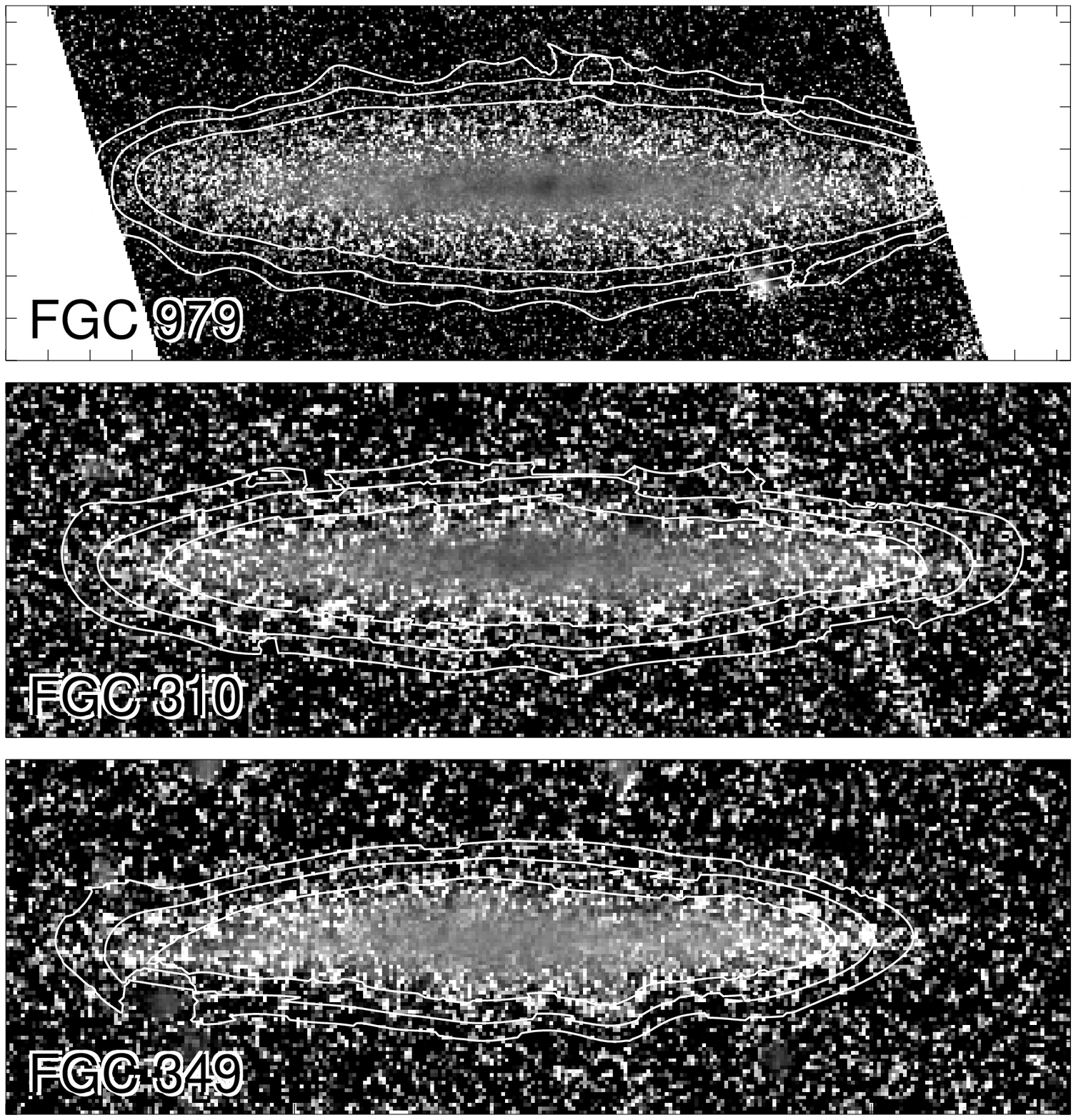}}
\caption{\footnotesize (continued)}
\end{figure*}

\setcounter{figure}{0}
\begin{figure*}[t]
\hbox{ 
\includegraphics[width=3.5in]{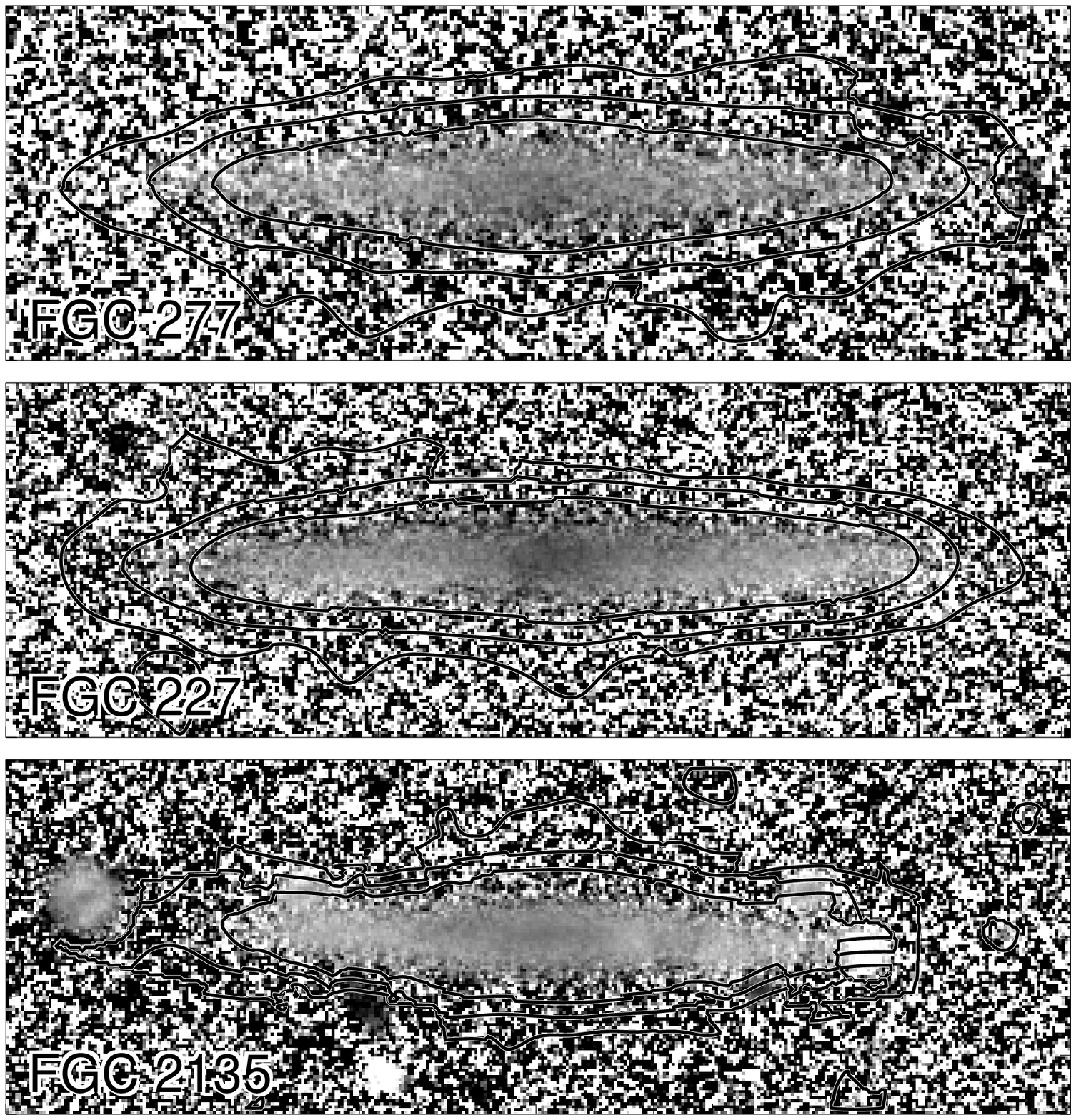}
\includegraphics[width=3.5in]{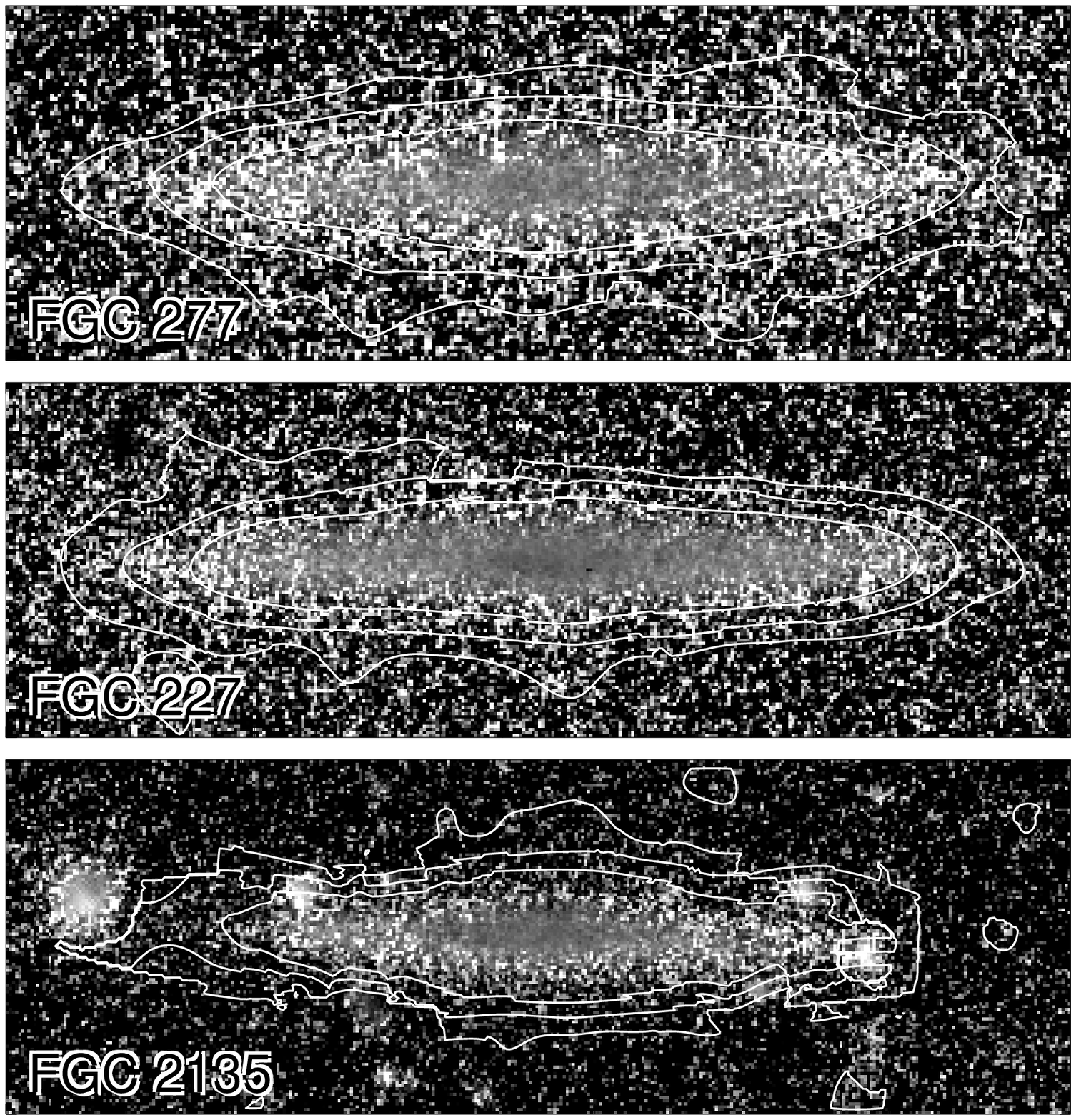}}
\vspace{0.1in}
\hbox{
\includegraphics[width=3.5in]{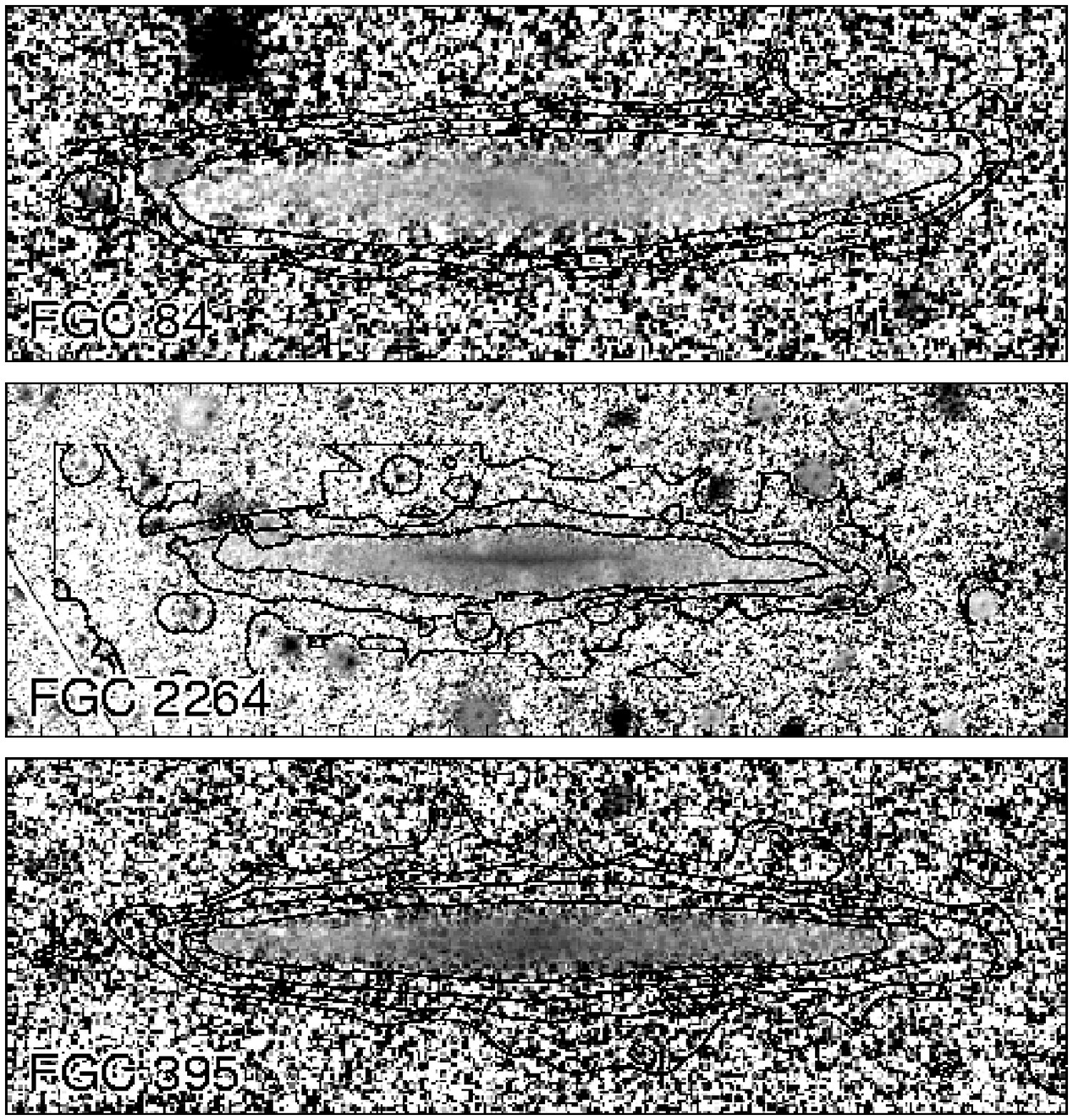}
\includegraphics[width=3.5in]{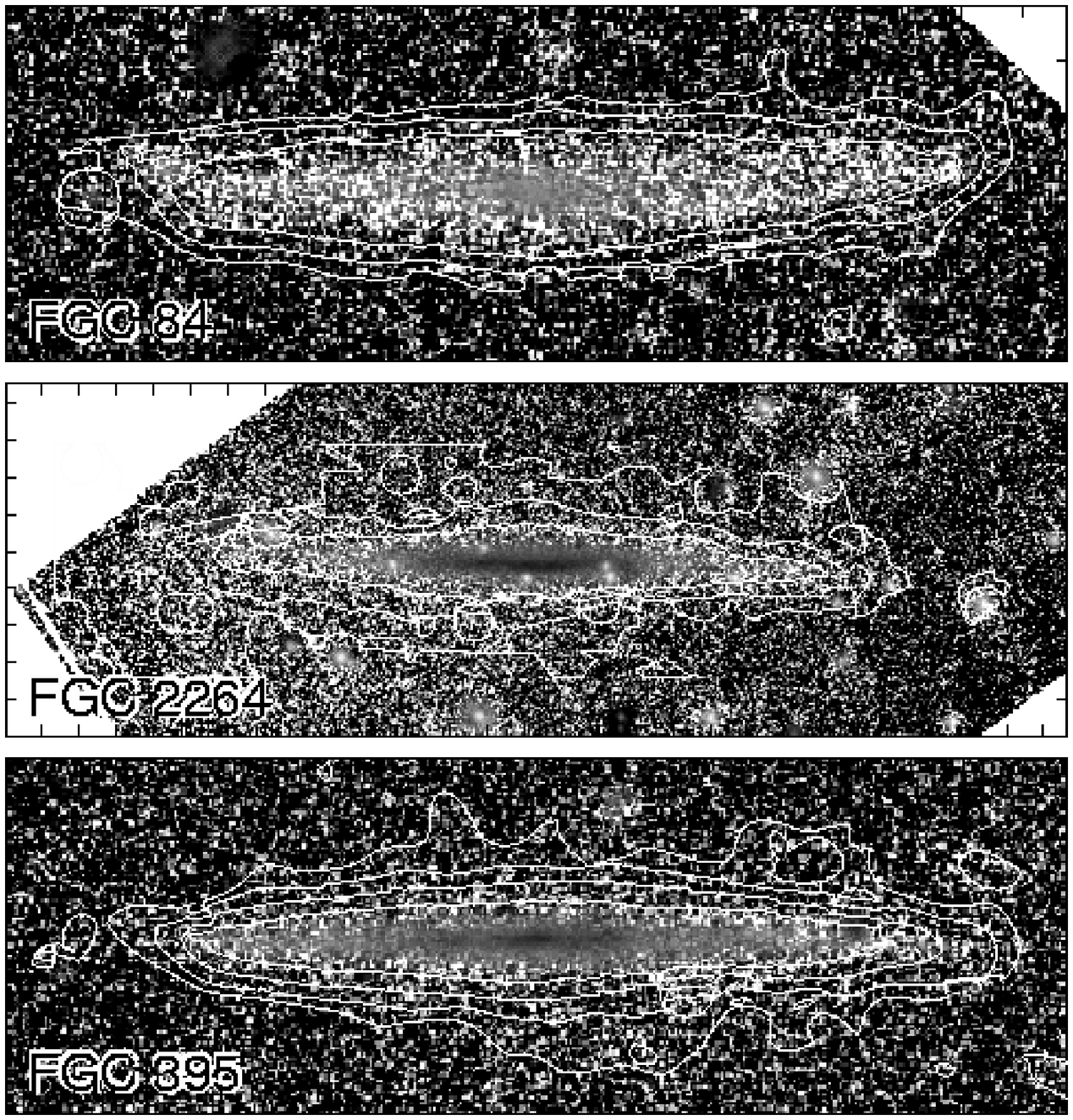}}
\caption{\footnotesize (continued)}
\end{figure*}

\setcounter{figure}{0}
\begin{figure*}[t]
\hbox{ 
\includegraphics[width=3.5in]{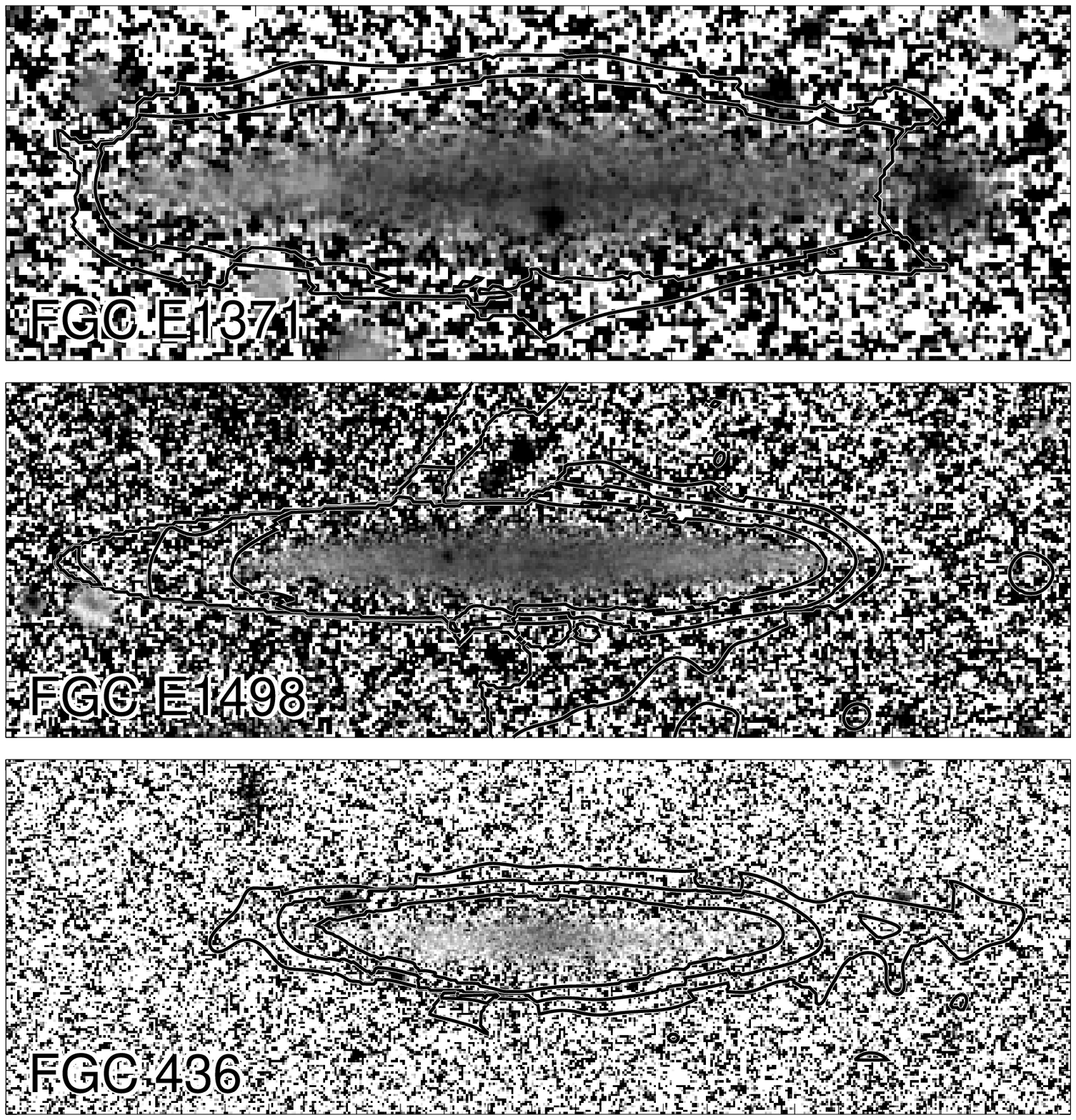}
\includegraphics[width=3.5in]{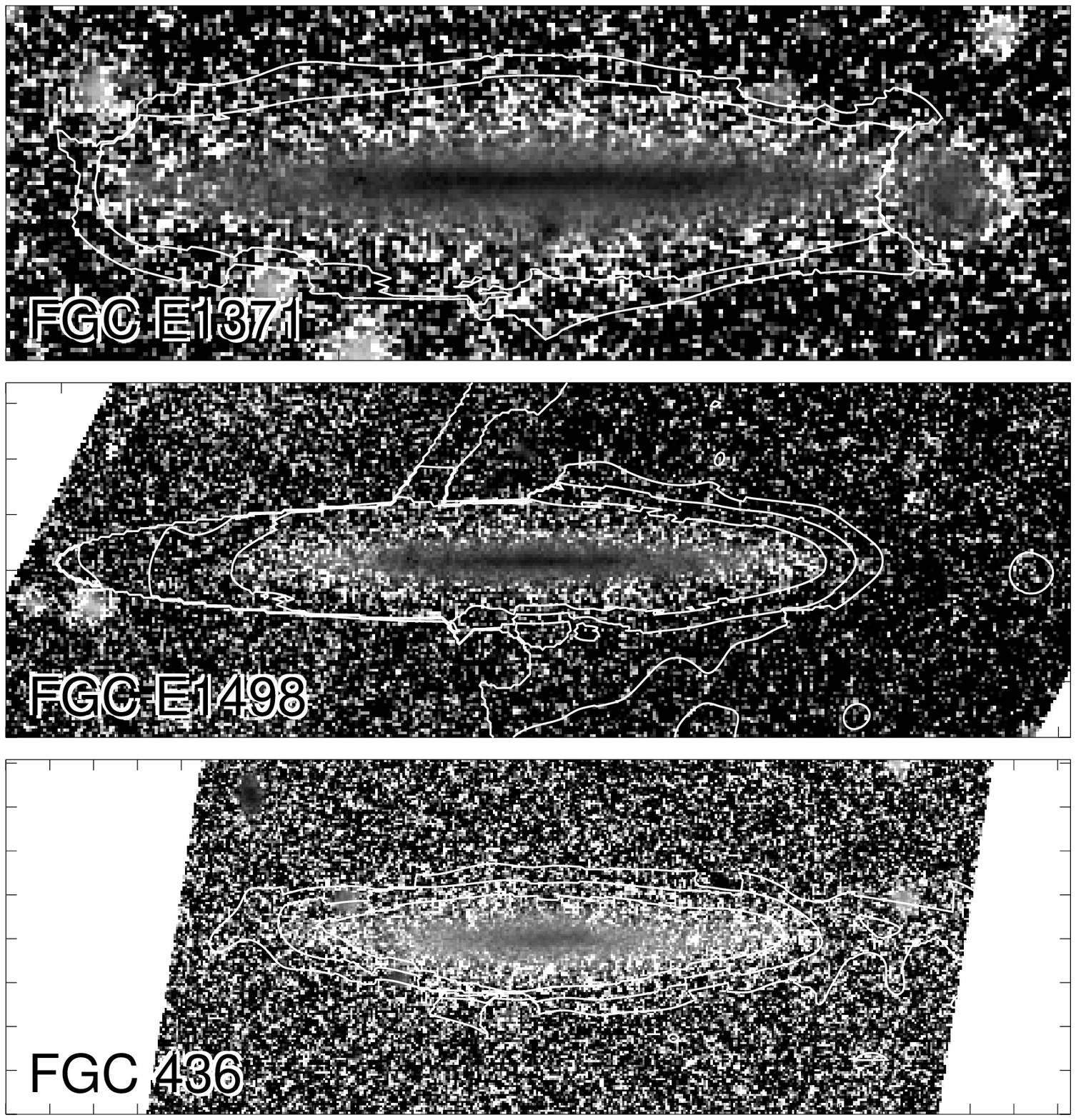}}
\vspace{0.1in}
\hbox{
\includegraphics[width=3.5in]{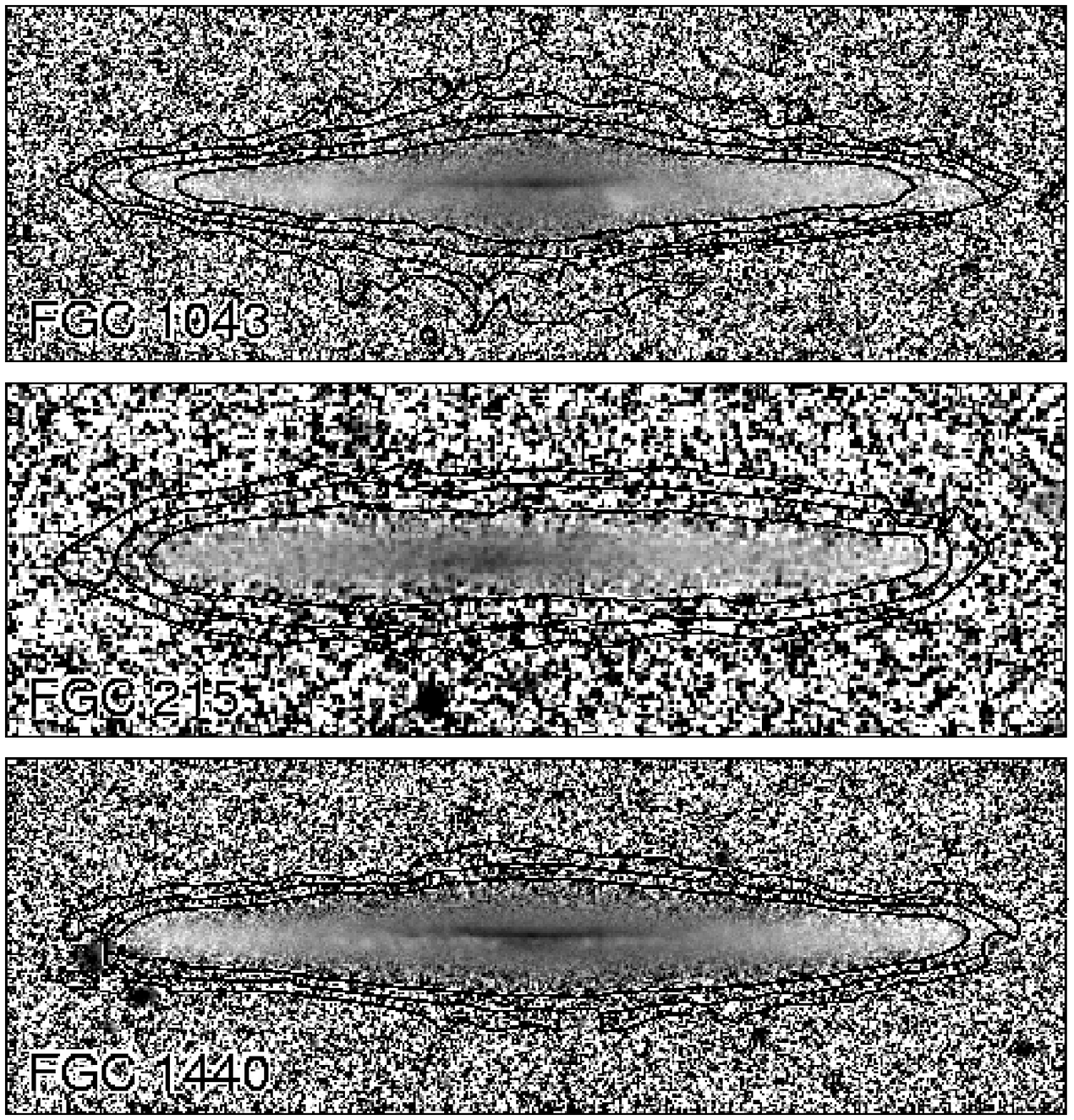}
\includegraphics[width=3.5in]{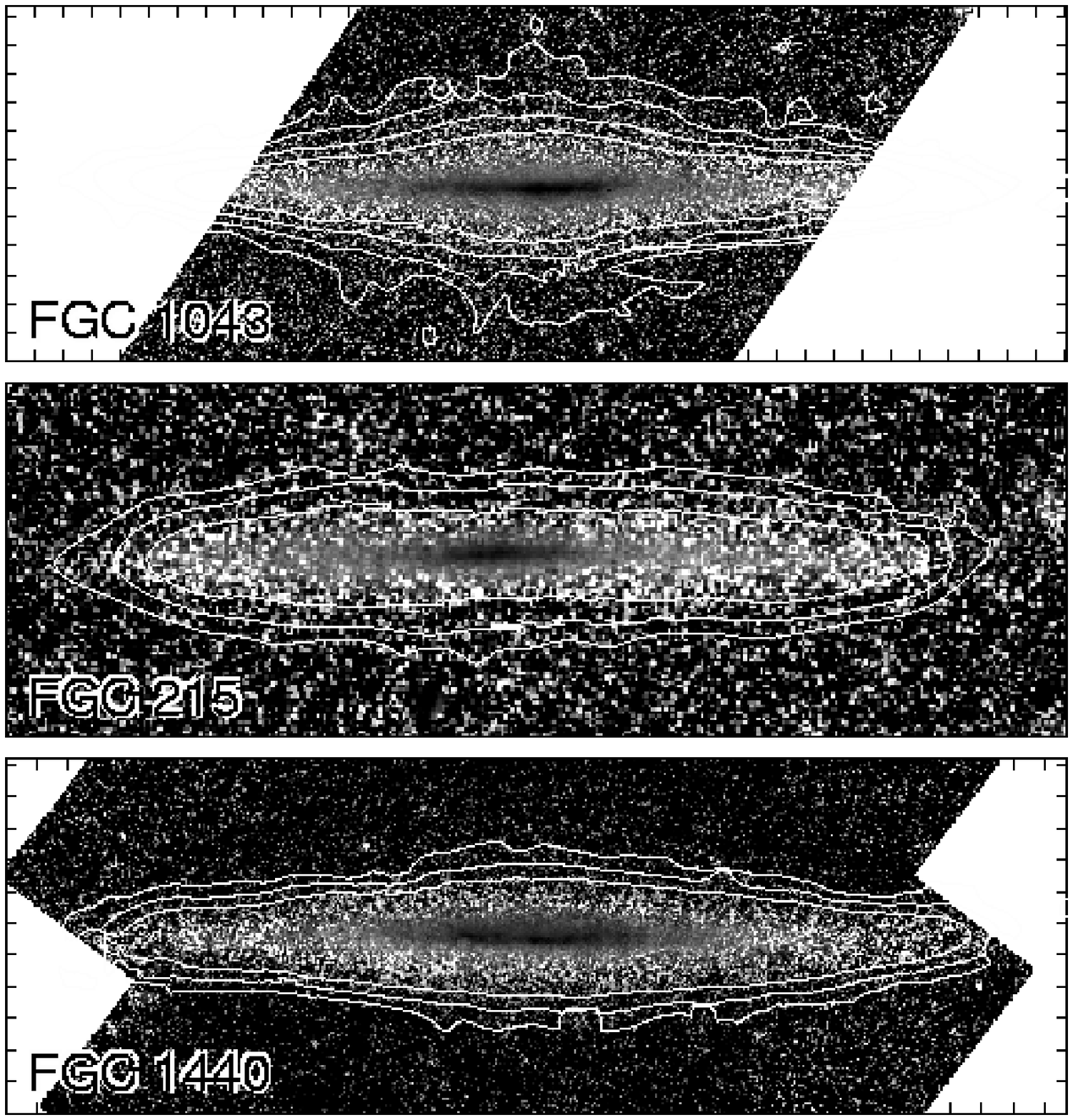}}
\caption{\footnotesize (continued)}
\end{figure*}

\setcounter{figure}{0}
\begin{figure*}[t]
\hbox{ 
\includegraphics[width=3.5in]{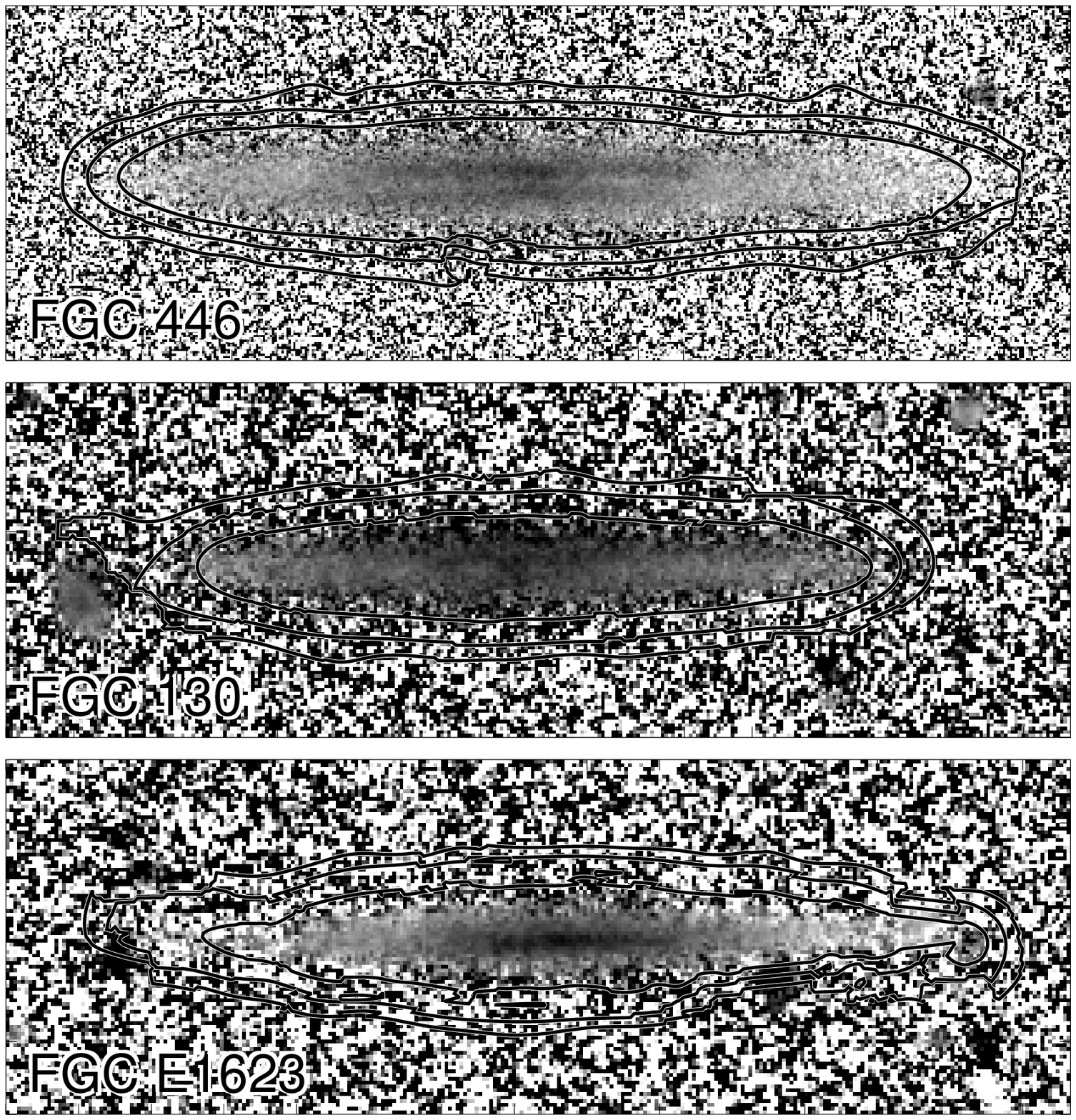}
\includegraphics[width=3.5in]{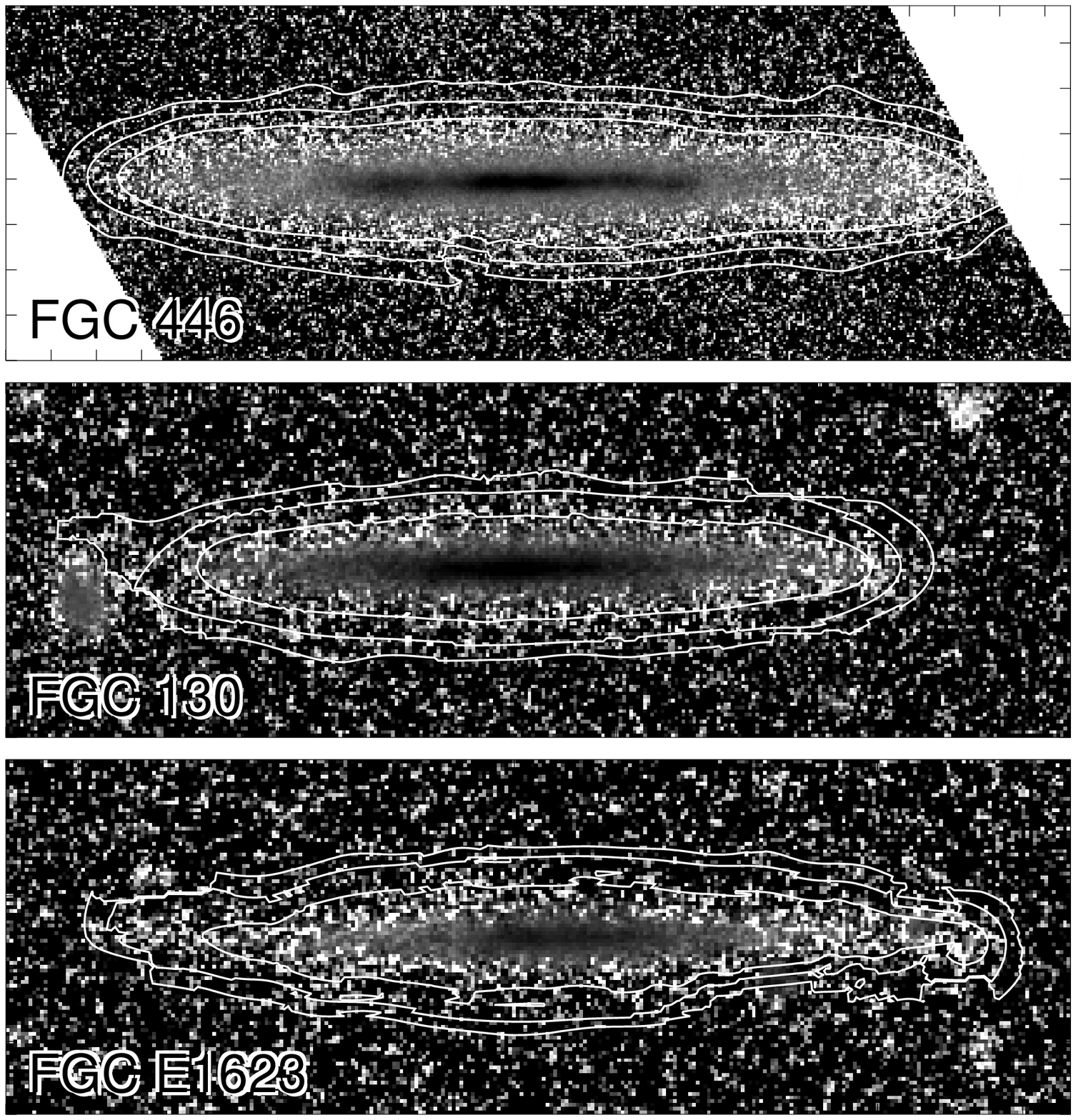}}
\vspace{0.1in}
\hbox{
\includegraphics[width=3.5in]{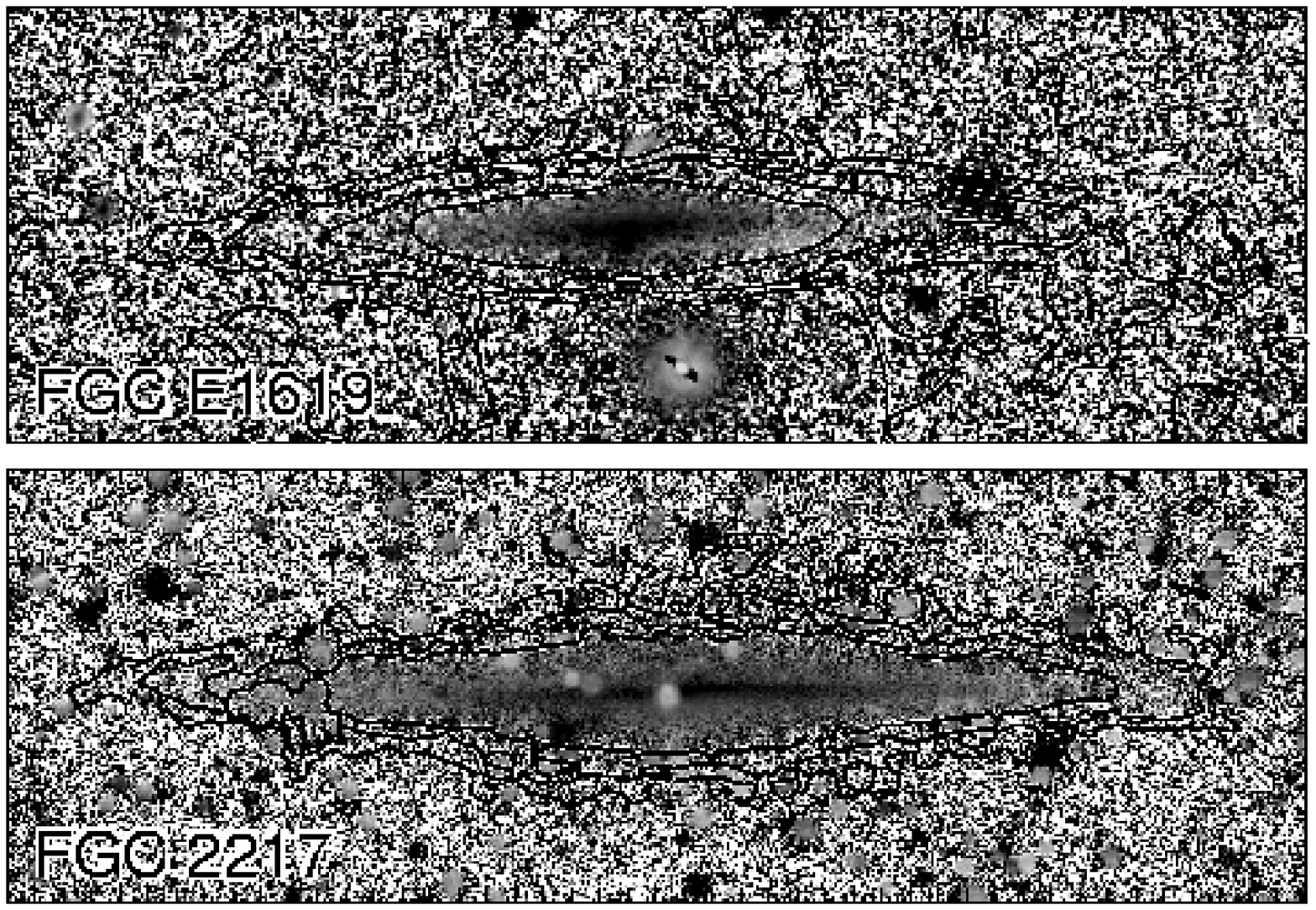}
\includegraphics[width=3.5in]{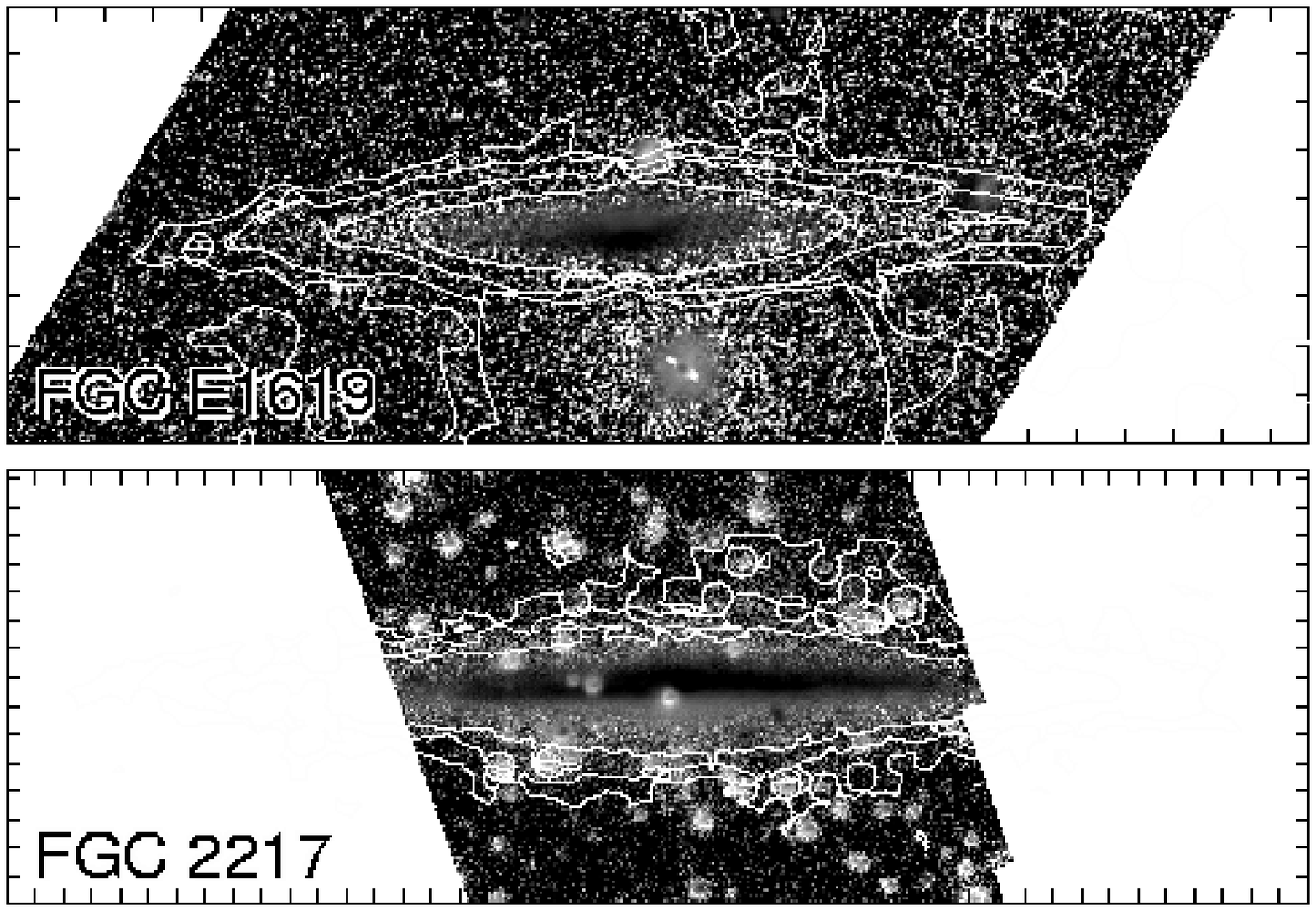}}
\caption{\footnotesize (continued)}
\end{figure*}
\clearpage

\subsection{The Presence of Dust}               \label{dustsec}

\begin{figure}[t]
\includegraphics[width=3.5in]{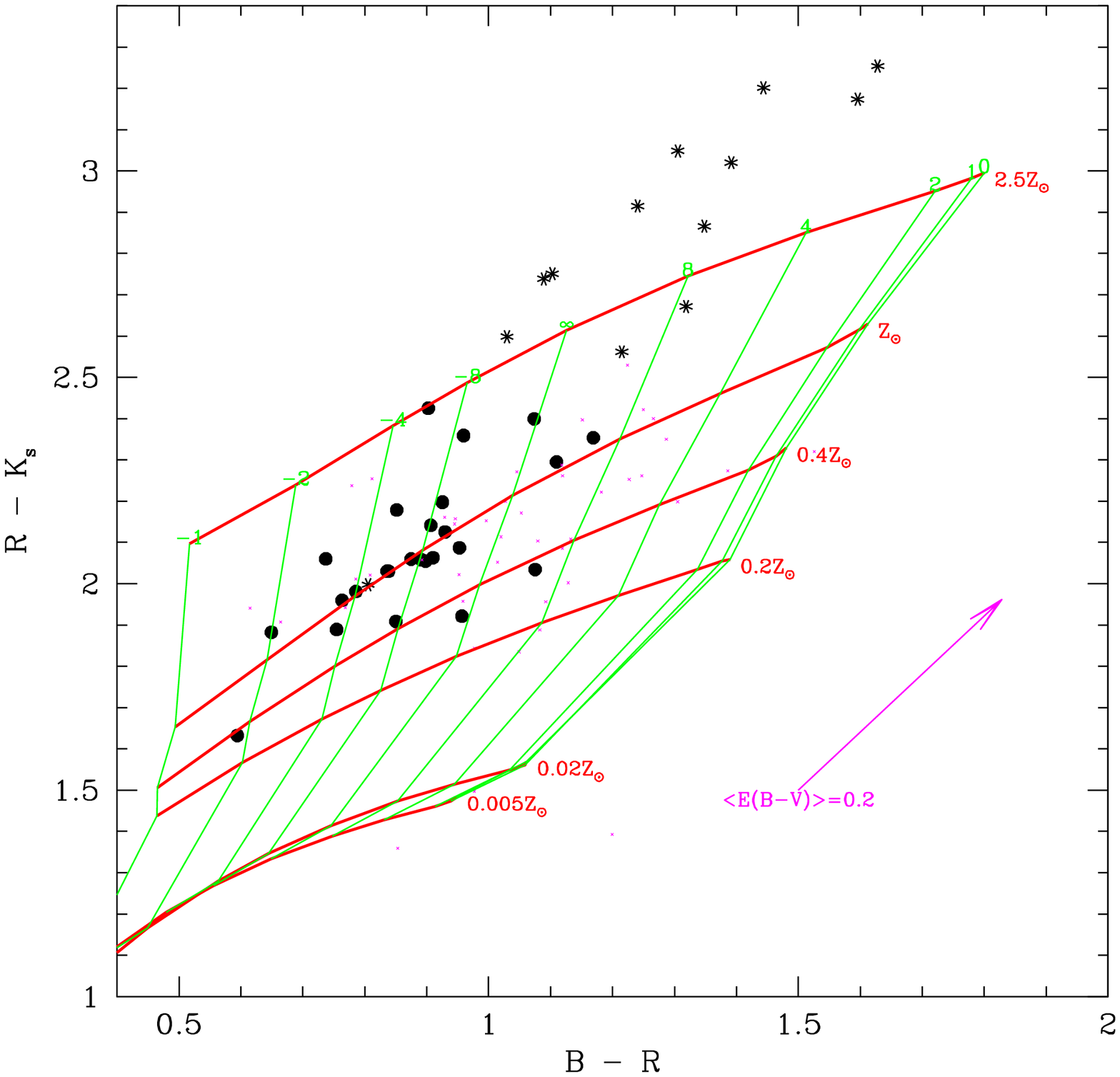}
\caption{\footnotesize 
$B-R$ and $R-K_s$ color-color plot of the
total magnitudes of the FGC galaxies with and without dust lanes
(asterix and solid circles, respectively).  The small dots are the
total disk colors of the face-on spirals from de Jong \& van der Kruit
(1994), assuming $K_s \approx K$.  The edge-on FGC galaxies
\emph{which do not have dust lanes} span a nearly identical range of
color as the face-on galaxies, suggesting that internal extinction
does not produce a substantial color change in the bulk of the edge-on
sample.  The overlayed grid (see \S\ref{stellarpopsec} and
Figure~\ref{gridgradfig} for details) is from Bruzual \& Charlot
(2001) models, for lines of constant metallicity (roughly horizontal
lines, [Fe/H] = -2.3, -1.7, -0.7, -0.4, 0, +0.4) and constant
exponentially declining star formation rates (roughly vertical lines,
$\tau = 0,1,2,4,8,\infty\Gyr$) assuming that star-formation began
$12\Gyr$ ago.  For reference, the effect of foreground screen
extinction with $E(B-V)=0.2$ is indicated by the vector in the lower
right.  FGC data points have been restricted to galaxies with
uncertainties in $B-R$ and $R-K$ of less than 0.25 mag and 0.5 mag
respectively.
\label{gridfig}}
\end{figure}

We can place some constraints on the {\emph{overall}} dust content of
the galaxies by comparing the global colors of the edge-on FGC sample
with comparable face-on galaxies.  We would expect that edge-on
galaxies with substantial obscuration would be reddened, and thus
offset in color from similar late-type disks viewed face-on.  In Figure
\ref{gridfig} we plot the $B-R$ vs $R-K_s$ colors of our FGC sample
(based on flux within the 25 $R\surfb$ isophote\footnote{These colors are
  slightly different than those derived from the magnitudes published
  in Paper I.  Here, we have eliminated masked regions from the
  determination of the color, whereas the magnitudes in Paper I have
  used models of the galaxies to attempt to recover light missing from
  the masked regions.}; solid circles) compared with the total disk
colors of the face-on spirals from de Jong (1996; small crosses),
measured using the derived disk central surface brightness and scale
lengths in $B$, $R$, and $K$, assuming that the magnitude zero point
offset between $K$ and $K_s$ ($\sim\!0.05$) is smaller than the
photometric errors.

The edge-on FGC galaxies with obvious dust lanes have been plotted
with an asterix, and, as expected, lie significantly redward of the
face-on disks.  However, the vast majority of the remaining FGC
galaxies have colors which are indistinguishable from the face-on
disks.  If anything, they are bluer.  This suggests that in the
galaxies without dust lanes, there either must be very little dust
overall, or the dust must be distributed in clumps which are optically
thick even in $K_s$, so that only unreddened light escapes the galaxy.
In either case, Figure~\ref{gridfig} suggests that the overall colors
we measure are only marginally affected by dust (even if the measured
luminosity is too low).  In \S\ref{stellarpopcaveatsec} below, we
discuss the role that dust may play on the derived gradients in more
detail.

\section{Extraction of Vertical Color Profiles} \label{extractionsec}

Of the qualitative trends discussed above, we concentrate hereafter on
quantifying and interpreting only the vertical color gradients.  We
begin in this section by extracting the surface brightness
profiles perpendicular to the galaxy midplanes.  

Because the galaxies in our sample span a wide range in physical
scales, we characterize the color gradients as a function of the
galaxies' scale length and height rather than fixed physical units
(i.e. kiloparsecs).  We derive a single scale height and midplane
position for each galaxy by fitting vertical profiles across the
galaxy.  For the analysis which follows, we have adopted a single
value for the location of the midplane $z_{cen}$ and for the vertical
scale height\footnote{For an exponential vertical surface brightness
distribution of scale height $h_z$, as is commonly assumed,
$h_z\approx 0.6z_{1/2}$.}  $z_{1/2}$ by averaging the values of
$z_{cen}(R)$ and $z_{1/2}(R)$ within one disk scale length on either
side of the center.  Over this range in radius, the values of
$z_{cen}$ and $z_{1/2}$ do not vary significantly in any systematic
way, as the galaxies possess no strong warps in their central regions,
and are nearly uniform thickness in the infrared.  We do this fitting
using the masked $K_s$ images, which are representative of the old
stellar population, less affected by dust, and less contaminated by
bright foreground or background objects.

The vertical profiles are derived in narrow bins of projected radius
$R$ by fitting a generalized ${\rm sech}^{2/N}$ profile $f(z) = f_0
2^{-2/N}{\rm sech}^{2/N}(\frac{N(z-z_{cen})}{2 z_0})$ (van der Kruit
1988) to the surface brightness distribution.  From these fits we
derive the midpoint $z_{cen}(R)$ of the disk and the vertical scale
height $z_{1/2}(R)$ such that half of the flux is contained within
$\pm z_{1/2}$.  In a few of the lowest mass galaxies, the $K_s$ band
surface brightness is sufficiently low that the fits do not converge;
in these handful of cases we adopt the $R$-band midplane and
scale height instead.  The resulting fits are integrated over $z$ to
derive the total ``vertical flux'' at each projected radius. Finally,
the disk scale length $h$ is identified by fitting an optically-thin
edge-on exponential disk profile: $f(R) = f_0 \frac{R}{h} K_1(R/h)$,
where $K_1(x)$ is a modified Bessel function (van der Kruit \& Searle
1981a).  A full analysis of these structural parameters will be
presented in a separate paper.

Using the adopted midplane location, we extract the mean flux from the
sky-subtracted $B$, $R$, and $K_s$ images in logarithmically spaced
bins parallel to the midplane.  We average data above and below the
plane to maximize our signal-to-noise at faint light levels, but
repeat the analysis independently on each side to quantify potential
systematic offsets (see below).  We mask out foreground and background
sources, and then calculate a mean height for each bin based on the
unmasked pixels.  The same mask is used in all three bandpasses, so
that colors are determined using identical apertures.  We have not
attempted to match the seeing between bands, because (i) the seeing
was almost always comparable in all observations (0.8-1.1$\arcsec$),
(ii) the scale heights of the disks are substantially larger than the
point spread function, and (iii) the gradients we are tracking vary
smoothly over several disk scale heights, and will not be affected by
10-20\% variations in the seeing (see also \S\ref{extractionsec} and
Figure~\ref{seeingfig} below).

The color in each bin is calculated from the flux through two separate
bandpasses, and is corrected for foreground extinction
(\S\ref{colormapsec}).  The errors on individual points are taken to
be the quadrature sum of the Poisson photon counting errors in the
flux and the overall uncertainty in the sky level determination as
measured in Paper I.  The read noise and photometric calibration
errors are negligible compared to photon counting errors and the sky
level uncertainty, and are not propagated.  We note that the resulting
error bars do \emph{not} represent uncorrelated Gaussian random
errors, because errors in the sky level determination will produce
correlated, systematic offsets in the overall color profile. These
correlated errors due to sky subtraction dominate the uncertainties in
the extracted profiles at large scale heights. We take the correlated
errors into account when assessing the errors in the color gradients
in \S\ref{quantifysec} below.
  
We limit our analysis to the central region of the galaxy, within
$\pm1$ disk scale length from the center, where we have sufficient
signal-to-noise for measuring the color of the galaxy out to very
large scale heights ($\sim 8 z_{1/2}$ in $B-R$).  The color gradients
are only weakly varying with radius at the large scale heights which
interest us (see Figure~\ref{colormapfig}).  Variation of the color
gradients with radius is therefore dominated by the well-known radial
color gradient of the thin stellar disk, not the fainter populations
on which we focus here.

The resulting surface brightness profiles and color gradients are
shown in Figure \ref{profilefig}, sorted in order of increasing
rotation speed.  The top panel shows the mean surface brightness
profiles in $B$, $R$, and $K_s$ as a function of the number of
vertical scale heights above the plane.  The middle and lower panels
show the $R-K_s$ and $B-R$ profiles, respectively.  The profiles have
been plotted using the average values above and below the plane,
plotting points in $B$ and $R$ where the uncertainty
$\sigma_{B-R}<0.3$ mag, and points in $K_s$ where $\sigma_{R-K}<0.5$ mag.

Figure~\ref{profilefig} shows that, as suggested by
Figure~\ref{colormapfig}, we can indeed make significant measurements
of the galaxies' colors well above their midplanes.  We do not believe
that this extraplanar light we have detected is an artifact of the
point-spread function (PSF).  First, the imaging data were taken in
excellent conditions of $1\arcsec$ or better.  Second, the du Pont
2.5m has a well-characterized PSF with little large angle scattering
(see Bernstein et al.\ 2002, Bernstein \& Crick 2002).  To briefly
verify that the faint light in our images is not due to scattered
light, in Figure~\ref{seeingfig} we plot the radial surface brightness
profiles of stars in several representative images in each bandpass,
including examples of the best seeing and the worst seeing.  In all
cases, the surface brightness of the stellar point-spread function
(PSF) drops rapidly, falling by $6\surfb$ at 2 times the full-width
half max (FWHM) of a Gaussian fit to the PSF.  In contrast, the
surface brightness of our galaxies fall by $\Delta\mu\sim 6\surfb$
over more than ten times the FWHM of the stars in the field (marked
with a vertical arrow in Figure~\ref{profilefig}, indicating a
distance of $1\arcsec$ from the midplane of the galaxy).  The
extraplanar light is therefore not due to the PSF.

In addition to large angle scattering from seeing, spurious color
gradients could be produced by inaccuracies in the sky-subtraction
and/or flat fielding.  Flat-fielding errors would produce erroneous
color gradients that differed from one side of the galaxy to the
other, but which were reproduced in all galaxies from a given night.
Sky subtraction errors would lead the outskirts of galaxies to be too
blue or too red depending upon the degree of under- or
over-subtraction in different filters.  This would tend to lead to
color gradients with random signs.  As discussed in Paper I, the
flat-fielding and sky subtraction of this data set has been done with
great care, and we have developed new techniques for quantifying the
errors associated with these steps.  Our measured uncertainties have
been propagated into the error bars in Figure~\ref{profilefig}, and
demonstrate that the gradients are not due to flat-fielding or
sky-subtraction errors.

As an additional test for these large-scale uncertainties, we explore
the possibility of systematic errors by comparing the color gradients
derived above and below the plane.  In Figure \ref{allprofilesfig},
the shaded regions indicate the difference between the colors derived
on each side of the midplane, plotted in order of increasing rotation
speed.  In general, the colors are quite consistent between the two
sides, and show gradients that are much larger than the systematic
differences.  Occasional deviations are seen at the outermost points
in the $B-R$ profiles beyond $5z_{1/2}$, where the flux is very low,
and in the inner parts of the more massive galaxies in $R-K_s$, where
the combined effects of dust lanes and slight deviations from an
exactly $90\degree$ inclination are important.  These latter cases do
not suggest errors in measurement but rather a well-understood effect
that has been previously used to analyze the opacity of disks (Jansen
et al.\ 1994, Knapen et al.\ 1991).


\begin{figure*}[t]
\hbox{ 
\includegraphics[width=2.95in]{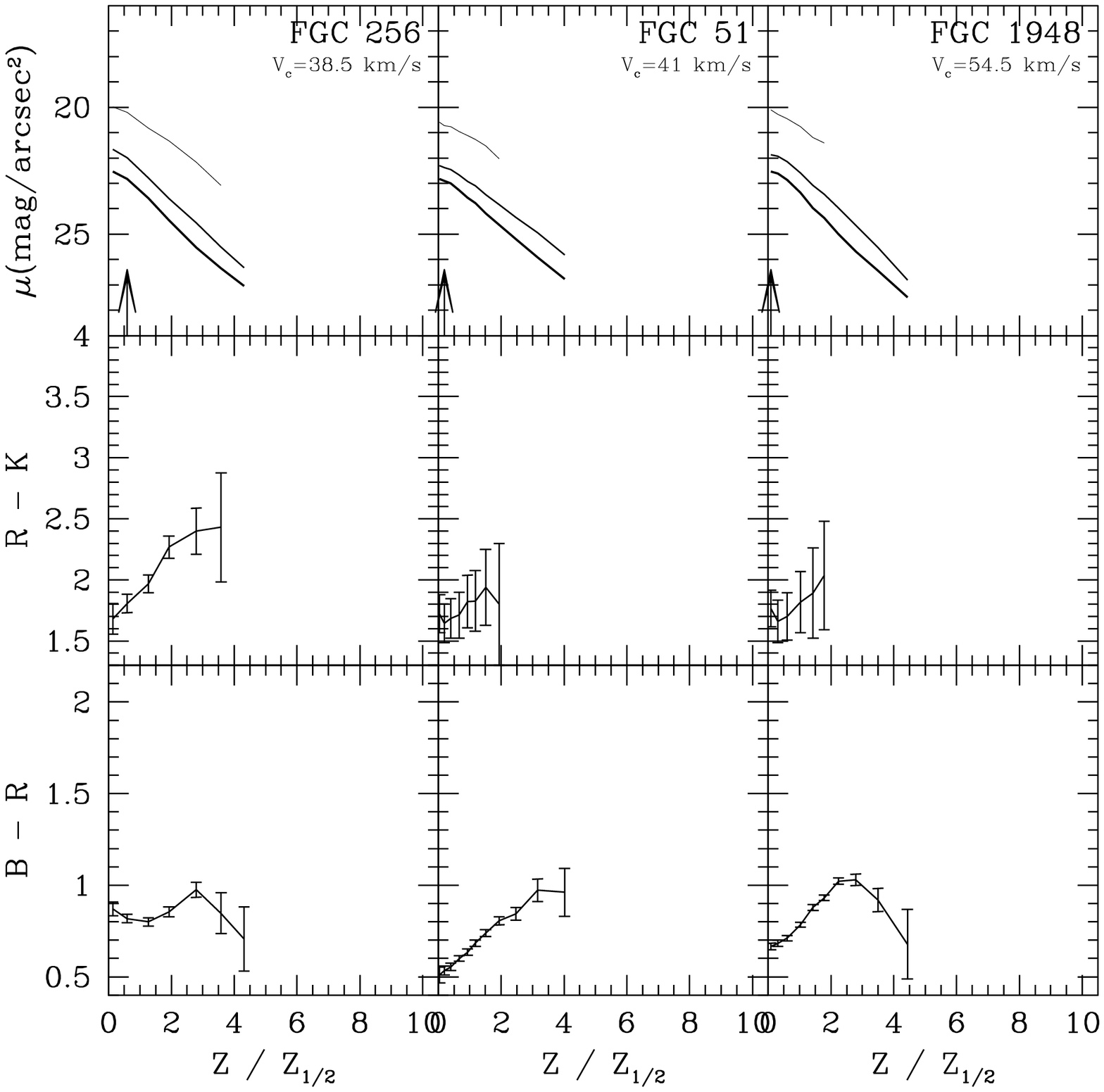}
\includegraphics[width=2.95in]{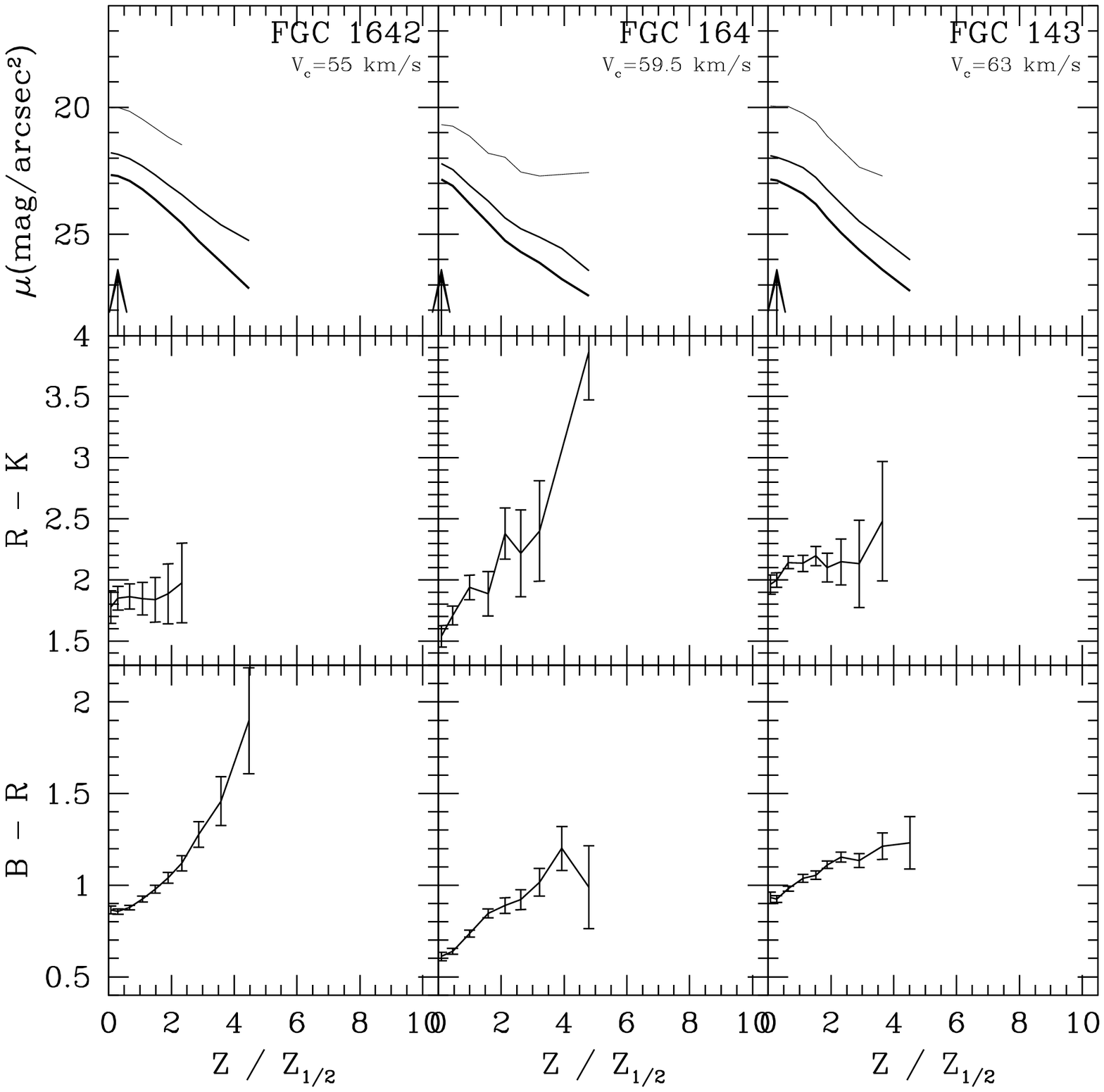}}
\vspace{0.1in}
\hbox{
\includegraphics[width=2.95in]{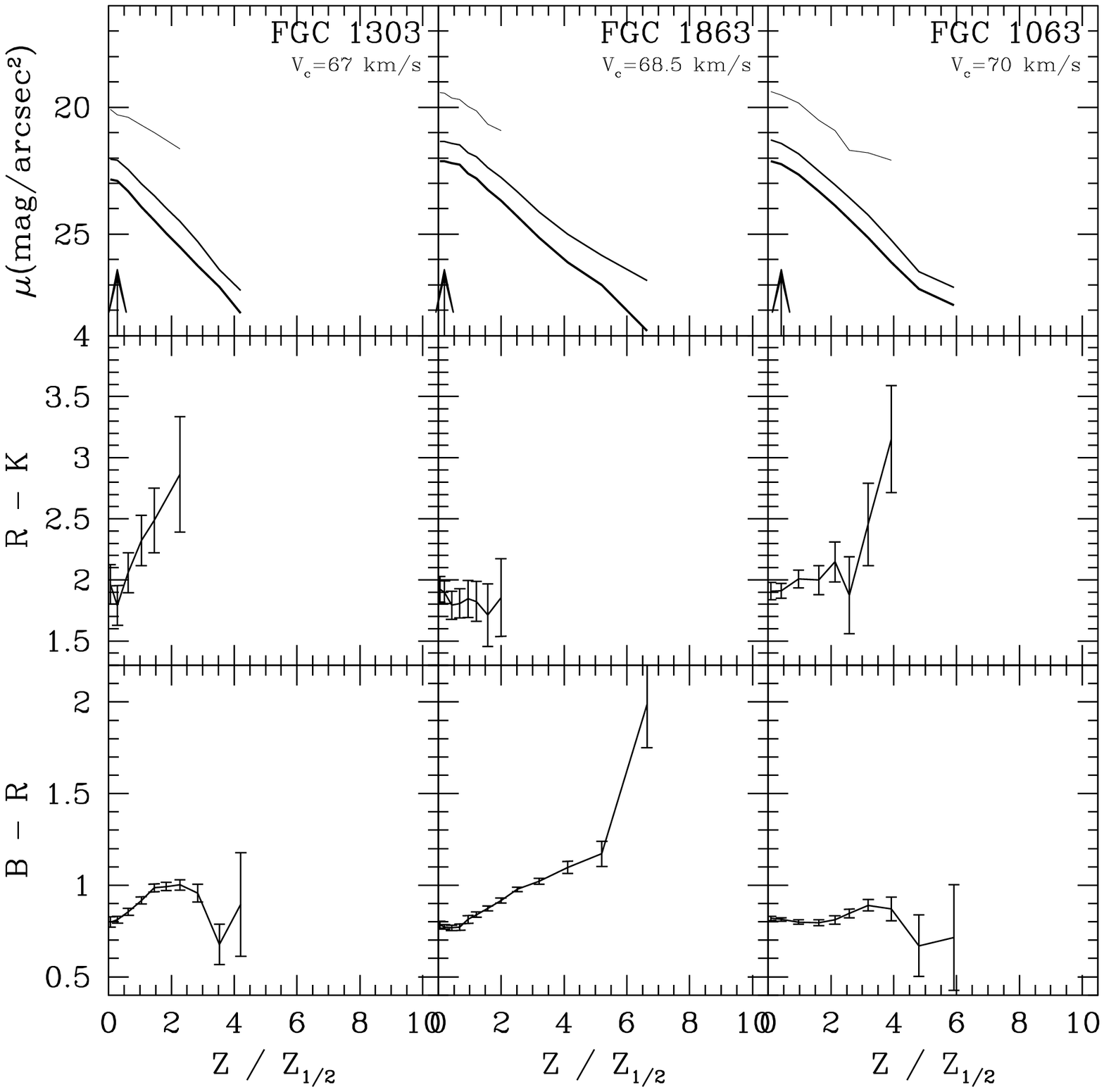}
\includegraphics[width=2.95in]{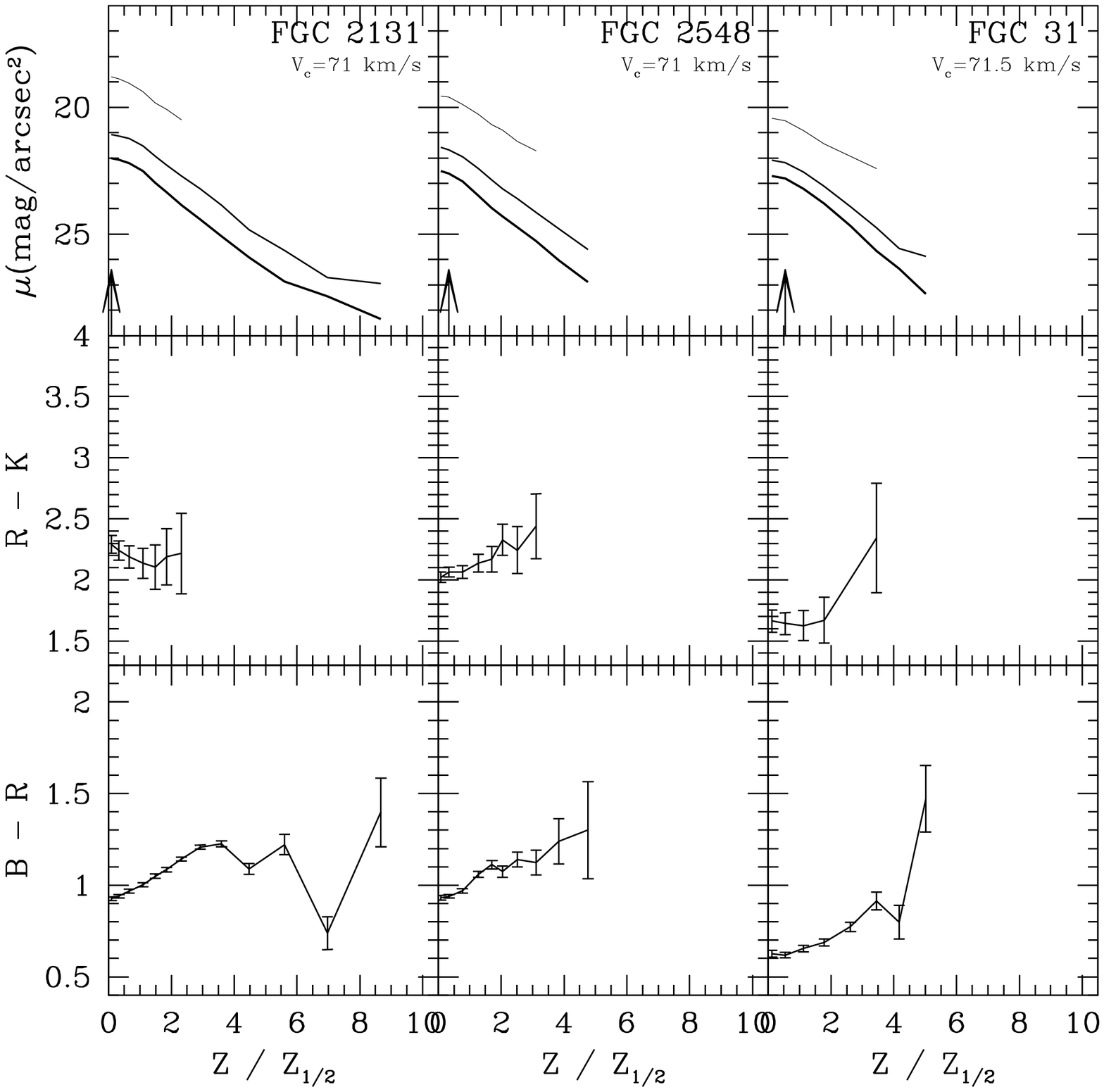}}
\caption{\footnotesize 
Vertical surface brightness and color profiles of FGC galaxies, in
order of increasing rotation speed.  The top panel shows the surface
brightness profiles in $B$ (heavy), $R$ (medium), and $K_s$ (light),
as a function of the number of vertical scale heights above the plane.
The middle and lower panels show the $R-K_s$ and $B-R$ profiles,
respectively.  The data plotted are the average values above and below
the plane, for $B$ and $R$ where the uncertainty $\sigma_{B-R}<0.3m$,
and for $K_s$ where $\sigma_{R-K_s}<0.5m$.  Profiles are measured
within $\pm3h_{K_s}$, where $h$ is the exponential scale length
derived from the $K_s$ data.  The arrow marks 1\arcsec, the typical
FWHM of the seeing disk.}
\label{profilefig}
\end{figure*}

\setcounter{figure}{2}
\begin{figure*}[t]
\hbox{
\includegraphics[width=2.95in]{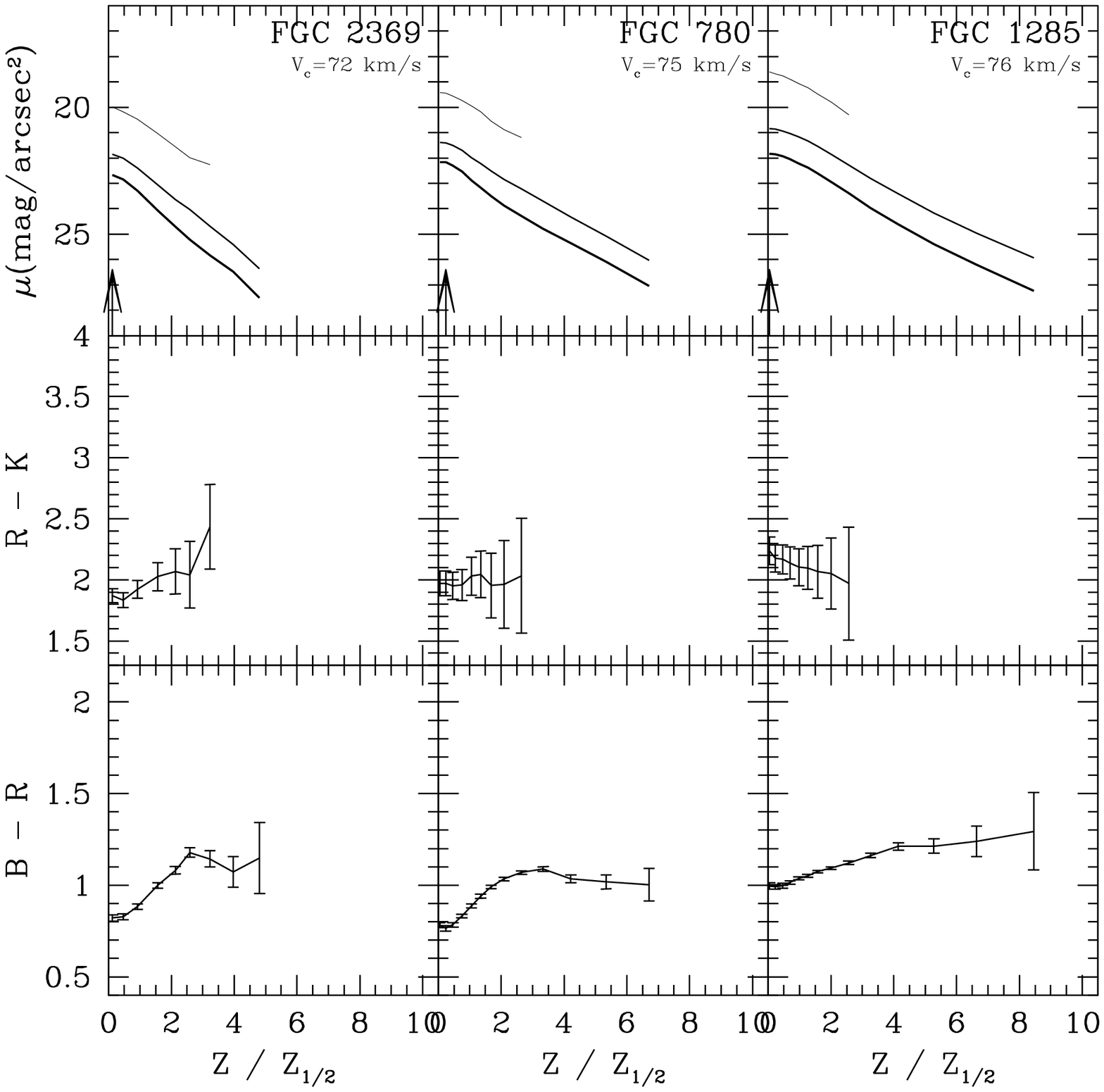}
\includegraphics[width=2.95in]{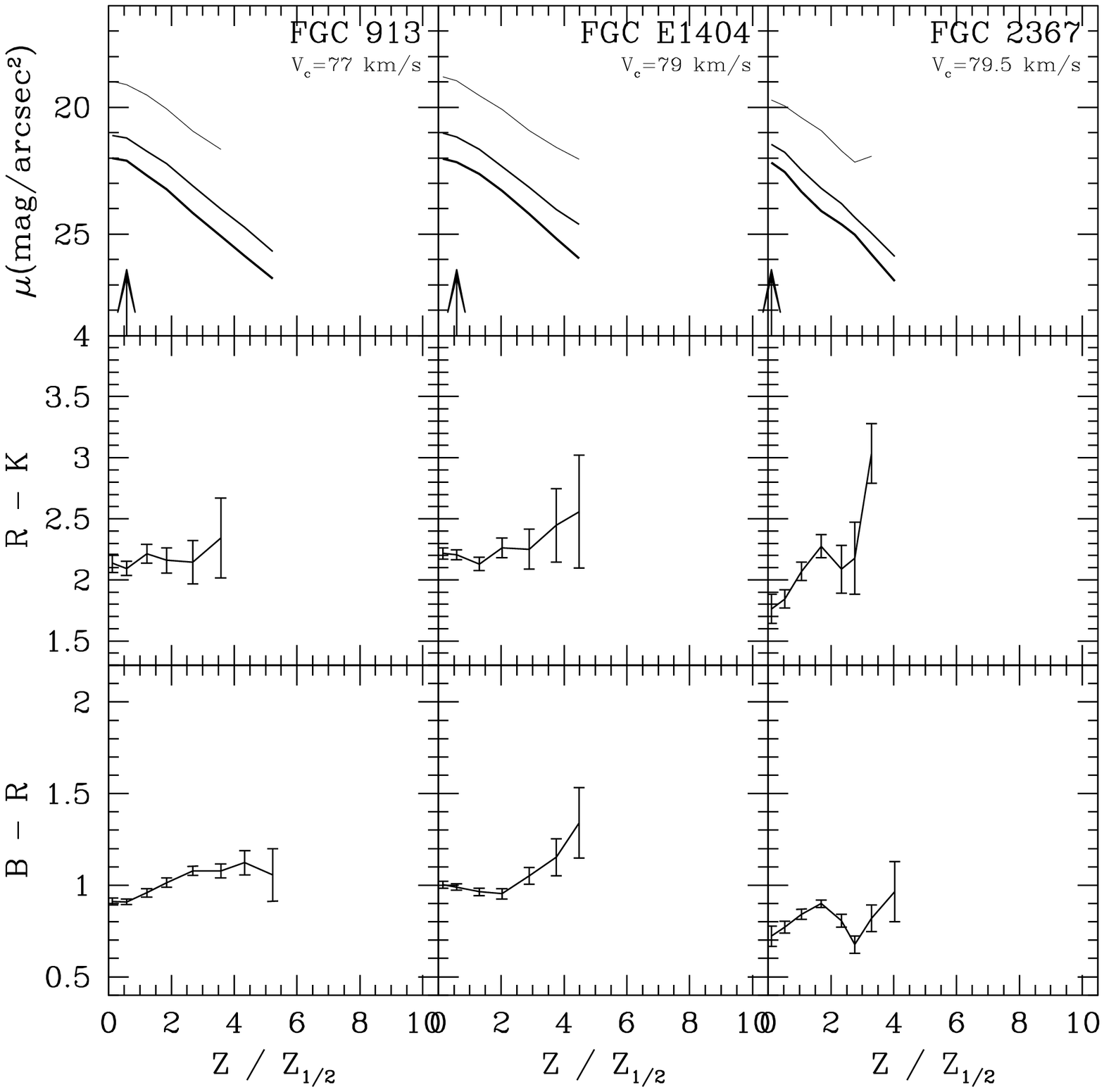}}
\vspace{0.1in}
\hbox{
\includegraphics[width=2.95in]{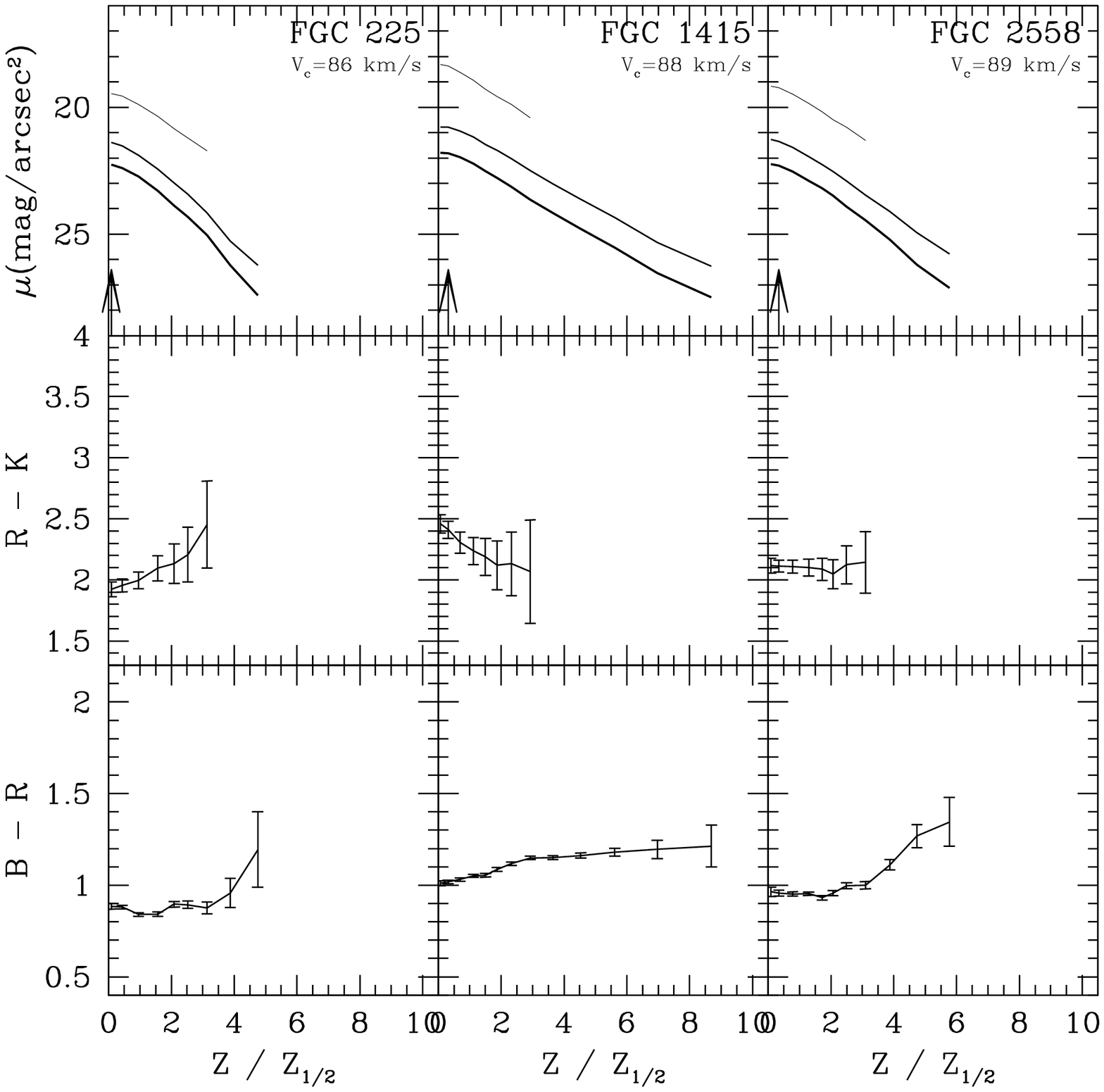}
\includegraphics[width=2.95in]{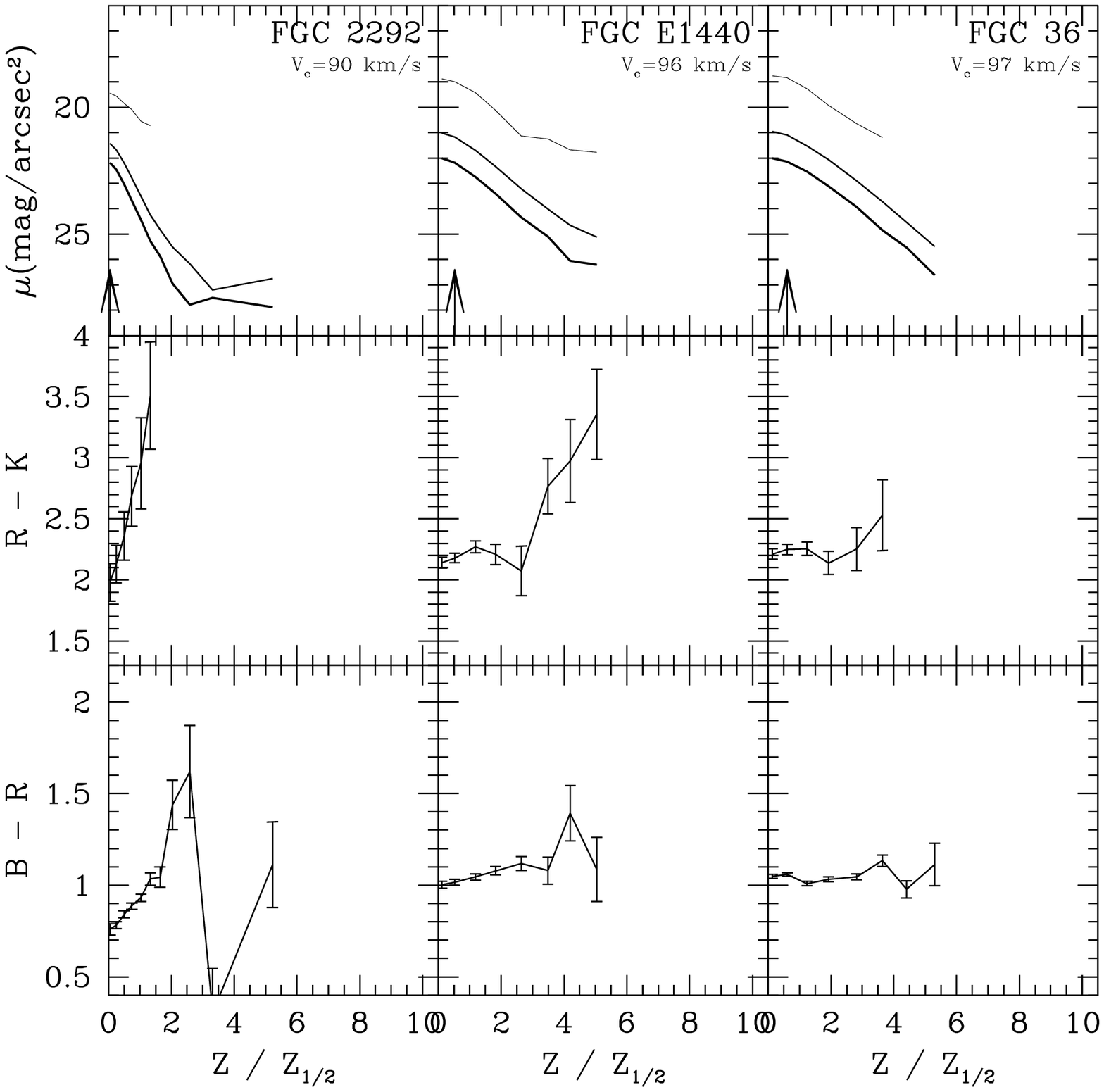}}
\caption{\footnotesize (continued)}
\end{figure*}

\setcounter{figure}{2}
\begin{figure*}[t]
\hbox{
\includegraphics[width=2.95in]{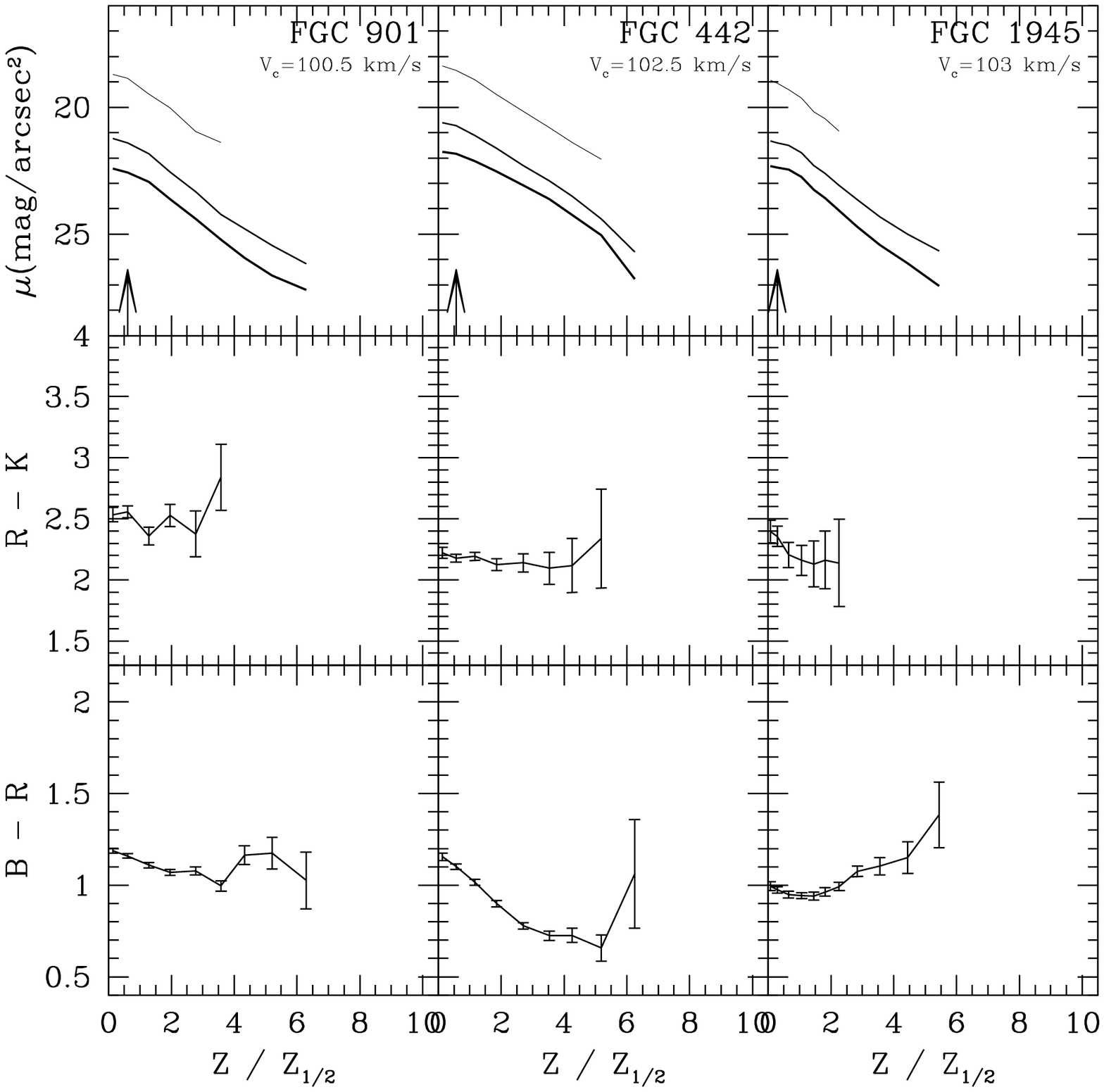}
\includegraphics[width=2.95in]{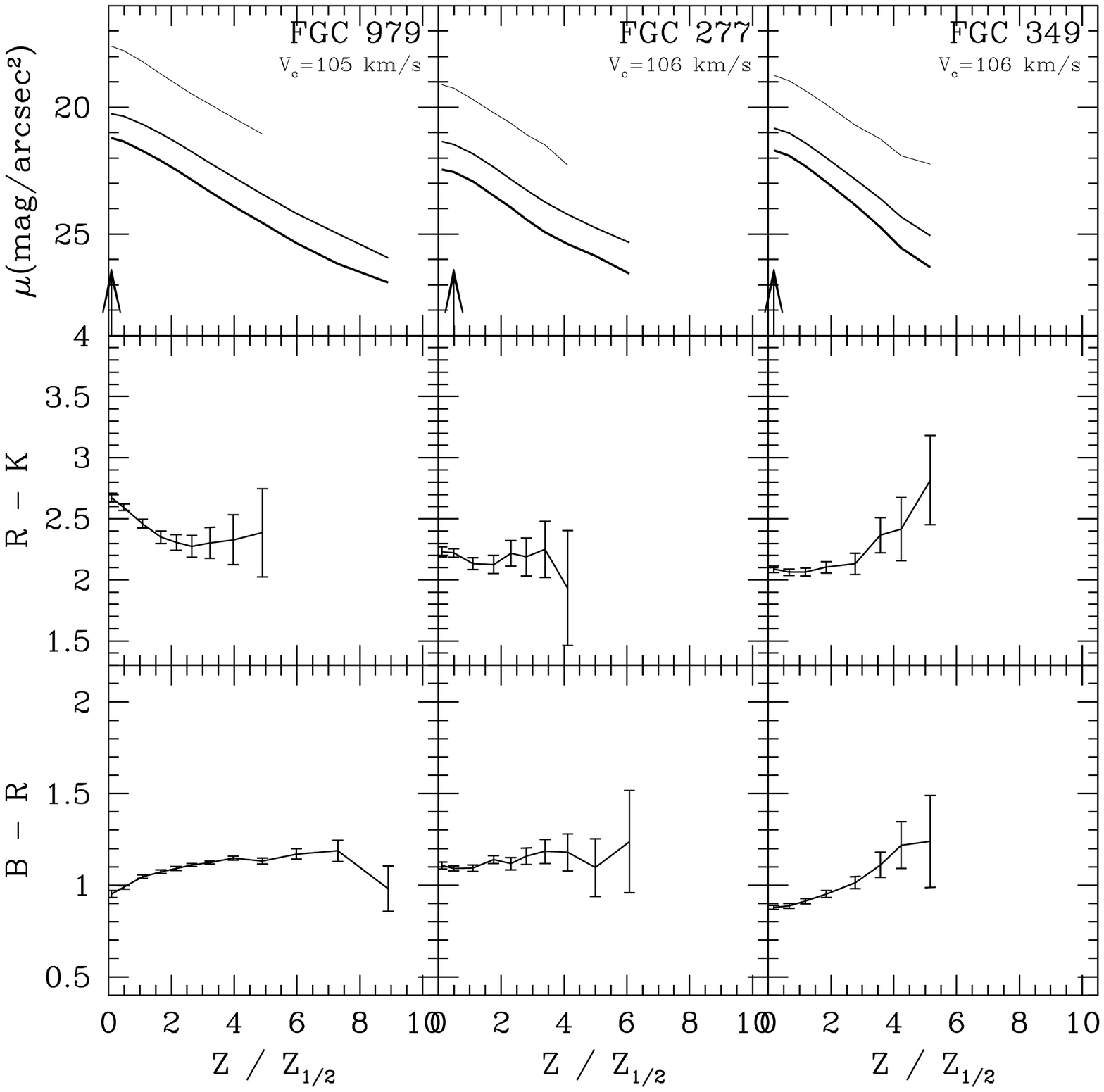}}
\vspace{0.1in}
\hbox{
\includegraphics[width=2.95in]{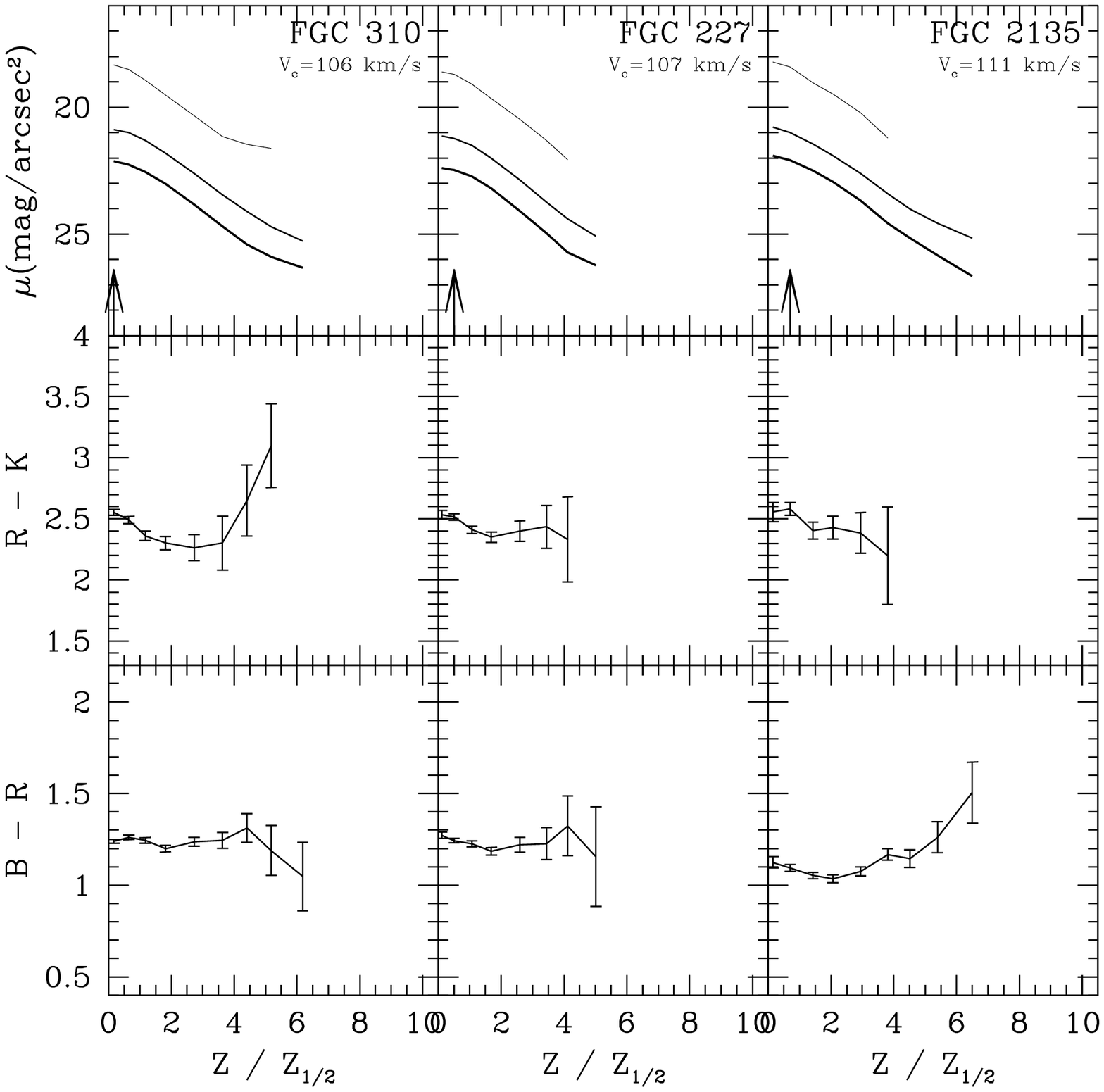}
\includegraphics[width=2.95in]{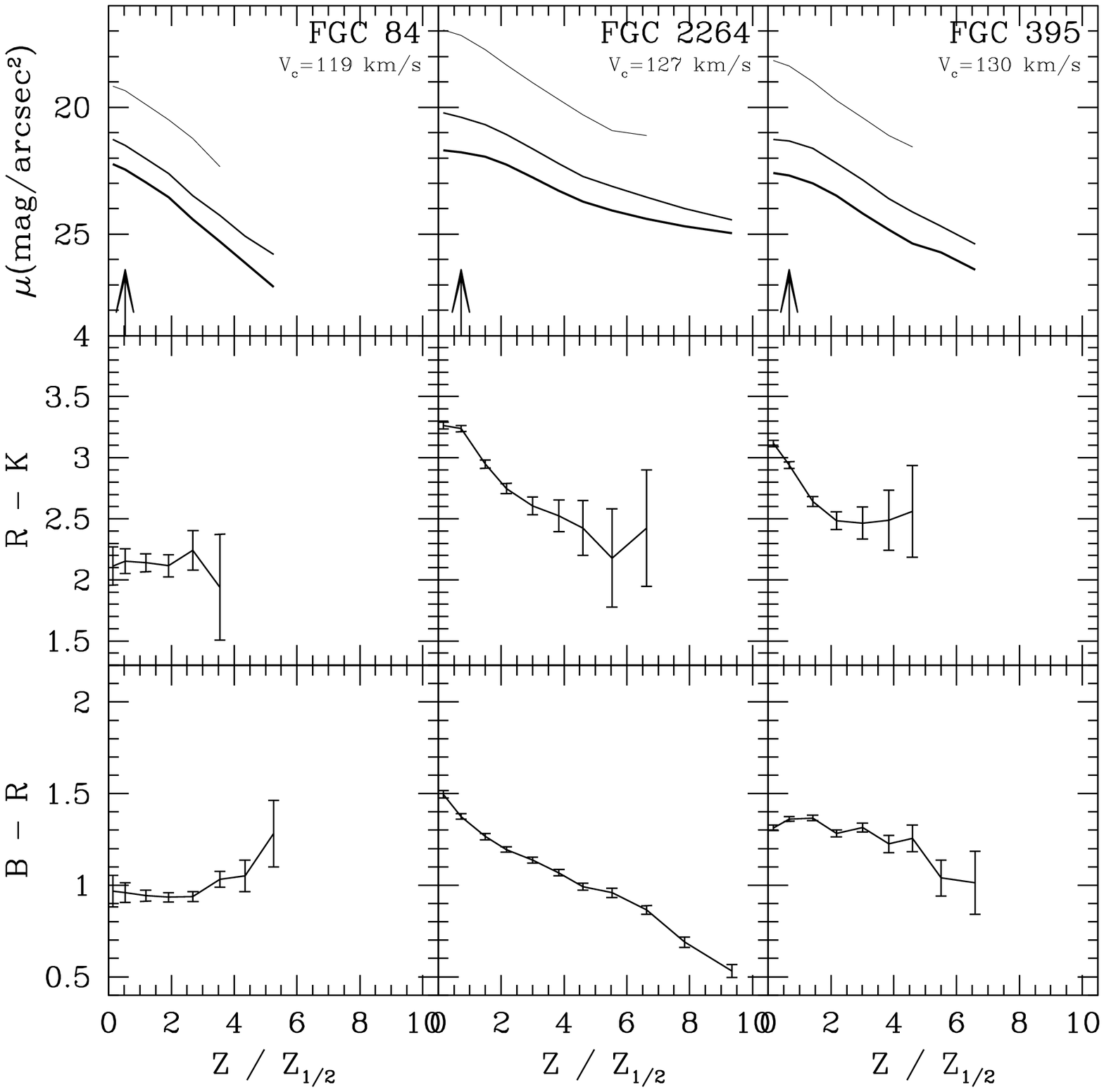}}
\caption{\footnotesize (continued)}
\end{figure*}

\setcounter{figure}{2}
\begin{figure*}[t]
\hbox{ 
\includegraphics[width=2.95in]{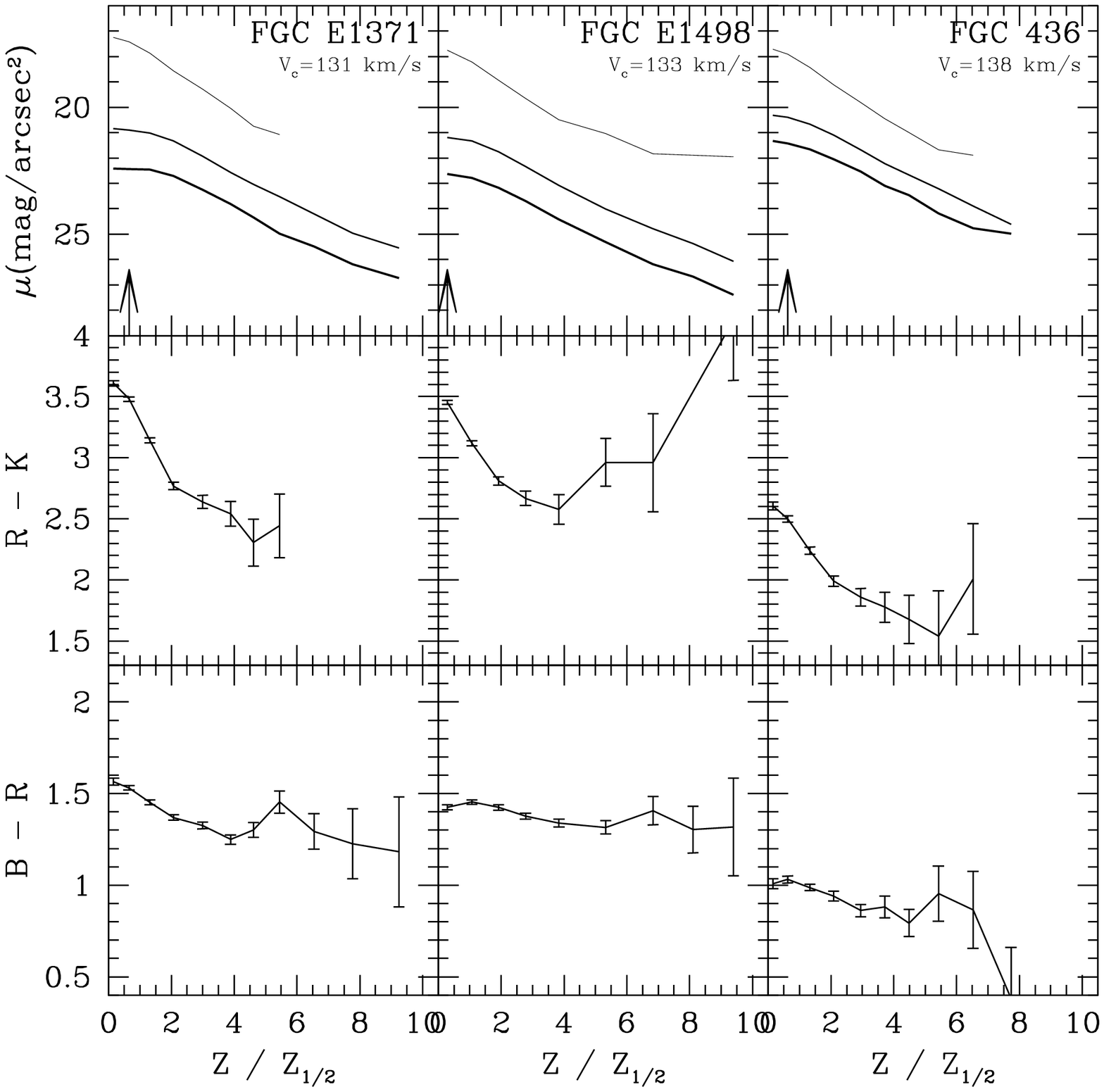}
\includegraphics[width=2.95in]{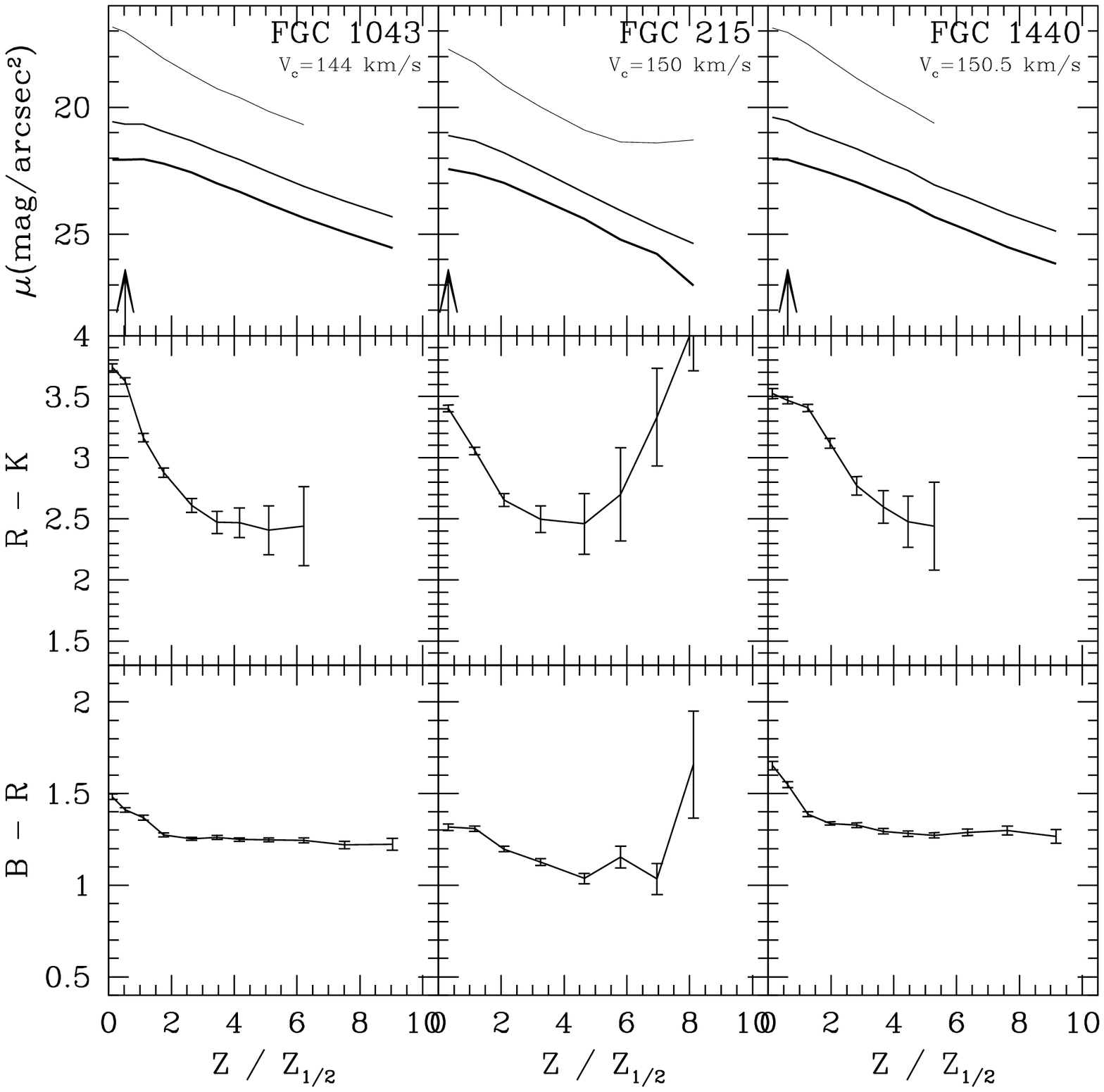}}
\vspace{0.1in}
\hbox{
\includegraphics[width=2.95in]{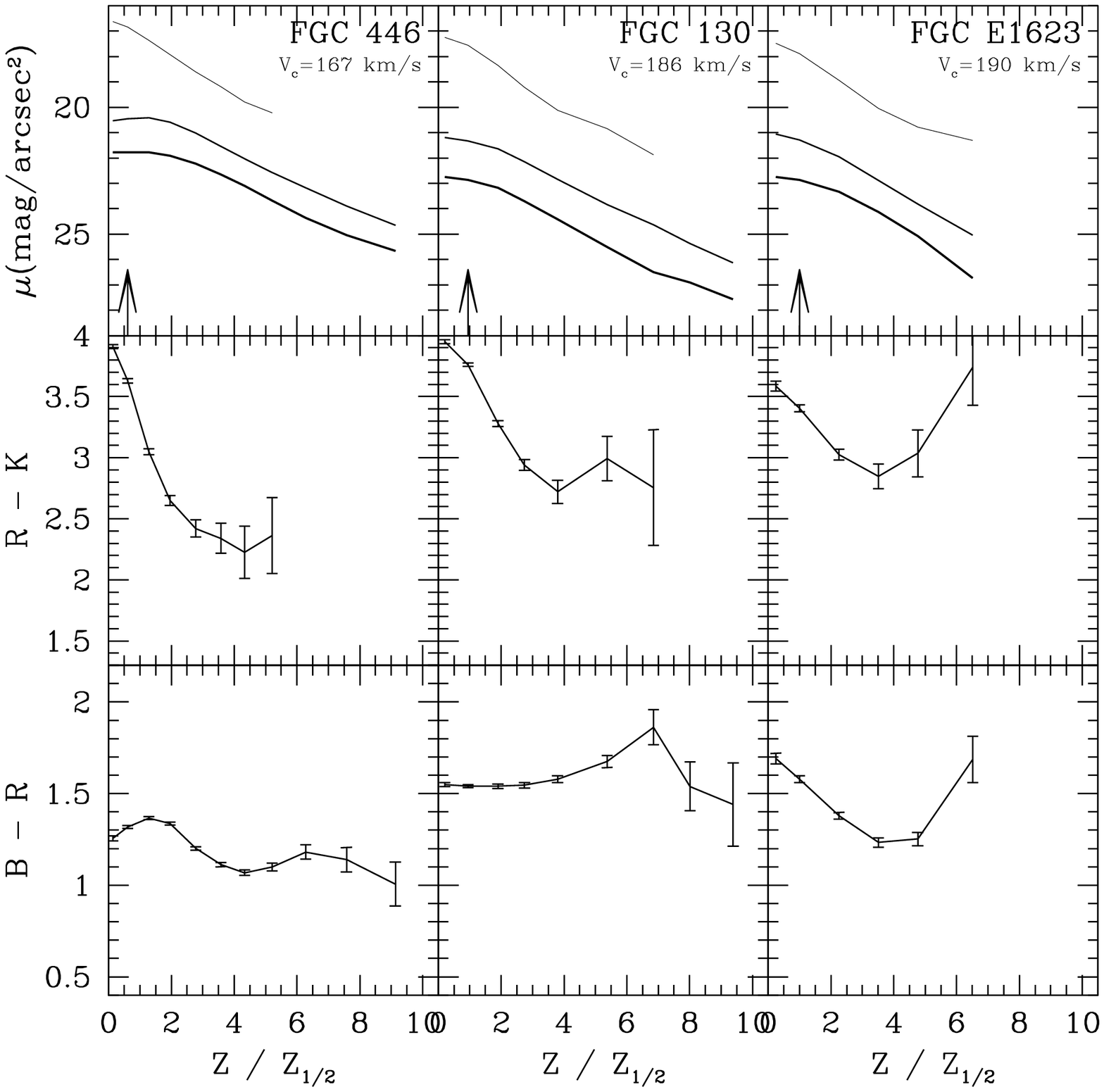}
\includegraphics[width=2.95in]{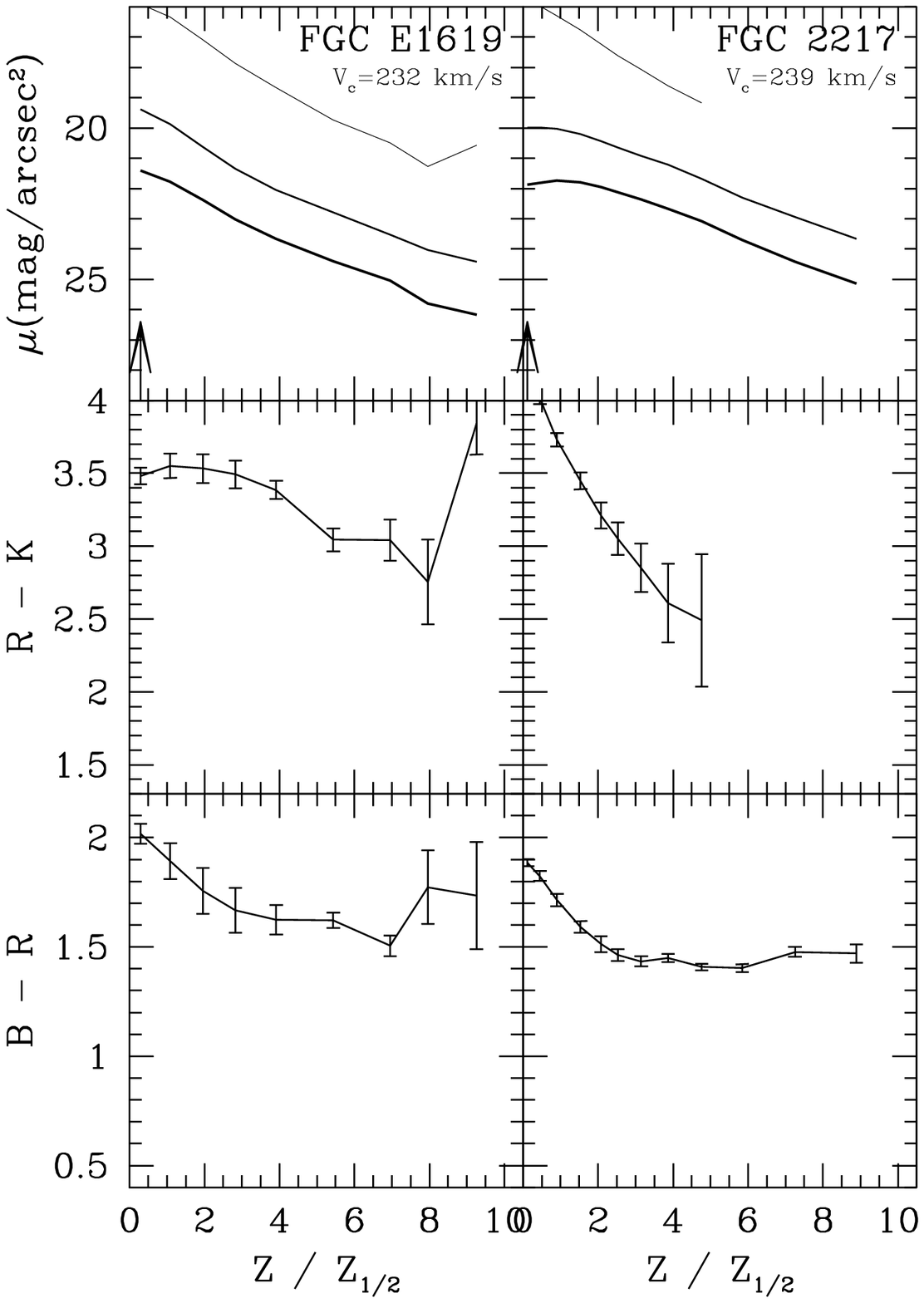}}
\caption{\footnotesize (continued)}
\end{figure*}

\section{Interpretation of the Vertical Color Gradients}  \label{interpsec}
\begin{figure}[t]
\includegraphics[width=3.5in]{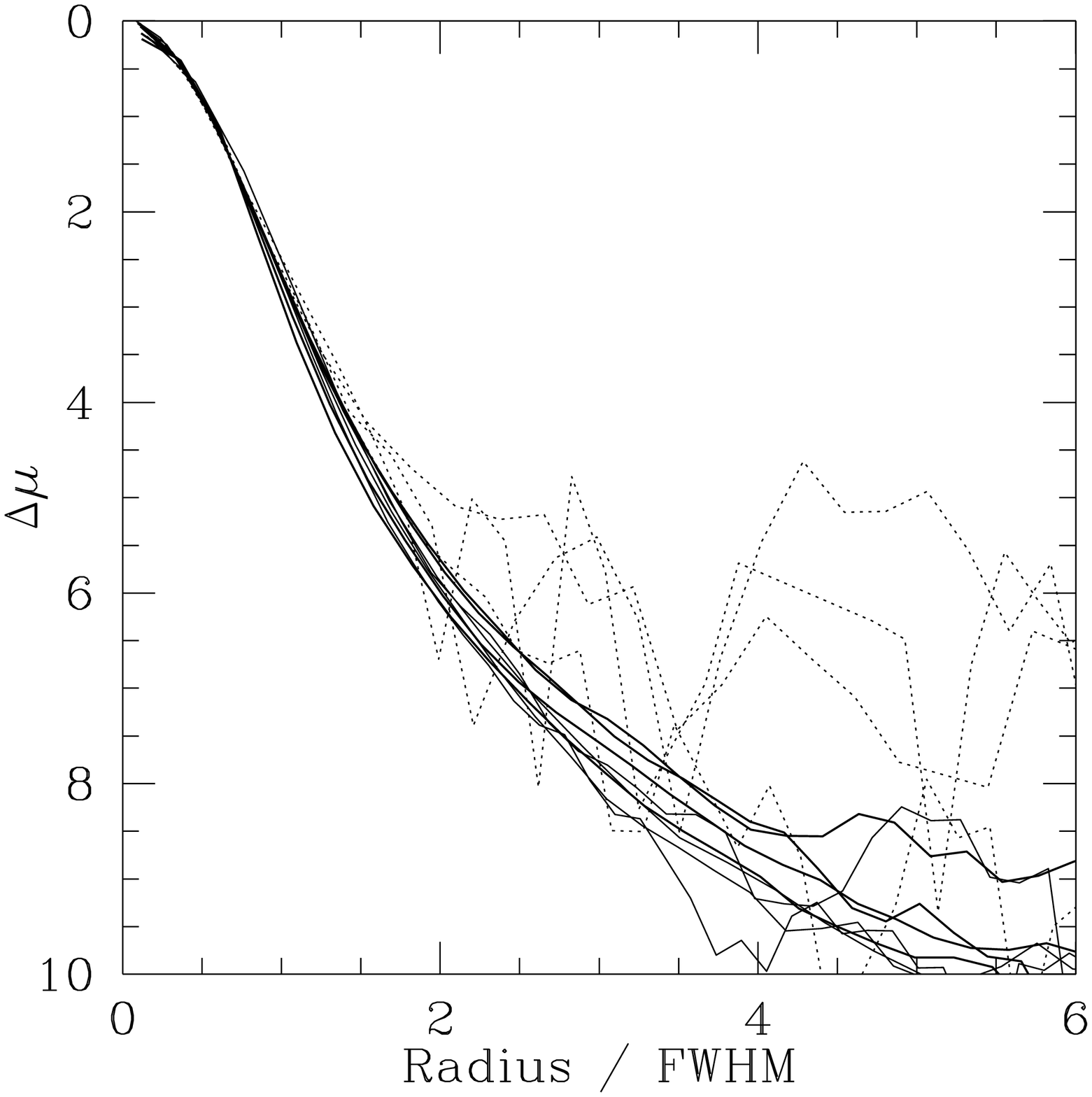} 
\caption{\footnotesize 
Stellar radial surface brightness
profiles, for the $R$-band (heavy solid lines), $B$-band (light solid
lines), and $K_s$-band (dotted lines), for a range of seeing
conditions.  The point spread functions are plotted after scaling by
the FWHM and central surface brightness of a Gaussian fit to the
profile.  In all cases, the profiles are nearly indistinguishable,
showing that the PSF (1) has uniform shape in a range of conditions,
(2) falls off rapidly with radius, and (3) has little light scattered
at large angles.  The noisier profiles in the $K_s$ band reflect the
higher level of sky noise in these images.
\label{seeingfig}}
\end{figure}

\begin{figure}[t]
\includegraphics[width=3.5in]{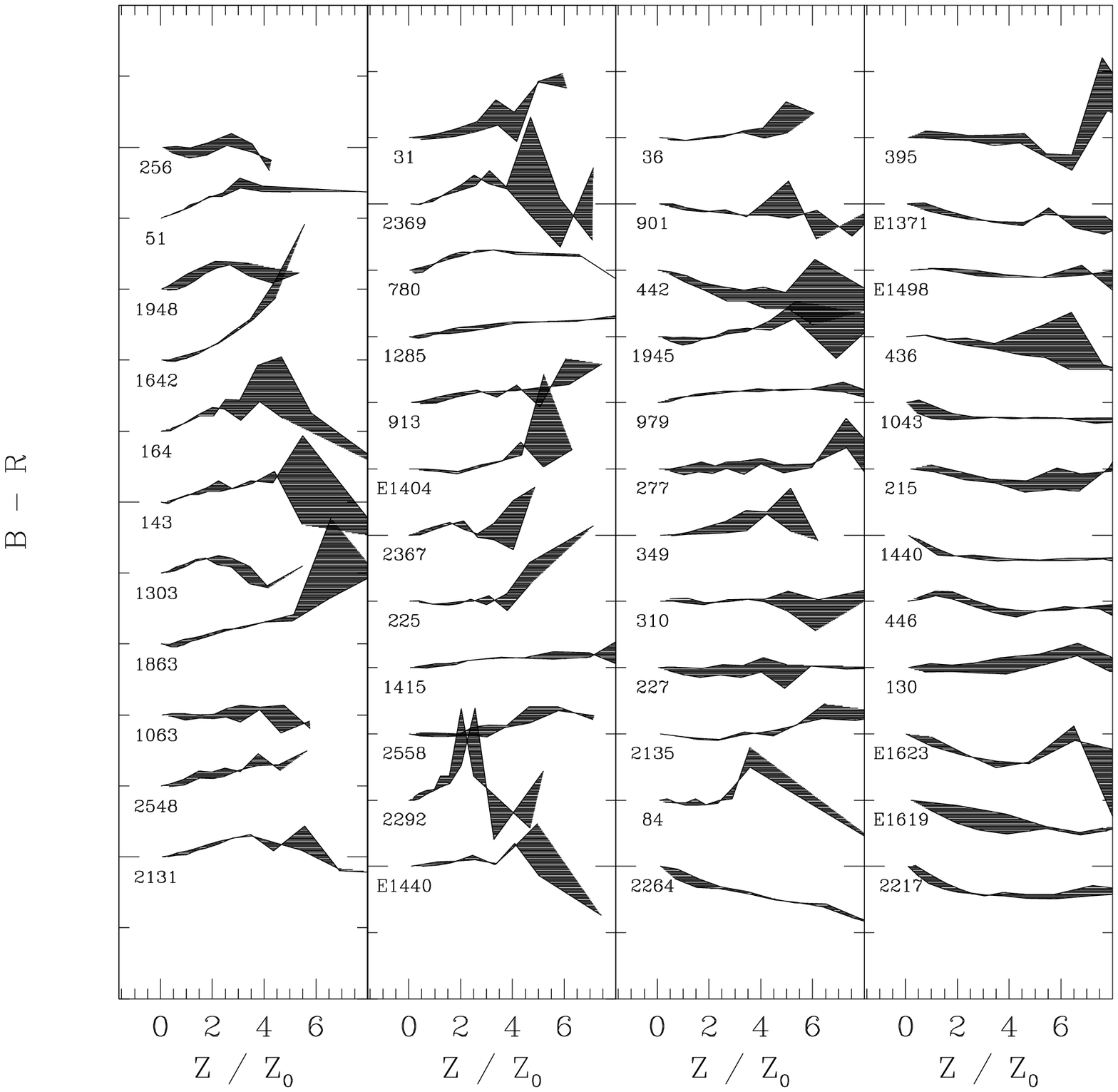}
\vspace{0.1in}
\includegraphics[width=3.5in]{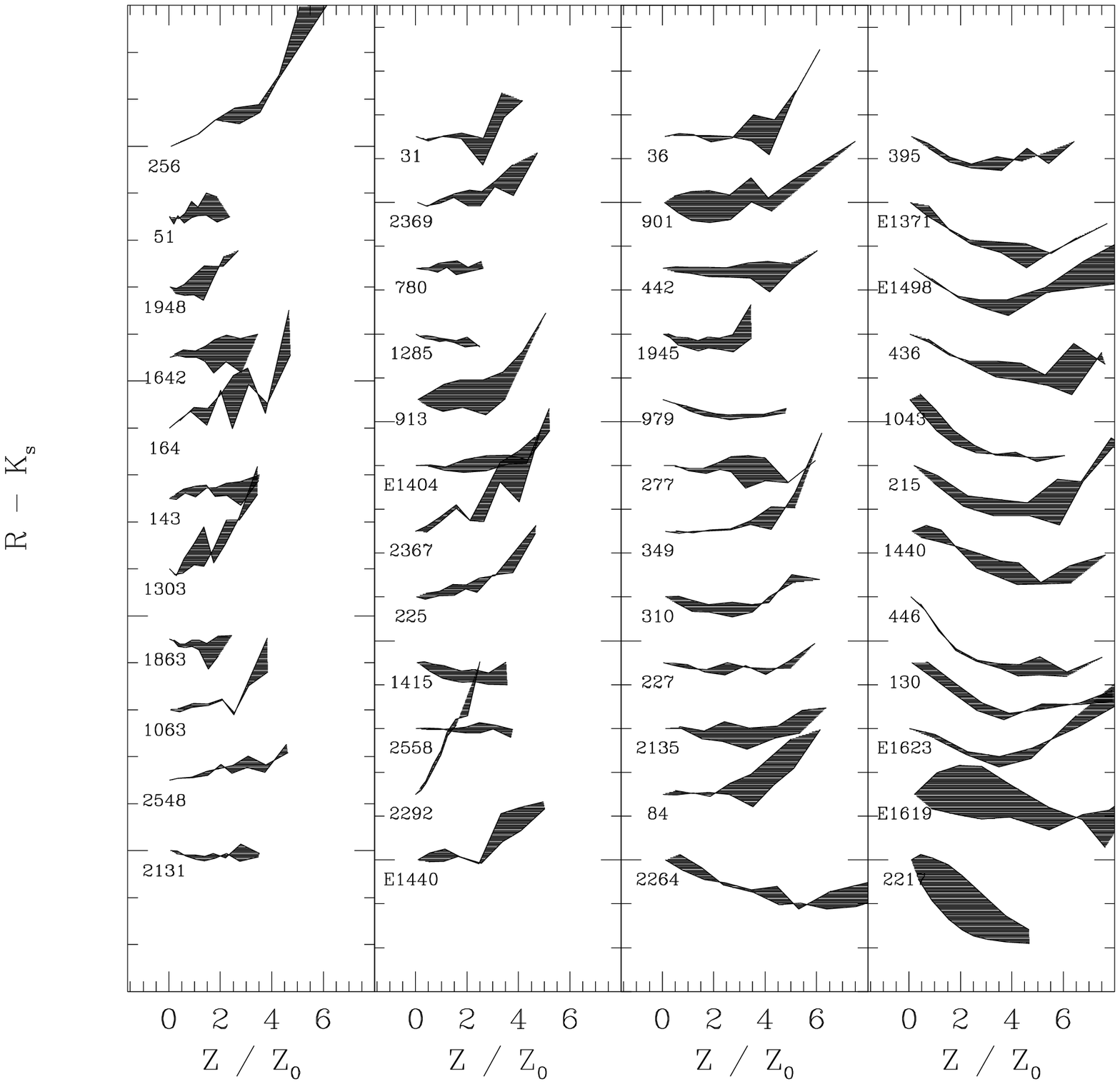}
\caption{\footnotesize 
Vertical color profiles
in $B-R$ (a) and $R-K_s$ (b) of the FGC galaxies, in order of
increasing rotation speed, in units of the vertical scale height.  The
shaded region represents the difference between the profiles derived
above and below the plane, and gives an empirical measure of the
systematic errors in the profiles at large scale heights.  In general,
the colors are very consistent between the two sides.  The exceptions
occur when the galaxy is not at exactly $90\deg$ inclination,
particularly among the more massive galaxies which have dustlanes and
higher overall extinction.  Only points where the color error is less
than $0.3m$ in $B-R$ and $0.5m$ in $R-K_s$ are plotted.  The midplane
colors of the galaxies have been offset by 1 mag for the $B-R$ profiles
and by 1.5 mag for the $R-K_s$ profiles.
\label{allprofilesfig}}
\end{figure}

The color gradients plotted in
Figures~\ref{profilefig}~\&~\ref{allprofilesfig} show strong trends
when sorted by rotation speed.  In the fast rotating, higher mass
galaxies with obvious dust lanes, we see color gradients which become
bluer with increasing height but then level out to roughly uniform
color.  In these galaxies, the stellar population is highly reddened
by dust within the plane, leading the galaxy to appear bluer at large
scale heights where the stellar population becomes unobscured.  In
somewhat lower mass galaxies the effects of dust lanes disappear and
the central color gradients flatten out in both $B-R$ and $R-K_s$.  At
even lower masses, the trends reverse, and the color gradient becomes
steep again, but such that the stellar populations become
{\emph{redder}} above the plane.  These steep gradients are visible in
both colors ($B-R$ and $R-K_s$).

\begin{figure}[t]
\includegraphics[height=3.25in,angle=-90]{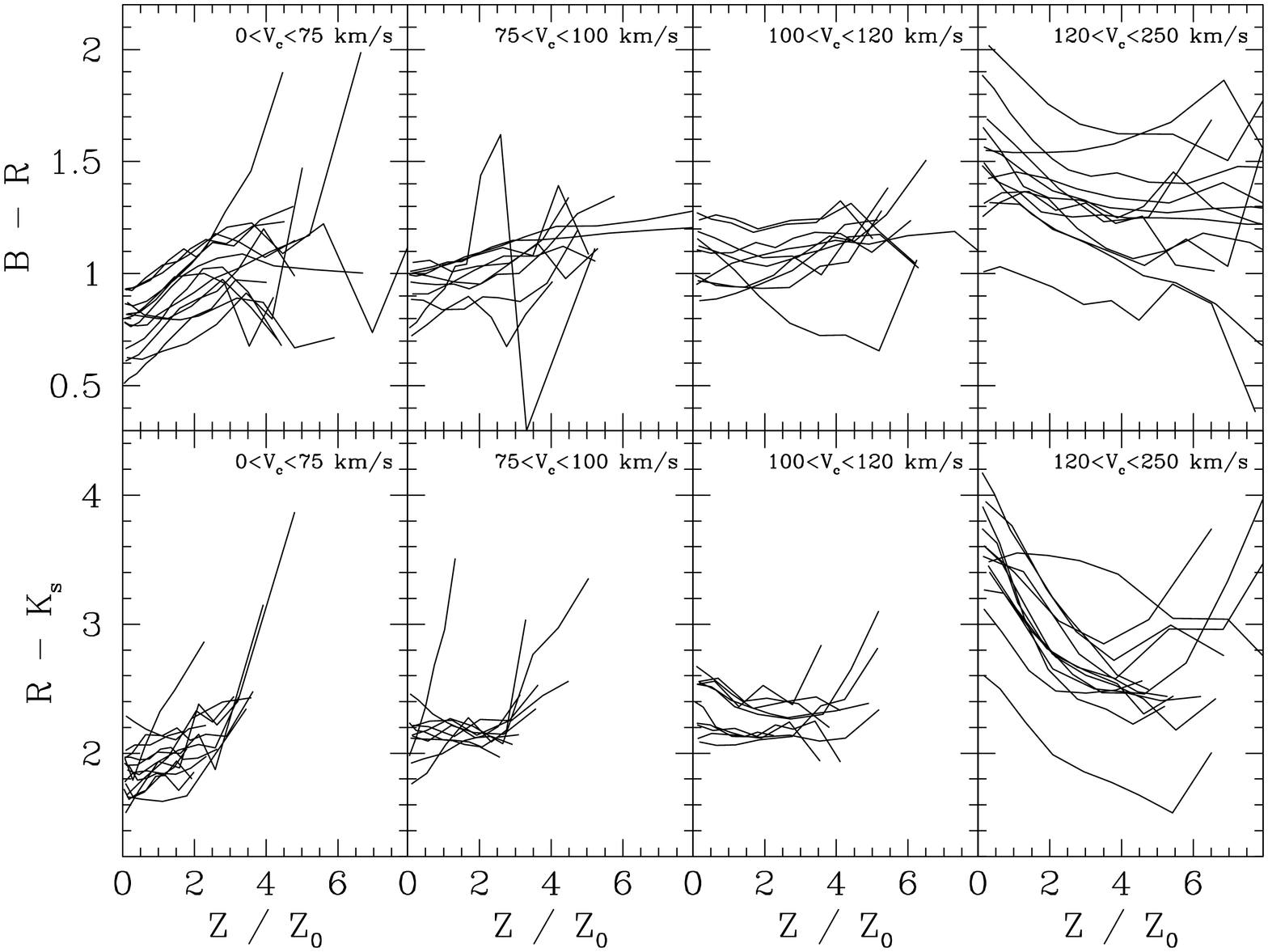}
\caption{\footnotesize 
Vertical color profiles of the FGC
galaxies (taken from Figure~\ref{profilefig}), sorted in bins of
circular velocity $V_c$.  Note that all profiles converge to similar
colors at increasing scale height, as also can be seen in
Figure~\ref{gridgradfig}.
\label{overlayproffig}}
\end{figure}

\begin{figure}[t]
\includegraphics[width=3.5in]{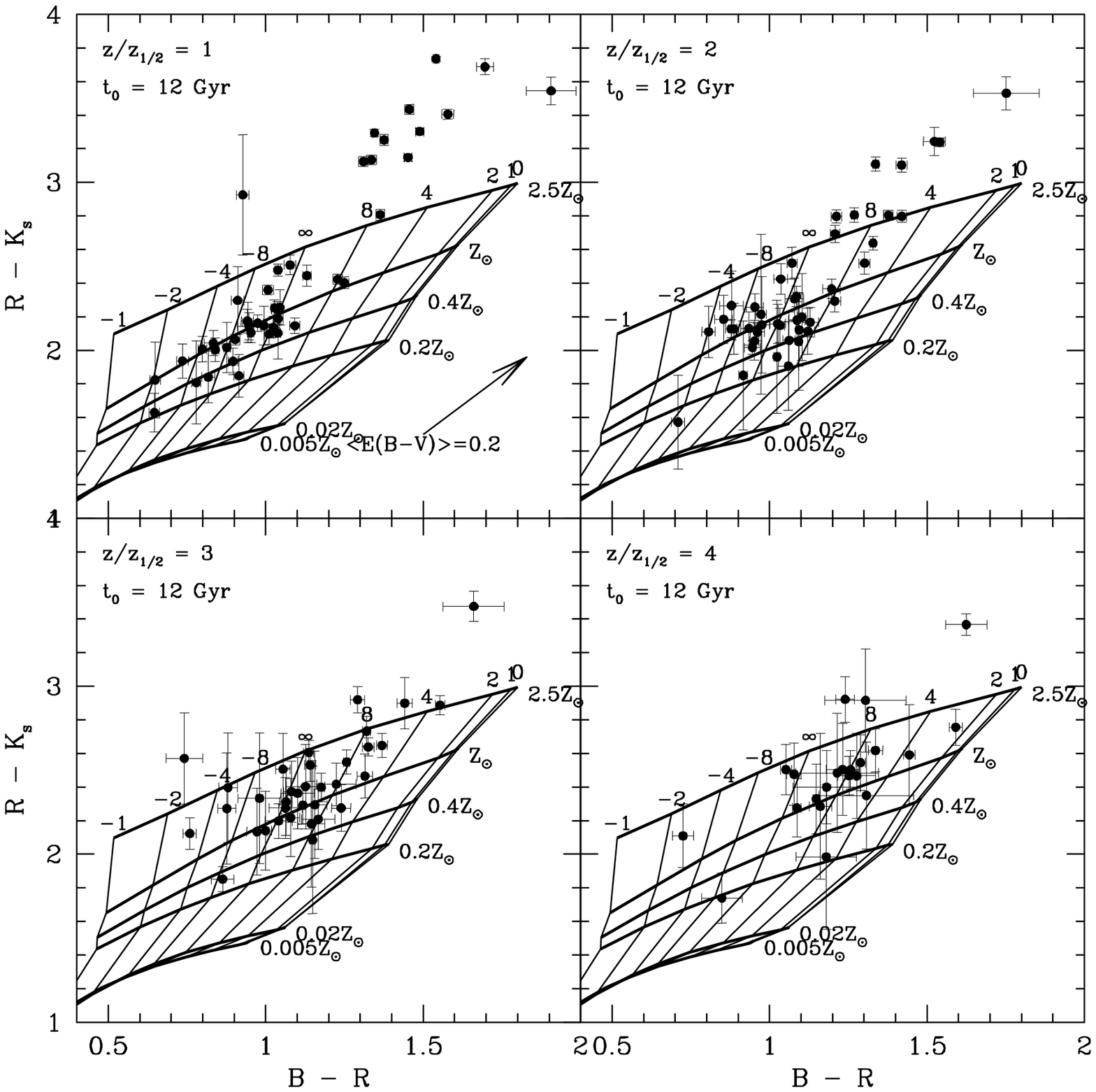}
\caption{\footnotesize 
The colors of FGC galaxies at scale
heights of $z_{1/2}$ (upper left), $2z_{1/2}$ (upper right),
$3z_{1/2}$ (lower left), and $4z_{1/2}$ (lower right).  Error bars
include both photon counting and sky subtraction uncertainties.
Superimposed are grids from Bruzual \& Charlot (2001) models, for
lines of constant metallicity (roughly horizontal heavy lines, [Fe/H]
= -2.3, -1.7, -0.7, -0.4, 0, +0.4) and constant exponentially
declining star formation rates (roughly vertical light lines, $\tau =
0,1,2,4,8,\infty,-8,-4,-2,-1\Gyr$; the negative values of $\tau$ are
for exponentials which rise to the present day, leading to mean ages
smaller than $t_0/2$).  We have assumed that star-formation began
$t_0=12\Gyr$ ago, giving mean stellar ages of
$<t>\approx12,11,10,8.6,7.5,6,4.5,3.4,2,1\Gyr$, for the corresponding
values of $\tau$ listed above.  Younger values of $t_0$ (not shown)
tend to move the right half of the grid downwards and to the left,
such that a given point corresponds to a younger age and higher
metallicity.
\label{gridgradfig}}
\end{figure}

To explore these trends further, in Figure~\ref{overlayproffig} we
overplot all of the color profiles from the galaxies in Figure
\ref{profilefig}.  While the galaxies start with a very wide range of
colors close to the midplane, they have a much smaller range of colors
well above the plane $z/z_{1/2}>3-4$, with $1.0 \lesssim B-R \lesssim
1.4$, and $2.0 \lesssim R-K_s \lesssim 2.6$; the same effect is
visible in Figure \ref{gridgradfig}, discussed below.

The simplest explanation for the above trends is that {\emph{the
    galaxy disks are all embedded in a faint stellar envelope whose
    observable properties vary little from galaxy to galaxy, while the
    properties of the embedded thin stellar disk varies systematically
    with mass}}.  In other words, the behavior of the vertical color
gradients shown in Figures~\ref{profilefig}~\&~\ref{allprofilesfig}
results primarily from variations in the properties of the disk at the
midplane, not from galaxy-to-galaxy variations in the surrounding
stellar envelope.  The high mass galaxies have red midplanes due
primarily to dust, and thus they get rapidly bluer with increasing
scale height.  The low mass galaxies have active star formation and
low metallicities (e.g.\ Stasinska \& Sodr\'e 2001,
Zaritsky et al.\ 1994), and thus extremely blue disks, with much
redder colors above the plane.  We further quantify the color gradients,
stellar populations, and envelope structure below.

\subsection{Quantifying Vertical Color Gradients}       \label{quantifysec}

In order to quantify the amplitude of the color gradients shown in
Figures~\ref{profilefig}~\&~\ref{allprofilesfig}, we measure the slope
of the color gradient as a function of scale height by fitting line
segments to the vertical color profiles which were extracted in
\S\ref{extractionsec}, in a series of scale height intervals.  We
restrict the fitting to $z/z_{1/2} = 1-2,2-4,4-6$, using the color
profiles plotted in Figure \ref{profilefig} averaged above and below the
plane.

\begin{figure}[t]
\includegraphics[width=3.5in]{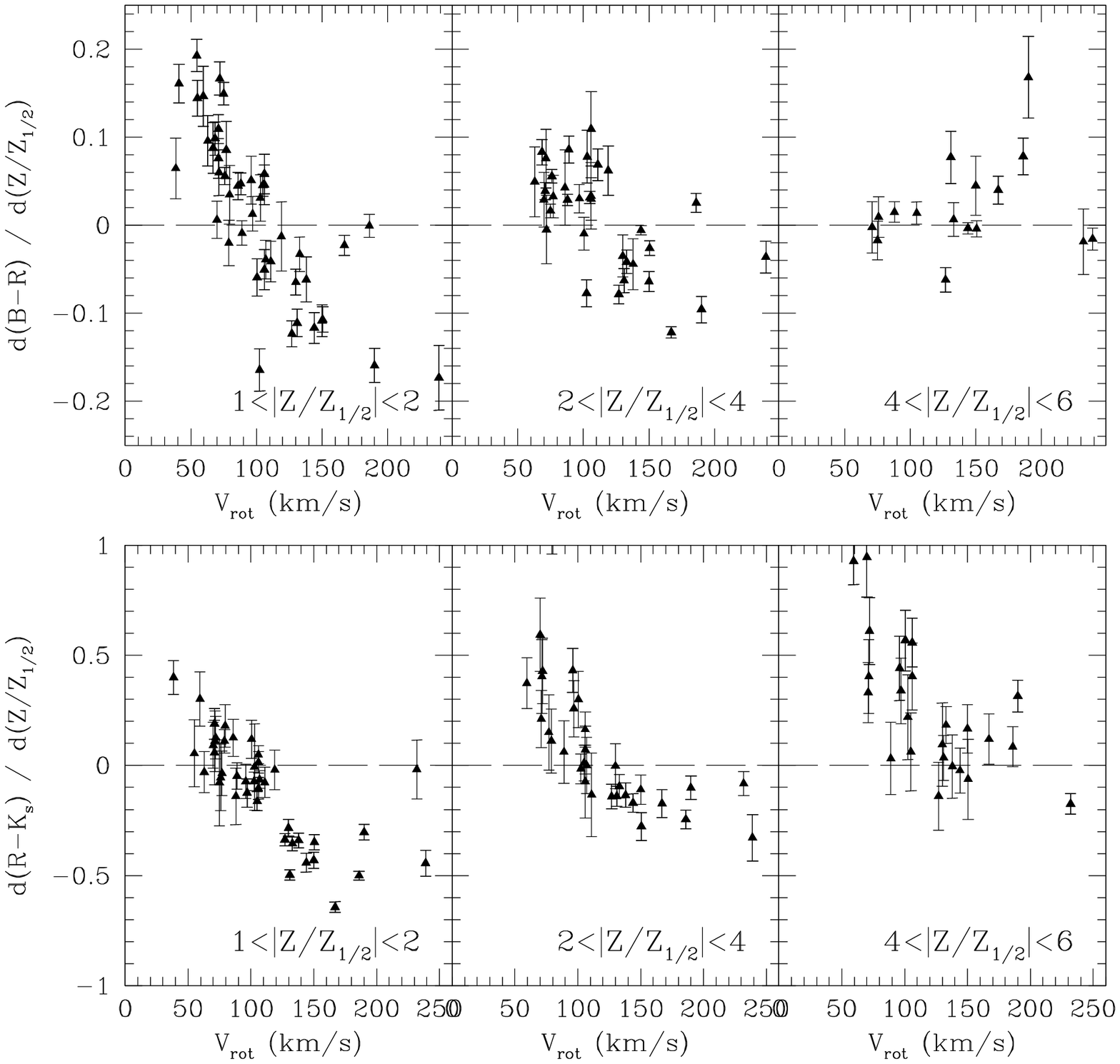}
\caption{\footnotesize 
The vertical gradient in $B-R$ (upper
panels) and $R-K_s$ (lower panels) as a function of rotation speed, in
height intervals of $1-2 Z_{1/2}$ (left panels), $2-4 Z_{1/2}$ (center
panels), and $4-6 Z_{1/2}$ (right panels), measured within a radius of
3 scale lengths.  Error bars are as described in the text.  Only
points with slope uncertainties of less than 0.05 in $B-R$ or 0.2 in
$R-K_s$ have been plotted.  The dashed line marks zero color gradient.
Points above the line get redder with increasing height above the
plane.
\label{slopevsVcfig}}
\end{figure}

Ascribing uncertainties to the measured slopes is somewhat
complicated.  Poisson photon-counting errors produce nearly Gaussian
random errors in the measured fluxes, which are uncorrelated as a
function of scale height.  However, uncertainties in the sky
subtraction can produce correlated errors in the surface brightness
profile, and thus systematic trends in the slope of the color
gradient.  To quantify the uncertainties in the slope, we follow the
method of Bell \& de Jong (2000) and run a series of Monte Carlo
simulations on the measurement of the slope.  In each filter, for each
point in the extracted surface brightness profile, we let the flux
vary as a Gaussian with a width set by the amplitude of the
photon-counting uncertainties; we ignore the photometric calibration
uncertainties, which are negligible.  We then choose a value for the
sky level drawn from a Gaussian distribution with a width set by the
sky uncertainties measured in Paper I.  We then use the new flux in
each radial bin and the new global sky level to recalculate the color
profile, and fit the slope.  We run 1000 trials for each galaxy, and
then measure the standard deviation of the resulting distribution of
slopes.  The measured slopes and their uncertainties are plotted in
Figure \ref{slopevsVcfig}, as a function of the galaxies' rotation
speeds.

The trends evident in Figures~\ref{profilefig}~\&~\ref{allprofilesfig}
are reproduced clearly in Figure \ref{slopevsVcfig}.  Near the plane
(i.e.\ the left hand panels), the amplitude and sign of the vertical
color gradient depend strongly upon a galaxy's circular velocity.  In
high mass galaxies, the galaxy becomes progressively bluer above the
plane as one rises above the dust lane, leading to negative color
gradients.  As the mass of the galaxy
decreases, dust becomes less important and the disk becomes bluer,
erasing the color gradient.  At low masses, galaxies behave in
the opposite manner; the galaxies becomes redder in both $B-R$ and
$R-K_s$ above the plane, as one rises above the presumably young,
metal-poor star-forming disk, producing positive color gradients. 

Within individual galaxies, the color gradient becomes shallower with
increasing height above the plane, as can be seen by comparing a
single row of panels in Figure \ref{slopevsVcfig}.  In other words,
the color becomes more uniform at larger scale heights.  In the high
mass galaxies, the amplitude of the color gradient flattens rapidly,
because the large color gradient is almost entirely due to dust
confined to a thin plane (typically $z_{dust}\lesssim z_{stars}/1.4$;
Xilouris et al.\ 1999).  In lower mass galaxies, the gradient is
statistically significant out to $\sim4z_{1/2}$. Beyond this, only one
or two galaxies have color gradients which are different than zero at
a $<2\sigma$ level, although their average trend as a group is towards a
halo which continues to redden outwards, although at a lower rate.
The only galaxy which gives a statistically significant negative
gradient in the range $z>4z_{1/2}$ is FGC 2264, which is in a crowded
field at low galactic latitude.  It is probable that the size of the
error bar for this galaxy does not reflect the true uncertainties due
to spatially variable extinction and to the unusually complicated
foreground source subtraction.

\subsection{Comparison with Previous Detections of Vertical Color Gradients in Other Galaxies}

A number of previous color gradient studies have focussed on
understanding the dust content and stellar populations of the young
thin disk, but those efforts have mostly concentrated on color
gradients near the midplane, rather than at large scale heights.  This
in part reflects the difficulties in acquiring data of sufficient
quality at very low count levels, even with modern CCDs.

Of previous work, we have identified only two studies which
specifically address color gradients at large scale heights above
disk-dominated galaxies.  The largest systematic study so far (de
Grijs \& Peletier 2000) finds a preponderance of gradients which
become redder in $B-I$ with increasing scale height in the region
1-3$h_z$ (corresponding to 0.6-1.8$z_{1/2}$), with a slight trend
towards larger gradients in late-type, lower mass galaxies.  However,
the amplitude of the detected gradients was typically quite small as
would be expected on the basis of our results, given the high masses
of the galaxies dominating their sample.  Their conclusions were much
also narrower in scope due to an inability to trace the colors beyond
$2z_{1/2}$.  Moreover, the majority of their sample contained bulges
(Sc or earlier) whose presence greatly complicates the analysis of the
disk colors alone, and possibly masks any faint stellar envelope.  A
similar problem affects the suitability of comparing our work to the
color gradients measured by Shaw \& Gilmore (1990) for four galaxies
with much more prominent bulges than in our sample.

More immediately applicable to our sample is the detailed study of the
nearby edge-on bulgeless galaxy UGC 7321, analyzed by Matthews et al.\
(1999).  The properties of this galaxy are comparable to those found
in our sample, including the detection of reddening of 
$\Delta(B-R)\!\sim\!0.3$ with increasing scale height.  More detailed
analysis of the dust properties of this galaxy (Matthews \& Wood 2001)
suggest that the actual color difference of the stellar population might
be even larger (by $\sim\!0.15$ mag).  Again, these results are similar
to the color gradients we have detected in comparable galaxies.

\subsection{Age \& Metallicity of the Stellar Envelope}  \label{stellarpopsec}

The vertical color gradients shown in
Figures~\ref{profilefig}~--~\ref{slopevsVcfig} and found in other
studies are almost certainly due to some combination of vertical
variations in the mean stellar population and reddening due to dust
extinction.  In the most massive galaxies ($V_c\gtrsim120\kms$), the
presence of strong dust lanes suggests that reddening dominates the
color gradient close to the plane.  However, other detailed work
modeling massive edge-on disks suggests that the dust is confined to a
thin plane, and is unlikely to affect the observed colors at large
scale heights (e.g.\ Xilouris et al.\ 1999).  In the less massive
galaxies without dust lanes, reddening is unlikely to play a dominant
role in shaping the color gradients (see \S\ref{dustsec}, and Matthews
\& Wood 2001).  The color gradients are therefore likely to be due to
true changes in the stellar population in the lower mass galaxies and
in the dust-free regions well above the plane of the high mass
galaxies (de Grijs \& Peletier 2000).

We now turn to stellar synthesis models to explore what changes in the
mean age and metallicity of the stellar population would be consistent
with the observed color differences between the thin disk and
surrounding envelope.  As in Bell \& de Jong (2000), we used the
combination of $B-R$ and $R-K_s$ colors to help separate the effects
of age and metallicity.  To do so, we have used Bruzual \& Charlot
(2001) models to calculate the colors of stellar populations for a
range in metallicity and for star formation rates which are either decreasing
or increasing exponentially from a starting time $t_0$ to the present
($\tau\!>\!0$ for $SFR\propto e^{-t/\tau}$, and $\tau\!<\!0$ for
$SFR\propto e^{(t_0-t)/\tau}$, respectively), assuming a Scalo initial
mass function (IMF).

In Figure~\ref{gridgradfig} we plot these stellar population grids
superimposed on the colors of the FGC galaxies at 1~--~4 scale heights
above the plane.  At $z/z_{1/2}\!=\!1$, the colors are spread out over
a very wide range, as shown earlier in Figure~\ref{overlayproffig}.
The reddest points ($R-K_s\!>\!2.4$) have clearly diverged from the
stellar population grids, suggesting that the colors cannot be due to
any reasonable stellar population, confirming our belief that dust is
strongly affecting the colors of these galaxies near their midplanes.
At increasing scale heights, however, the color distribution narrows,
with the bluest galaxies becoming redder, and the reddest, extincted
galaxies become bluer.  Note, however, that some of the intrinsically
bluest galaxies are sufficiently faint in $K_s$ that their $R-K_s$
colors cannot be traced beyond $2\!-\!3z_{1/2}$, and thus convergence
to the mean color is an artifact in some cases (see also
Figure~\ref{overlayproffig}).

\begin{figure*}[t]
\hbox{ 
\includegraphics[width=2.9in]{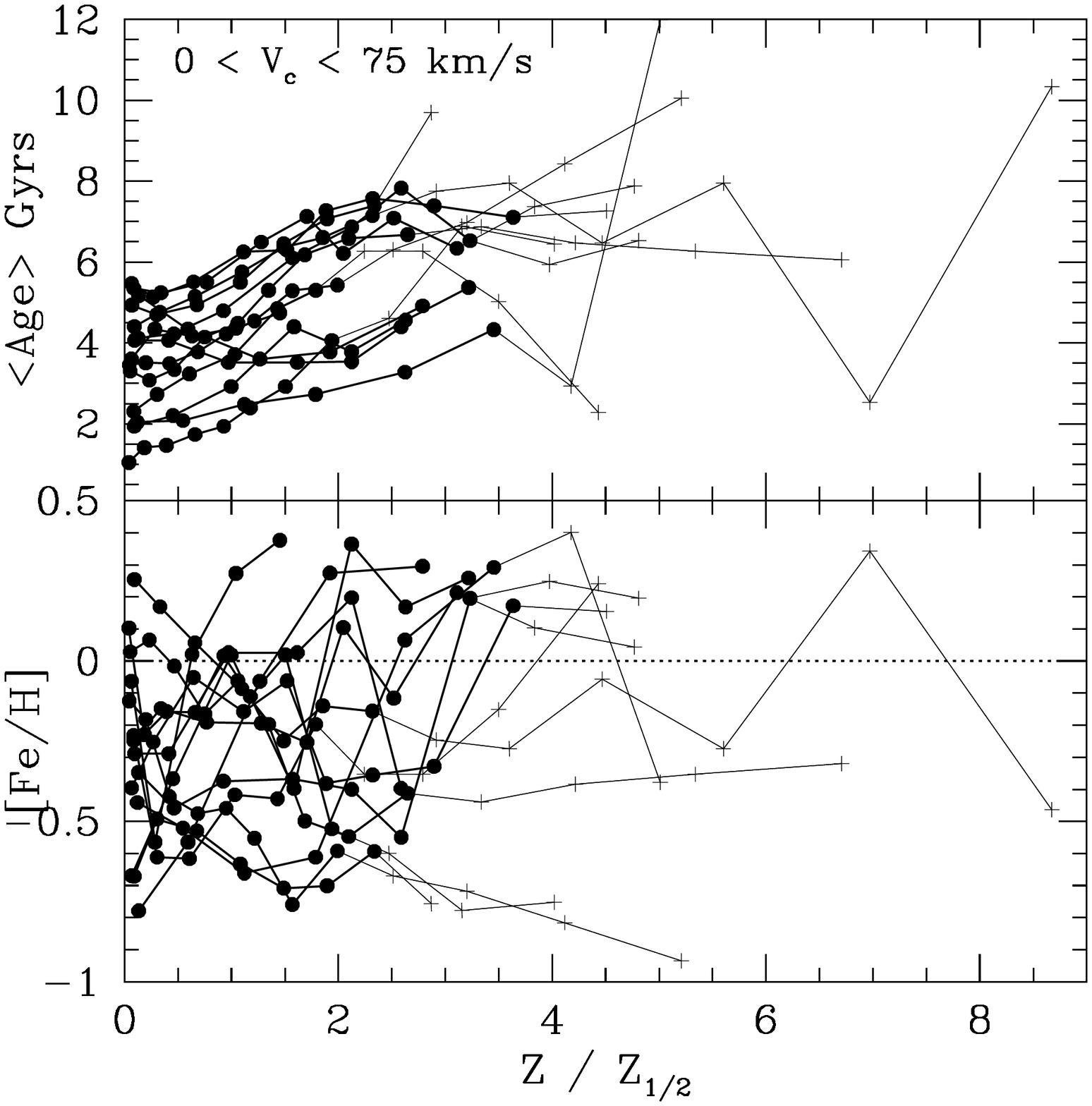}
\includegraphics[width=2.9in]{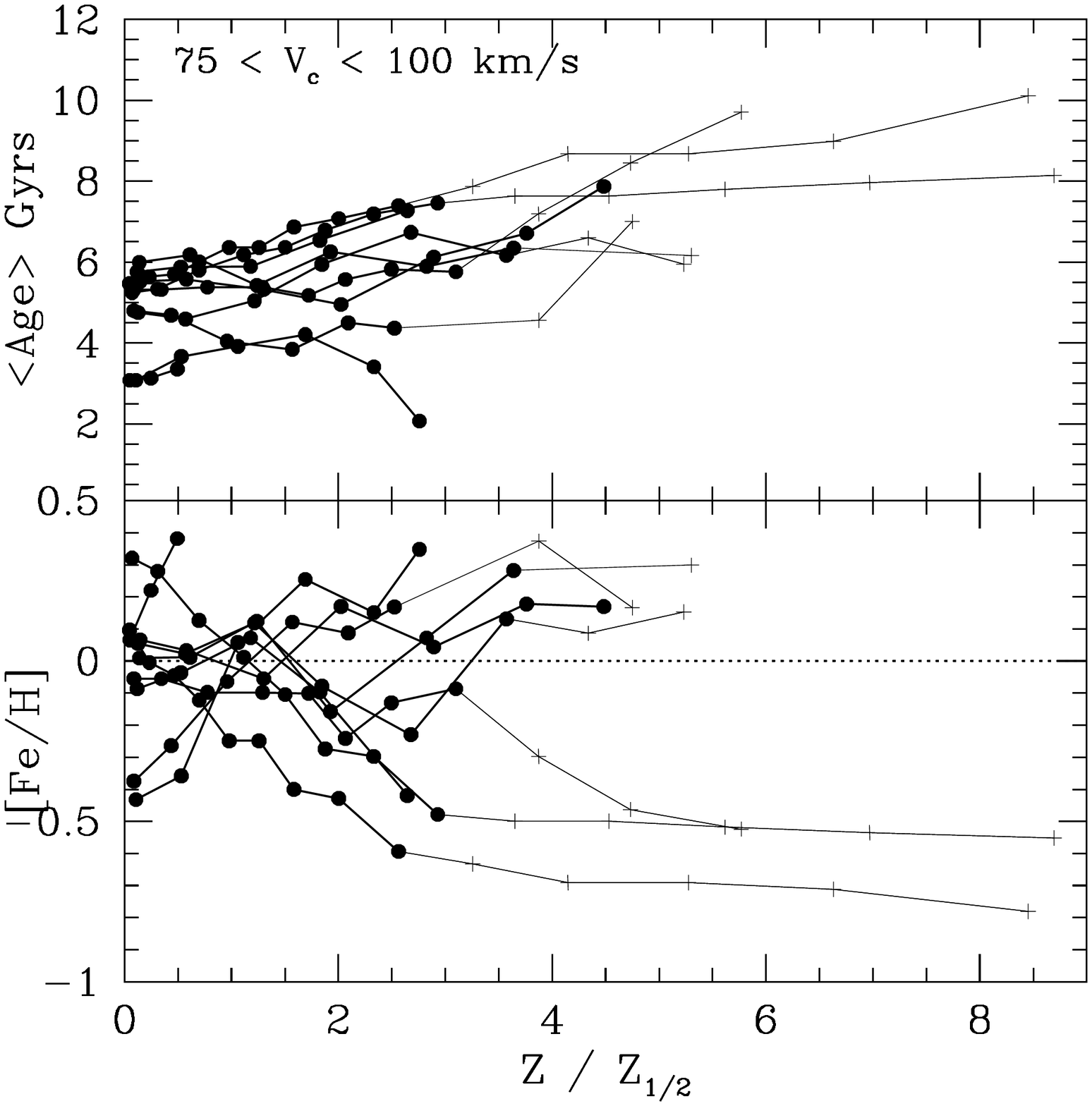}}
\vspace{0.1in}
\hbox{
\includegraphics[width=2.9in]{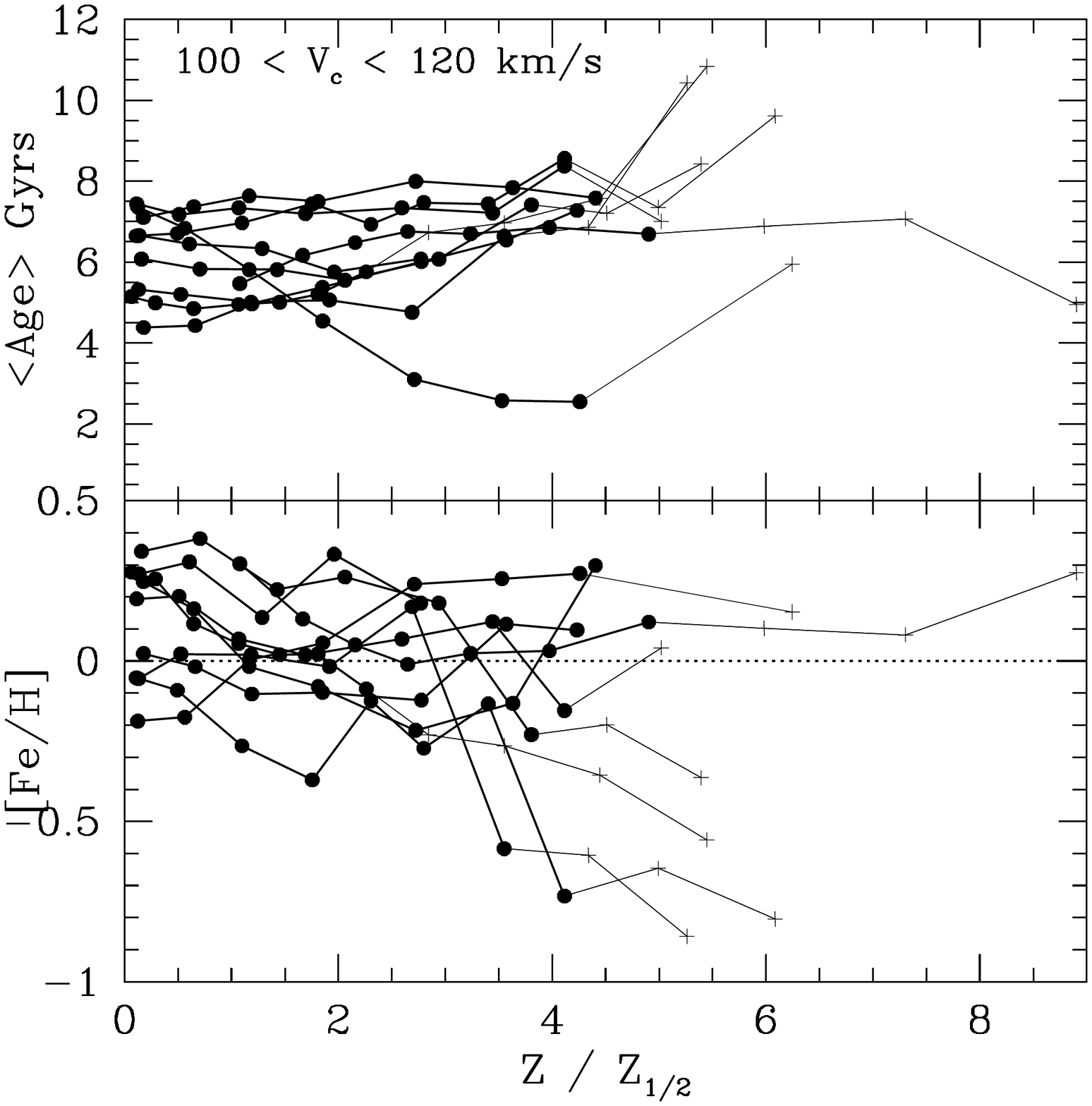}
\includegraphics[width=2.9in]{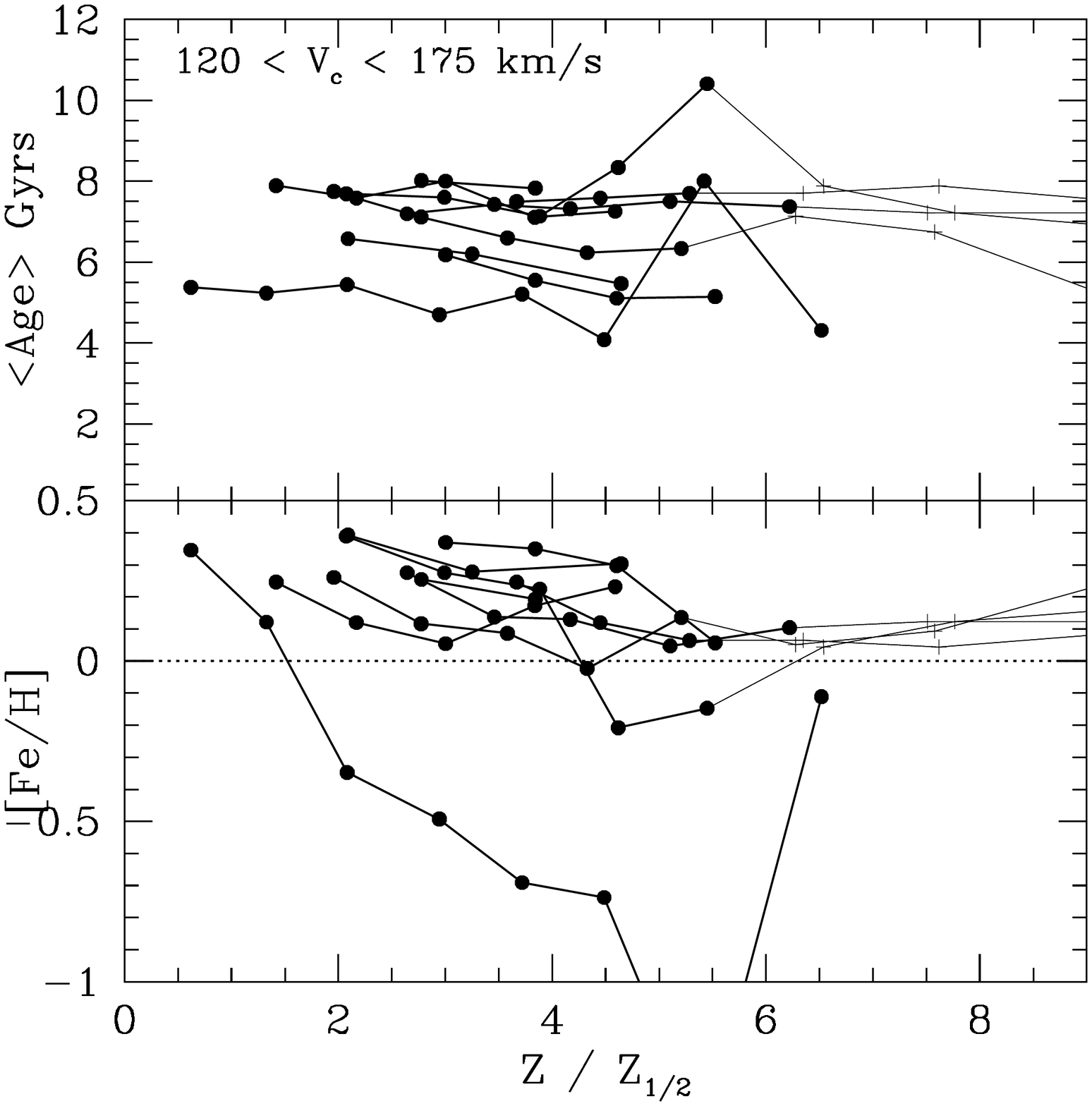}}
\caption{\footnotesize 
Mean light-weighted ages and metallicities as a
function of scale height, derived from the color profiles in
Figure~\ref{profilefig} by interpolating stellar population grids
(Figure~\ref{gridgradfig}) of age $12\Gyr$.  Profiles have been sorted
into 4 ranges of circular velocity $V_c$, to show trends in the
population as a function of mass.  At scale heights where there is no
$R-K$ data, the last reliably measured value of $R-K$ is used; points
which use these extrapolated values are marked with crosses.  No data
is plotted when the $B-R$ and $R-K$ points fall outside of a grid (for
example, if the stellar population is younger than any of the grid
points or if there is substantial extinction; see
Figure~\ref{gridgradfig}).
\label{ageZfig}}
\end{figure*}

In addition to having a narrower range of colors, the stars well above
the plane are consistently red.  As can be seen from the overlayed
grid, this color shift parallels tracks of constant metallicity and
thus is likely driven largely by an increase in the typical age of the
stellar population at increasing scale height.  We may examine this
behavior in more detail by interpolating the color profiles of
individual galaxies onto the age-metallicity grid.  We derive the
mean stellar age and metallicity profiles for each galaxy as follows.
First, we take the $B-R$ and $R-K_s$ color profiles shown in
Figure~\ref{profilefig}.  Then, because the $B-R$ colors can be traced
to much larger scale heights than the $R-K_s$ colors, we extrapolate
the $R-K_s$ color using the last reliably measured value as an
estimate of the missing $R-K_s$ data.  We then interpolate the colors
onto a series of stellar population grids.  We have allowed the
starting time of the star formation to vary ($t_0=2,5,\&12\Gyr$) to
explore the sizes of the systematic uncertainties inherent in
interpreting the broad-band colors\footnote{Note that for a single
  assumed value of $t_0$, there is a maximum possible age gradient
  between the disk and the envelope, corresponding to an age
  difference of $\Delta t = t_0$.  Thus, the apparent age gradients
  for the $t_0=2\Gyr$ models are quite small.  The largest age
  gradients are possible if different values of $t_0$ apply to the
  disk and the surrounding red envelope (i.e.\ $t_0\sim2\Gyr$ for the
  thin disk and $t_0=12\Gyr$ for the envelope).}.  In
Figure~\ref{ageZfig} we show the resulting profiles for the $t_0=12\Gyr$ case.

Figure~\ref{ageZfig} shows several trends as a function of the typical
mass of the galaxy.  At very low masses ($V_c<75\kms$), the thin disks
have young luminosity weighted ages ($\sim1-5\Gyr$), and sub-solar
metallicities.  With increasing mass, the typical age of the thin disk
increases to the point where it becomes comparable to the age of the
old stellar envelope.  The mean stellar metallicity of the thin disk
increases with increasing mass as well.  In contrast to these mass
dependent trends in the properties of the thin disk, we see no
significant age variations in the much older stellar envelope.  In
almost all cases the stellar envelope reaches mean ages of $6-8\Gyr$,
even when the thin disk is very young.  This conclusion is not highly
dependent on our extrapolation of the $R-K_s$ colors; Inspection of
the stellar population grids shows that the well-measured $B-R$ colors
alone are a reasonably strong age indicator, with only a weak
dependence of age on $R-K_s$.  The observed reddening in $B-R$ at
large scale heights could have only implied a younger stellar envelope
if the surrounding stars were much more metal rich, a possibility we
consider unlikely.

It is more difficult to make secure claims about the luminosity
weighted metallicity of the red stellar envelope, and the relative
metallicity of the envelope and thin disk.  The measurement of the
metallicity depends sensitively on the $R-K_s$ color, but due to the
low surface brightness of the envelope and the difficulty of
performing near-IR surface photometry at low light levels,
we are limited in our ability to constrain even the sign of the
metallicity gradient in many of the galaxies.  This limitation is most
severe in the lower mass galaxies due to a very strong correlation
between mass and infrared surface brightness within our sample;
the lowest mass galaxies are nearly invisible in the near-IR (see
the $K_s$-band images in Paper I).  Comparing the size of the typical
$R-K_s$ errors to the stellar population grids in
Figure~\ref{gridgradfig} suggests that the interpolated metallicities
are uncertain by at least 0.5 dex for the bluest galaxies.  In some of
the higher mass, higher surface brightness galaxies, there are
reliable indications of a declining metallicity towards increasing
scale height, although in the very highest mass galaxies with clear
dust lanes, there may be lingering concerns about the presence of
dust, even at these very large scale heights.  The trends towards
lower metallicity with increasing scale height are most clear in the
most spatially well resolved galaxies, and can seen as a bluing of the
$R-K_s$ colors towards larger scale heights in
Figure~\ref{colormapfig} (see FGC 780, 1285, 913, 1415, E1440, 1945,
979, 277, 1043, \& 1440 for example).  We do not see any significant
gradients in the spatially well resolved lower mass galaxies, because
of both the lower signal-to-noise and the lower metallicity of the
thin disk itself.  We place little faith in metallicity gradients
based upon the extrapolated values of $R-K_s$.  Inspection of the
grids in Figure~\ref{gridgradfig} shows that moving to redder $B-R$
colors at constant $R-K_s$ on the stellar population grids
automatically implies lower metallicities. Only a slight increase in
$R-K_s$ would be needed to erase the metallicity gradient.

Although absolute ages are difficult to characterize, vertical
gradients in age and metallicity are more coherent and show notable
trends.  To see this, we first derive $d\langle Age\rangle /d(B-R)$,
$d\langle Age\rangle /d(R-K_s)$, $d{\rm [Fe/H]}/d(B-R)$, and $d{\rm
  [Fe/H]}/d(R-K_s)$ for arbitrary color.  We then transform the
measured color gradients to age and metallicity gradients using
$d\langle Age\rangle /d(z/z_{1/2}) \approx (d\langle Age\rangle
/d(B-R))\times(d(B-R)/d(z/z_{1/2})) + (d\langle Age\rangle
/d(R-K))\times(d(R-K)/d(z/z_{1/2}))$ (and similarly for [Fe/H]).

\begin{figure}[t]
\includegraphics[width=3.5in]{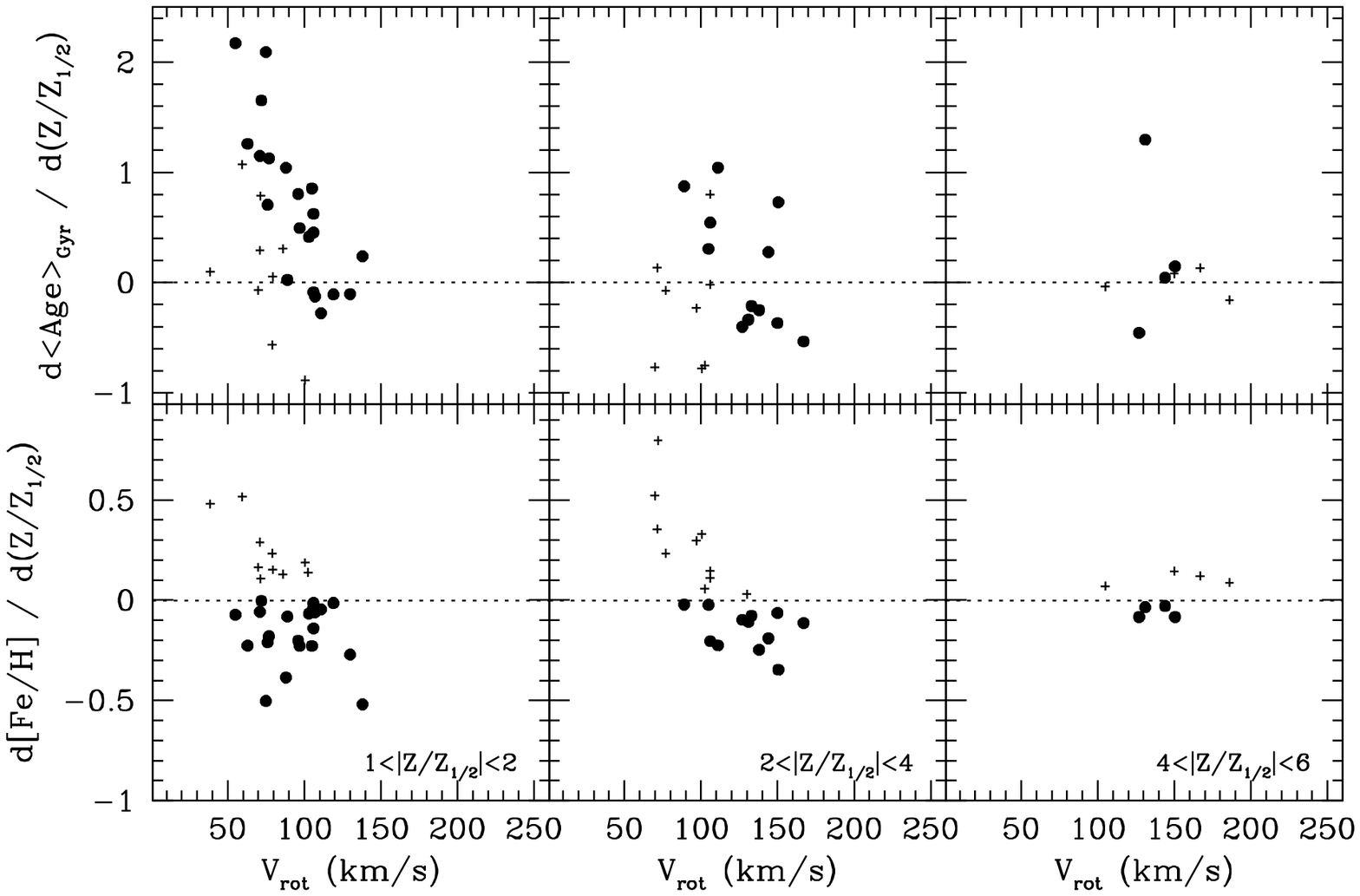}
\vspace{-1in}
\caption{\footnotesize 
The vertical gradient of the mean age (in
Gyr, upper panels) and metallicity ([Fe/H], lower panels) as a
function of rotation speed, in height intervals of $1-2 Z_{1/2}$ (left
panels), $2-4 Z_{1/2}$ (center panels), and $4-6 Z_{1/2}$ (right
panels), measured within a radius of 3 scale lengths.  Points were
derived using the $B-R$ and $R-K$ color gradients shown in
Figure~\ref{slopevsVcfig}, and translated into age and metallicity by
interpolating the gradient of the stellar population grid with
$t_0=12\Gyr$.  Solid points are for galaxies where the metallicity
decreases with increasing scale height.  Only points with slope
uncertainties of less than 0.05 in $B-R$ or 0.2 in $R-K_s$ have been
plotted.  The dashed lines marks zero gradient.
\label{agezslopevsVcfig}}
\end{figure}

The resulting gradient amplitudes are plotted in
Figure~\ref{agezslopevsVcfig} as a function of the host galaxy's
rotation speed, analagously to Figure~\ref{slopevsVcfig}.  As
expected, the age gradients are typically positive out to $4-6$ scale
heights, with typical slopes of $\sim0.5-1\Gyr$ per scale height.  The
age gradients are steepest for the lowest mass galaxies
($\sim1-2\Gyr/z_{1/2}$) with the youngest star forming disks.
The age gradients become more shallow with increasing scale
height, as the light in the profile begins to be dominated by a
presumably more uniform, older stellar envelope.  We have also plotted
the metallicity gradients, but given the uncertain influences on these
values the lack of coherence is not surprising and specific values for
individual galaxies should not be over-interpreted.  Based upon both
theory and Milky Way studies, we have a bias towards believing that
the metallicity of the stellar envelope should be less than or equal
to that of the disk\footnote{We can envision scenarios where the
  envelope does have higher metallicity than the disk, however.  For
  example, if the metals created in the initial formation of the
  envelope are blown out, then late-time infall of unpolluted gas
  could lead to a metal poor disk embedded in a metal enriched halo.
  This scenario would be more viable in the lower mass galaxies.
  Indeed, the measured metallicity gradients are largely positive in
  the low mass galaxies, but we feel that this probably
  reflects the difficulty in measuring the metal-sensitive $R-K_s$
  color in these very low surface brightness systems.}, but our
infrared data is not sufficiently reliable to find this trend in all
but the most massive, highest surface brightness galaxies.

\subsubsection{Caveats on the Derived Age \& Metallicity Gradients} \label{stellarpopcaveatsec}

We have argued that the interpretation of the $B-R$ color gradients is
sufficiently robust to indicate a definite age gradient towards older
stellar populations at larger scale heights (in low mass galaxies in
particular).  However, the exact {\emph{amplitudes}} of the age
gradients in Figure~\ref{ageZfig}~\&~\ref{agezslopevsVcfig} are rather
uncertain, even if their sign is not.

One of the principal limitations in deriving the age and metallicty
gradients is our reliance on stellar population synthesis models
to translate the observed broad-band colors into ages and
metallicities.  We have calculated the gradients assuming that all
star formation began $t_0=12\Gyr$ ago in both the envelope and the
thin disk, and proceeded according to a declining or increasing
exponential, with a Scalo initial mass function (IMF) between
$0.1\msun$ and $100\msun$.  There is no reason for these exact
assumptions to be true.  First, while the old stellar envelope could
have begun star formation $12\Gyr$ ago, the thin disks may not have
begun forming stars until much later.  We have inspected profiles similar to
those in Figure~\ref{ageZfig}, but for younger $t_0$, and find that a
more recent onset of star formation in the thin disk would imply even
steeper gradients in the mean age than we have derived for the
same observed color gradient.  Even the galaxies with apparently
constant age gradients (for a single assumed value of $t_0$, e.g. FGC
227, 395, 1043) might indeed have significant age differences between
the thin disk and the envelope, if the appropriate value of $t_0$
increases with increasing scale height.  Second, it is also possible
that the stellar envelope is better modelled by a truncated star
formation history, rather than a smoothly declining one.  If the
envelope formed from rapid thickening of a previously thin disk, then
star formation presumably stopped when the thickening took place.
Finally, there is always the possibility that star formation in
unusual environments does not produce stars with a standard IMF, such
that the colors of the stellar population of the envelope are significantly
different than one would expect for a given age and metallicity (e.g.\ see
observations of NGC 5907's halo by Zepf et al.\ 2000).

Another potentially significant uncertainty in the derived age
gradients is that we have neglected the effects of dust.  While we are
not currently making a detailed accounting of the contribution of
dust to the measured age gradient, we can instead argue that if
significant amounts of dust are present, they are distributed in such
a way that the true age gradient is probably even steeper than we have
derived.  Any reddening due to dust would likely be largest in the
plane, such that the true midplane colors would be even bluer than we
measured (see \S\ref{notsec} below for a fuller discussion of dust
distributions).  This would increase the color difference between the
midplane and surrounding envelope, increasing the age difference.
Thus, if the galaxies were reddened by significant amounts of dust,
then the true age gradients would be even steeper than derived above.
Although significant amounts of dust could lead us to underestimate
the age gradient in our galaxies, the $R-K_s$ color maps suggest that
dust plays little role in all but the most massive galaxies in our
sample (\S\ref{dustsec}).  Thus, we do not believe that dust is a
major uncertainty in the derived age gradients of the lower mass
galaxies.  Detailed modelling by Matthews \& Wood (2001) of an edge-on
system similar to those in our sample (UGC 7321, $V_{c}\sim100\kms$)
confirms that the overall reddening due to dust is small, and, moreover,
that any residual dust only weakens the measured color
gradients, suggesting a small revision towards a steeper derived
age gradient.

Even if the dust component is significant in an individual galaxy, its
distribution can drastically affect the exact degree of reddening
(Disney et al.\ 1989, Bianchi et al.\ 2000, Misiriotis \& Bianchi
2002).  In some geometries, large amounts of dust may in fact have
little effect on measured color gradients or the inferred mean ages
and metallicities.  Dust that is distributed in optically thick clumps
would not actually change the colors of the galaxies.  Only if the
clumps were optically thin in at least one of the filters would the
colors change, and then only in color combinations involving the
optically thin filter.  Given the wavelength sensitivity of dust
extinction to wavelength, color combinations involving only optical
wavelengths might in fact be less reddened than optical-infrared
colors (i.e.\ there might be more reddening in $R-K_s$ than in $B-R$,
surprisingly, if the dust clumps were optically thick in both $B$ and
$R$, but not $K_s$).  Indeed, inspection of Figure~\ref{gridfig} shows
that bringing the colors of galaxies with dustlanes (marked with
asterii) into alignment with the face-on de Jong (1996) disks would
require a much larger change in $R-K_s$ than in $B-R$.  If this is
true in general for our sample, then $R-K_s$ is strongly reddened, and
the inferred metallicities are too high, while the $B-R$ colors and
derived age are little affected.  This is additional evidence that the
inferred age gradients are unlikely to be strongly affected by
dust.

\section{Shapes of the Outer Isophotes}   \label{isophotesec}

In the previous sections we have outlined evidence for an old stellar
envelope surrounding the disks in our sample.  In addition to the
above analysis of the stellar population of the envelope, we can
measure the structure of the envelope by tracing the faintest
isophotes of the galaxies' surface brightness distributions.  At these
faint levels, the color maps suggest that the red envelope dominates
the light, and thus the shape of the isophote constrains the shape
of the stellar envelope.

We quantify the shape of the envelope by fitting ellipses
to the $R$-band isophotes plotted in Paper I (and partially reproduced
in Figure~\ref{colormapfig}).  We fit the isophotes using ellipses
with major and minor axes $a$ and $b$ and orientation $\theta$.  We
have also included a $\cos(4\theta)$ term of amplitude $C_4$ in the
isophote shape to allow for ``boxiness'' and ``diskiness'' in the
isophotes. Our definition of $C_4$ differs somewhat from what
is frequently used for elliptical galaxies.  First, we are fitting
changes in the {\emph{shape}} of the isophote, not changes of
intensity along a purely elliptical track.  Second, we generate the
perturbed ellipse by first generating a circle with $\delta
r(\theta)/r = 1 + C_4 \cos(4\theta)$ and then compressing it along the
minor axis by a factor of $b/a$.  This moves the ``nodes'' of the
boxiness or diskiness away from $45^\circ$ such that the
$\cos(4\theta)$ term produces pertubations more appropriate to our
extremely flattened distributions.  

\begin{figure}[t]
\includegraphics[width=3.5in]{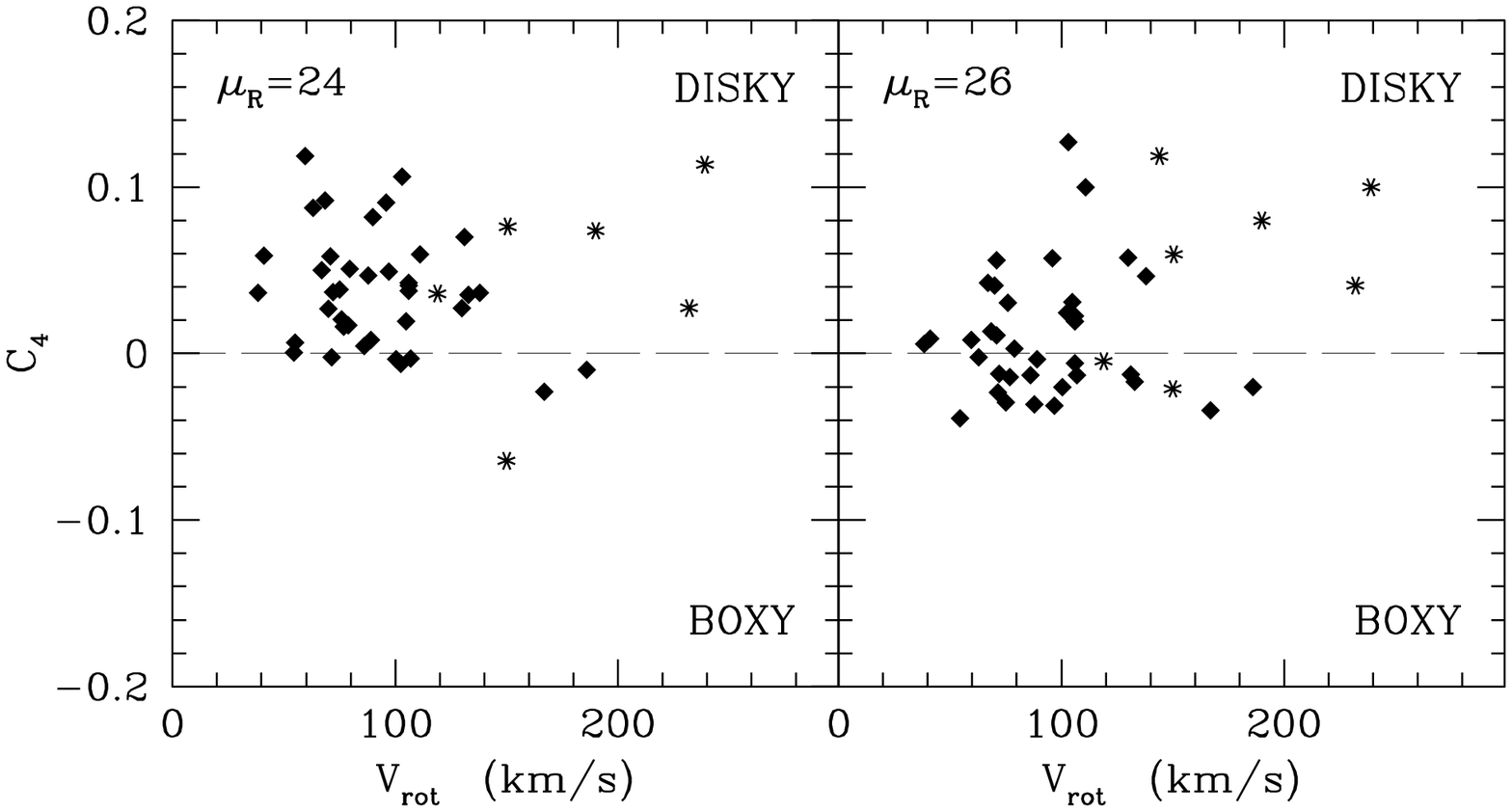}
\vspace{-1.5in}
\caption{\footnotesize 
The amplitude $C_4$ of $\cos4\theta$
deviations of ellipses fitted to $R$-band isophotes at
$\mu_R=24\surfb$ (left panel) and $\mu_R=26\surfb$ (right panel), as a
function of galaxy rotation speed $V_{rot}$.  The brighter inner
isophotes are disky ($\langle C_4 \rangle_{median}=0.037$), whereas
the fainter outer isophotes are as likely to be boxy as disky
($\langle C_4 \rangle_{median}=0.008$), particularly among the lower
mass galaxies where the faint stellar envelope is more dominant.  This
suggests that the radial scale length of the stellar envelope is not
tightly correlated with the scale length of the stellar disk.
\label{C4fig}}
\end{figure}

Figure~\ref{C4fig} shows the shape parameter $C_4$ for inner and outer
isophotes of the galaxies in the sample.  The brighter isophotes are
dominated by the thin disk and indeed they show that the galaxies are
almost uniformly disky in their brighter inner regions.  However, the
fainter outer isophotes dominated by the stellar envelope are as
likely to be boxy as disky.  This suggests that the radial extent of
the stellar envelope is somewhat decoupled from the disk.  When the
characteristic radial scale length of the envelope is longer (shorter)
than that of the disk, it leads to boxy (disky) outer isophotes.  It may
also be that the stellar envelope is prolate, and the variation in
boxiness/diskiness reflects differences in viewing angle.

\begin{figure}[t]
\includegraphics[width=3.5in]{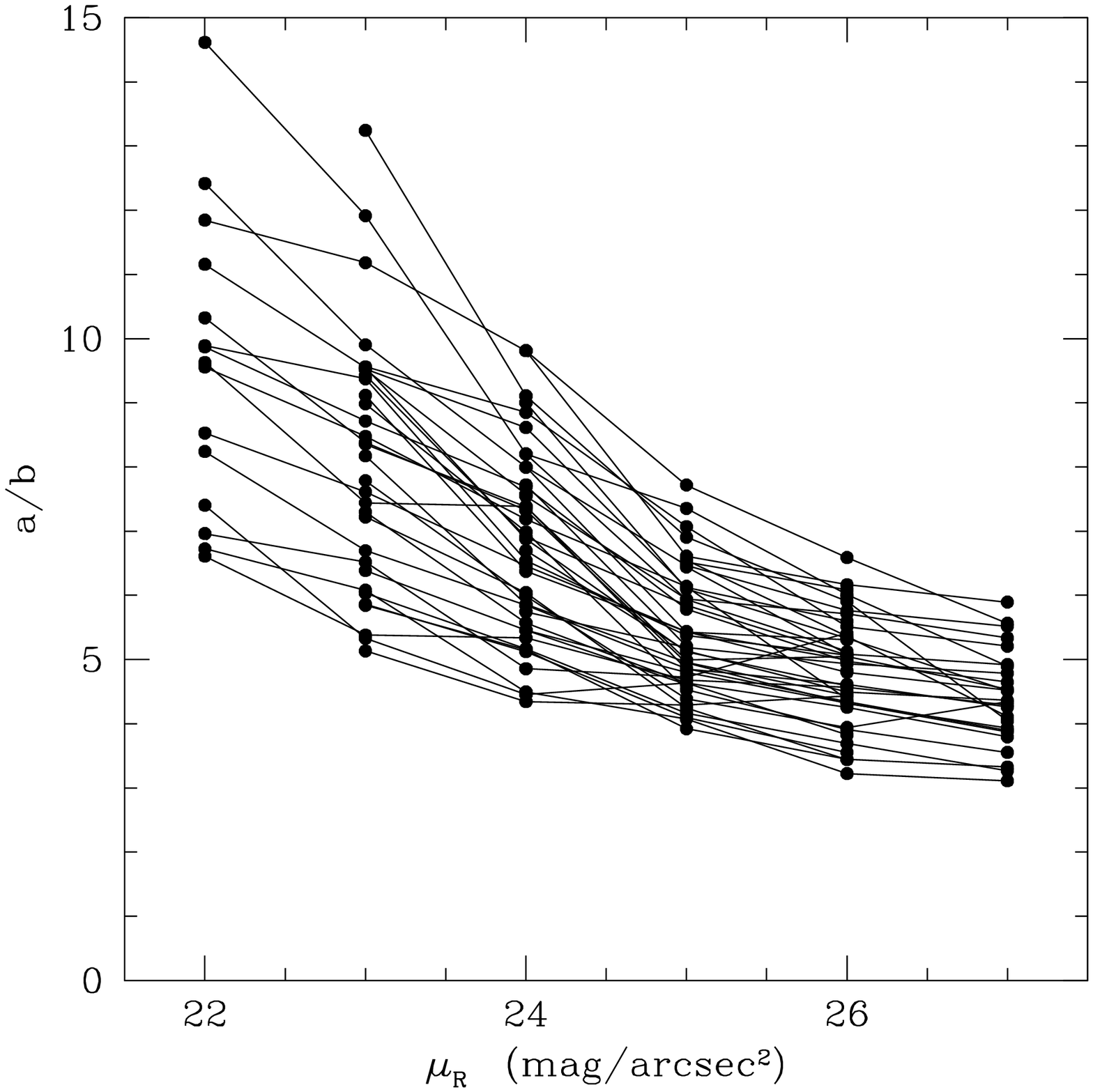}
\caption{\footnotesize 
Axial ratio $a/b$ of ellipses fitted to
$R$-band isophotes of surface brightness $\mu_R$, showing the tendency
of the edge-on galaxies to become progressively rounder at faint
isophote levels.  Axial ratios of a single galaxy are connected with
line segments, starting with the maximum axial ratio.  On average the
galaxies become rounder by a factor of 2.
\label{axialratiofig}}
\end{figure}

In addition to the difference in scale length between the envelope and
the embedded disk, we also see a significant change in axial ratio.
In Figure~\ref{axialratiofig} we plot the axial ratios of
our isophote fits as a function of the surface brightness of the
isophote.  The outer isophotes are clearly systematically rounder than
those tracing the inner disk, as can also been seen directly in the
faint $R$ band isophotes superimposed on the color maps plotted in
Figure~\ref{colormapfig}.  Note that the galaxies in our sample were
originally selected to have very thin axial ratios, with $a/b\!>\!8$.
Indeed, at the depth of the Palomar Sky Survey plates, the galaxies
all share a needle-like appearance.  However, at the faint isophotes
detectable in our deep, well-flattened images, the galaxies are
typically much rounder, with axial ratios closer to
$a/b\!\sim\!3\!-\!6$; at $\mu_R\!=\!27\surfb$, the mean axial ratio is
$\langle a/b \rangle \!=\! 4.4 \pm 0.7$.

Before the analysis in this paper, the reason for the thickening of
the disk at faint isophotes had been unclear, with possibilities being
radial truncation of the thin stellar disk (e.g.\ de Grijs et al.\ 
2001, Barteldrees \& Dettmar 1994, Pohlen et al.\ 2000, Kregel et al.\ 
2002), or a simple manifestation of steady vertical heating (i.e.\ the
age-velocity dispersion relation found in Milky Way disk stars; Weilen
1977).  However, as our data indicate, the transition between the
color of the thin disk and the surrounding red envelope is abrupt (at
least in the lower mass galaxies), suggesting that it does not arise
from a smooth change in the stellar population (as would have resulted
from any vertical heating not caused by a sudden merging event).
Instead, the galaxies with the deepest, most well-resolved images show
that the rounder outer isophotes trace a distinctly different stellar
population and not a continuation of the thinner disk.

\section{Stellar Envelopes in Massive Galaxies?}   \label{massivegalaxysec}

In contrast to in the lower mass galaxies ($V_c\lesssim100\kms$), the
distinction between envelope and thin disk stellar populations is less
obvious in the more massive galaxies.  While the outer isophotes of
the massive galaxies are as round as in the lower mass
systems\footnote{The axial ratio of the outer isophotes is completely
  uncorrelated with rotation speed, with a Spearman correlation
  coefficient of 0.2 at $\mu_R\!=\!26\surfb$ and 0.01 at
  $\mu_R\!=\!27\surfb$.} they do not have particularly strong color
gradients outside of the central dust lane.  However, there are a
number of reasons why we might not expect to easily detect a distinct
low surface brightness envelope in these systems, even if one exists.
First, the thin disks of the massive galaxies are higher surface
brightness overall than the lower mass galaxies.  This makes the thin
disk surface brightness more prominent at a given number of scale heights
above the plane, and can more easily mask a secondary population if
one exists.  Second, the thin disks of the massive galaxies are redder, and
thus there is no obvious color change to mark the transition between
the disk and the envelope.  Third, if the stars in the massive, high
surface brightness disks are also old (as would be consistent with
their observed red colors and with previous studies of face-on
galaxies, e.g.\ Bell \& de Jong 2000), then steady vertical heating of
the disk has had more time to create a thicker disk, pushing stars
from the thin disk to proportionally larger heights where they could
more easily mask a faint stellar population.

In summary, while Figure~\ref{colormapfig}
does not show obvious red stellar envelopes in the most massive
galaxies, it likewise does not provide strong evidence against them.
For now, we will defer to Occam's Razor: it would be
somewhat contrived to assume that the envelope population ceases
to exist exactly where it would become most difficult to detect, and so
we make the simpler assumption that the stellar envelope is likely to
be a feature of all disk-dominated galaxies, though a far less obvious
one around more massive, older disks.  Certainly the existence of the
thick disk of the Milky Way supports this supposition.

\section{Interpretion of the Red Stellar Envelope}  \label{interpretationsec}

In the above sections we have outlined several lines of evidence that
lead us to believe that we are detecting a relatively old, red stellar
envelope surrounding the disks in our sample.  We briefly summarize
the evidence as follows:

\begin{itemize}

\item Color maps of the galaxies (Figure~\ref{colormapfig}) show
clearly that the thin stellar disk is surrounded by a red envelope,
particularly in the lower mass galaxies where the the thin disk is
sufficiently blue to provide a strong color contrast.  In these
disks the transition between the disk and the envelope is abrupt,
and the difference in shape between the thin disk and the much rounder
red population is readily apparent.

\item Comparing the optical and infrared colors of the disks to the
  surrounding envelopes suggests that the redness of the stellar
  envelope is due primarily to age.  The age differences between the
  bluest disks and their surrounding envelopes seem to be at least
  $4\!-\!6\Gyr$, which would suggest that the envelope was in place by
  at least $z\sim0.5-0.9$ for flat, low-$\Omega_m$ models with
  $H_0=75\kms/\Mpc$, assuming an (unrealistic) age of zero for the
  thin disk.  If we instead adopt the absolute ages of 6-8$\Gyr$
  derived in Figure~\ref{ageZfig}, then the envelope was in place at
  redshifts of 0.9-1.5.  Uncertainties in the infrared color, the dust
  content, and the stellar synthesis models are all such that the
  likely formation epoch of the red envelope is even earlier.

\end{itemize}

We have also identified two main limitations to this work:

\begin{itemize}

\item The uncertainty in the measured $R-K_s$ colors limits our ability
to judge the relative metallicity of the disk and the envelope.  

\item In massive galaxies, the high surface brightness, red color, and
  greater thickness of the thin disk prohibits us from either
  confirming or ruling out the presence of the stellar envelope.

\end{itemize}

\subsection{What the Stellar Envelopes are Not}   \label{notsec}

The question now arises as to what processes have lead to the creation
of this relatively old and somewhat flattened stellar distribution
surrounding the thin disk.  Before we discuss the likely orgins for
this red stellar envelope, we briefly consider several alternative
explanations for the observed color maps (in order of increasingly
likelihood) and explain why we consider them unlikely.

{\sc High Latitude Dust:} One possible explanation for the redder
colors at large scale heights is an increasing dust content above the
plane of the galaxy.  This explanation would require there to be more
reddening above the plane of the thin disk than within it.  We find
this solution to be intractable for a number of reasons.  First, the
strongest colour gradients are in the lowest mass systems, which
should have the lowest metallicities (e.g.\ Stasinska \& Sodr\'e 2001,
Zaritsky et al.\ 1994), and thus the least raw material for forming
dust.  Second, only small amounts of extraplanar dust have been
identified in a few nearby edge-on galaxies (e.g.\ Howk \& Savage
1999), and the amounts have been dwarfed by the dust content closer to
the plane.  Moreover, few of the galaxies studied have this extra
component of dust, and those that do also have extraplanar ionized gas
and high star-formation rates as measured by their IRAS far-infrared (FIR)
luminosities.  Only three galaxies in our sample were detected in the
FIR, suggesting that the majority of our sample is unlikely to have the
extraplanar dust component.  Finally, detailed modelling by Matthews
\& Wood (2001) of the multi-color morphology of UGC 7321 suggests that
the dust in this galaxy (which is a thin bulgeless disk like the
galaxies in our sample) is sufficiently concentrated towards the plane
that it reddens the midplane slightly without changing the color of
the disk at high latitudes.

{\sc Warping and/or Current Accretion:} It is possible that some of
the high-latitude stars are due to transient phenomena such as warps
in the thin stellar disk or on-going accretion of very low surface
brightness material (such as the ring in NGC 5907; Xheng et al.\
1999).  However, there are a number of arguments against such
processes being widespread in this sample.  First, the existence of
the very tight relationship between a galaxy's rotation speed and the
amplitude of its color gradient (Figure~\ref{slopevsVcfig}) argues
strongly against a stochastic and/or transient origin for the
gradient.  Instead, the gradient must arise through a physical process
which is tightly coupled to a galaxy's mass, and whose after effects
are long-lived.  Second, the gradients in this paper have been
measured within one disk scale length of the center, where the effects
of warping are smaller than they would be on the outskirts.  Third,
while we do see some warping at large radii in a few galaxies, the
warps are quite small; only two galaxies have faint isophotes which
are fit by ellipses tilted by more than $2^\circ$ from their inner
regions, and none are tilted by more than $5^\circ$.  Finally,
Figure~\ref{allprofilesfig} shows that the color gradients are
typically quite symmetrical, which would not necessarily be expected
if warps or accreting satellites were responsible for the gradients.

{\sc Vertical Heating of a Thin Disk:} Among the more plausible
scenarios for creating the high-latitude red stellar population are
the many mechanisms capable of steadily increasing the vertical
velocity dispersion (and thus the scale height) of stars within a thin
disk.  Possible mechanisms include scattering off of giant molecular
clouds (Spitzer \& Schwarzschild 1951, 1953, Lacey 1984) and spiral
density waves (Carlberg \& Sellwood 1985).  In these models stars are
born with the velocity dispersion of the thin gaseous disk, and then
steadily ``heat'' vertically, gradually gaining velocity dispersion
and travelling to larger distances above and below the galactic plane.
Eventually these heating processes saturate, as the stars spend the
majority of their time outside of the thin disk, in regions where the
heating is least effective.  These mechanisms nicely explain the age
vs. velocity dispersion relation seen for disk stars in the Milky Way
(e.g. Wielen 1977, Quillen \& Garnett 2000, G\'omez et al.\ 1997,
Haywood et al.\ 1997); the most recent determinations find that the
vertical velocity dispersion of Milky Way stars rises steadily with
stellar age, but saturates at around $20\kms$ for stars more than
$3-6\Gyr$.  The combination of continuous star-formation and vertical
heating has also been used to explain the non-isothermal surface
brightness distribution seen in other edge-on disks (e.g. Dove \&
Thronson 1993).

If the vertical heating seen in the Milky Way exists in all disks,
then it would produce a color gradient which has the same sense as we
observe in our sample: the older, redder stellar populations would
have larger scale heights than the younger, bluer ones (ignoring
metallicity effects for the moment).  However, while this process may
be operating to some degree within our sample, it is difficult to
invoke it as the sole origin of the red envelopes, particularly for
the lowest mass disks in our sample, which have the strongest color
gradients and red envelopes which are 3-7 times thicker than the blue
thin disks at the outermost isophotes in Figure~\ref{colormapfig} (see
for example FGC 780).  There are several difficulties with producing
these envelopes via vertical heating.  First, creating such large
thicknesses through steady disk heating is nearly impossible, due to
the saturation of most heating processes.  Second, while the data 
requires that the vertical heating be strongest in the
lowest mass galaxies, all of the processes that are thought to
be responsible for vertical heating are probably {\emph{weakest}} in
the low mass disks.  These disks all have very low surface densities
($\Sigma_0\approx5-100 \msun/\pc^2$, based upon the deprojected $K_s$
band surface brightnesses, assuming $M/L_{K_s}\sim0.4$) and their
face-on counterparts show neither spiral structure nor copious amounts
of molecular gas (if any).  Thus, they are unlikely to support the
large gravitational pertubations necessary for scattering stars out of
the plane.  Moreover, even when vertical heating is known to have been
efficient, such as in the disk of the Milky Way, the expected
amplitude of the resulting color gradient is relatively small.
Calculations by de Grijs \& Peletier (2000) show that the expected
$B-R$ color gradient should be less than 0.03 mag per scale height, when
the dependencies of metallicity and age with velocity dispersion are
included.

In addition to the above more theoretical arguments, there is more
empirical evidence that vertical heating alone is not responsible for
the stellar envelopes.  In particular, \S\ref{isophotesec} and
Figure~\ref{C4fig} suggest that the radial scale length of the
envelope is not tightly correlated with the radial scale length of the
disk, as we would expect for heating scenarios.  Instead, the faint
isophotes which trace the red envelope are equally likely to be boxy
as they are disky, with no dependence on the properties of the galaxy.
Furthermore, although the thin disks have a very wide range of color
(indicating a spread in mean age, stellar metallicity, and dust),
surface density, gas content, and rotation speed, the properties of
the stellar envelopes are quite similar from galaxy to galaxy; the
colors of the surrounding envelopes span a much more limited range
than their host disks, and their axial ratios are similar, varying by
$\pm$16\% at faint isophotes ($\mu_R=26,27\surfb$).  The strength of
most processes thought to drive vertical heating of stars should
vary tremendously within our sample, and thus it would be very
surprising for all the disks in our sample to create similar envelopes
through heating alone.

{\sc An Analog of the Stellar Halo of the Milky Way:} Another
plausible explanation for the old red stellar
envelope is that it is an analogue of the old metal poor stellar halo
of the Milky Way.  However, we find three strong reasons why the
stellar envelopes we have detected are significantly different than
the MW stellar halo.  Namely, as we discuss below, our stellar
envelope population is too red, too flat, and too bright to be an
immediate analogue of the MW stellar halo.

The first point of conflict is that, unlike the observed stellar
envelopes, the stars and globular clusters which are kinematically
identified as part the MW halo are generally very metal poor
([Fe/H]$\lesssim$-1.7, e.g.\ Chiba \& Beers 2000; see review by van
den Bergh 1996 for full references).  For the ages we infer for the
stars in the stellar envelope, this metallicity would imply a typical
$R-K_s$ color bluer than 1.7, much bluer than we observe for the red
envelopes in our sample.  To show this, in Figure~\ref{colorcompfig}
we have plotted both the colors of the red envelopes and an
approximate color for the MW halo.  We have generated the halo color
by assuming a $12\Gyr$ old population with a range of metallicities
that brackets 50\% of MW halo stars ($-1.2 > $[Fe/H]$ > -2$; Carney et
al.\ 1994).  Although the bluer color of the MW halo is compatible at
the $3\sigma$ level with the color of any individual envelope (due to
the limitations of accurate sky subtraction in the infrared), it is an
entire magnitude bluer than the median high latitude color we observe
(see Figure~\ref{overlayproffig}).  Barring a gross systematic error
in our analysis, it is highly unlikely that the typical stellar
envelope observed in our sample has a metallicity nearly as low as
that of the MW stellar halo.

The second discrepancy is in the shape of the stellar envelope.
Although kinematic and star count analyses suggest that the MW stellar
halo is flattened, it seems to be flattened by never more than a
factor of two, even in the inner halo (see Chen et al 2001, Chiba \&
Beers 2000, Yanny et al 2000, Layden 1995, Larsen \& Humphreys 1994,
Kinman et al 1994 for recent determinations using a variety of
methods).  In contrast, the isophotes tracing the stellar envelopes in
our sample seem to have converged to an axial ratio which is twice as
flat as the inner MW halo ($\langle a/b\rangle\!\sim\!4.4$;
Figure~\ref{axialratiofig}), with no galaxy showing an axial ratio
flatter than 3:1.  If anything, we would expect that any halo analogs
in our sample should be {\emph{rounder}} than in the MW.
The flattening of the MW's inner halo is probably
due to gravitational compression by the thin massive disk (e.g.\ Bekki
\& Chiba 2001, Flores 1980).  However, the galaxies in our sample have
disks that are much less massive and lower surface density than the MW
and as a result should be even less efficient at flattening their
stellar halos than the MW.  Thus, it is very unlikely that the flatter
stellar envelopes we see in our sample are more extreme versions of
the MW's halo.

The third major difference between the MW stellar halo and the stellar
envelopes is in their relative surface brightness.  While the stellar
envelopes we have detected are extremely faint, they are still much
brighter than we would expect based on the relative density of the MW
halo and disk.  Most studies of the Galaxy's stellar halo suggest that
its density at the solar circle represents only a tiny fraction of the
total density ($\lesssim\!0.15$\%; see compilation in Chen et al.\
2001).  Assuming for the moment that the MW halo and disk have
comparable mass-to-light ratios, the relative surface brightness of
the two components at $R\sim2h_r$ should be
$\sim\!\Delta\mu\sim7\surfb$ at the midplane.  The galaxies in our
sample have {\emph{maximum}} surface brightnesses of
$\mu_B\sim22\!-\!23\surfb$, implying a {\emph{peak}} surface
brightness for the halo component of nearly $30\surfb$ in $B$, and an
even fainter surface brightness above the plane.  More realistically,
the likely surface brightness difference between the disk and halo is
even larger, given the low mass-to-light ratio of the young disk
compared to the higher mass-to-light ratio of the older envelope.  The
situation might not be quite as bleak in the center of the galaxy
along the minor axis, where the steeper density profile of the MW halo
(roughly $r^{-3}$) relative to the disk ($e^{-r}$) leads the halo to
become proportionally more important.  However in detailed models of
the MW, Morrison et al (1997) find that, even in the center, the light
from the halo only just becomes comparable to the light in the thick
disk at the very limits of detectability ($\mu_R\sim28\surfb$).  In
short, we do not expect true analogs of the MW stellar halo to be
detectable in our sample galaxies, unless through direct analysis of
resolved stellar populations (e.g.\ M31; Sarajedini \& van Duyne 2001,
Holland et al.\ 1996).


\subsection{The Red Stellar Envelopes: Universal Thick Disks}

Having eliminated the above scenarios, we are left with one 
viable explanation for the stellar envelope.  Namely, we believe that
the envelopes which we have detected are likely to be analogs of the
Milky Way's old thick disk (for detailed properties of the thick disk,
see recent reviews by Wyse 2000, Norris 1999, van den Bergh 1996,
Majewski 1993).  The observed colors of the envelopes are consistent
with their being relatively old and somewhat metal-poor, like the
Milky Way thick disk, and their geometry suggests a similar axial
ratio as well.

\subsubsection{Comparison with the Milky Way Thick Disk}

\begin{figure}[t]
\includegraphics[width=3.5in]{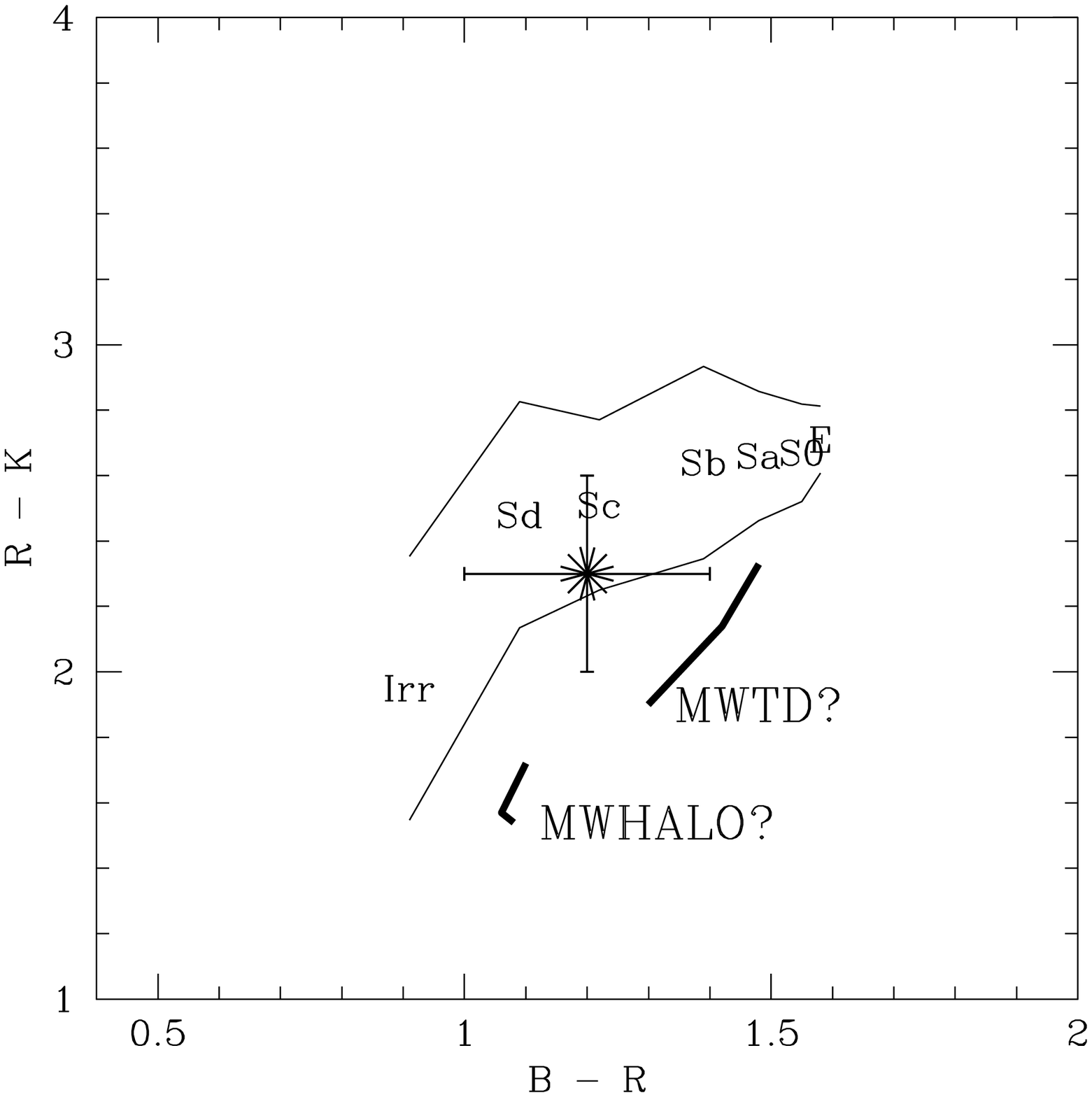}
\caption{\footnotesize 
Comparison of the $B-R$ and $R-K$ colors
  of the thick disks detected in our sample (large asterix with error
  bars) with the mean colors of galaxies (denoted with their
  morphological type, with thin solid lines idicating the range of
  observed colors; Mannucci et al.\ 2001), the Milky Way thick disk
  (assuming [Fe/H] between -0.4 and -1.0), and the Milky Way stellar
  halo (assuming [Fe/H] between -1.2 and -2.0; Carney et al 1994).
  The colors of the Milky Way thick disk and stellar halo have been
  estimated using the stellar population grids in
  Figure~\ref{gridgradfig} and a fixed age of $12\Gyr$, and are thus
  only approximate.
\label{colorcompfig}}
\end{figure}

In detail, the typical metallicity of the Milky Way thick disk is
somewhat metal poor, with a mean [Fe/H]$\!\approx\!-0.7 - -0.5$
(Gilmore \& Wyse 1985, Carney et al.\ 1989, Gilmore et al.\ 1995,
Layden 1995, Robin et al.\ 1996).  It is also thought to be relatively
old, with a typical age comparable to the metal rich globular clusters
($\sim\!12\Gyr$; Gilmore et al.\ 1995). Based on the $12\Gyr$ old
stellar grids in Figure~\ref{gridgradfig}, the likely age and
metallicity of the MW thick disk would correspond to a typical color
of $R-K_s\!\sim\!1.8-2.3$ and $B-R\sim1.3-1.5$ or redder.
Figure~\ref{colorcompfig} shows the locus of $B-R$ and $R-K_s$ colors
of a $12\Gyr$ old population with the range of metallicities observed
for the MW thick disk stars ($-0.4 > $[Fe/H]$ > -1$), along with error
bars representing the range of colors spanned by our sample at high
latitudes.  The fiducial $R-K_s$ color of the MW thick disk is
slightly bluer than the envelope colors observed in our sample at high latitude
(by a few tenths of a magnitude at most), and slightly redder in $B-R$
(again by a few tenths).  Given the systematic uncertainties in the IR
colors of stellar population models (i.e.  Charlot et al.\ 1996) and
in the low mass end of the IMF, the IR color difference is probably
not significant, and thus we consider the optical and IR colors of the
red stellar envelopes to be consistent with the likely color of the MW
thick disk.  However, the more accurately measured $B-R$ colors may
suggest a slightly younger mean age for the stellar envelopes in our
sample than for the MW thick disk, as would be consistent with the
overall younger age of the thin disks in the majority of our sample.

In addition to the color similarities,
the red stellar envelopes have roughly the same geometry as the MW thick disk.
Recent studies using star counts and proper motions along many
different sightlines within our galaxy (most recently Chen et al.\ 
2001 (SDSS), Ojha 2001 (2MASS), Buser et al.\ 1999, Robin et al.\ 
1996) suggest that the MW thick disk population has a scale height
$\sim\!600\!-\!900\pc$, roughly 2-4 times thicker than the old thin
disk.  Recent determinations of its radial scale length are in the range
$2.8\kpc$  (Robin et al.\ 1996) to $4.5\kpc$ (Ng et al.\ 1997), giving an
overall axial ratio for the MW thick disk in the range of 3:1 - 7:1.
These axial ratio values agree well with the range of observed axial
ratios traced by the red envelope (Figure~\ref{axialratiofig}), as do
the apparent relative thickness of the thin blue disk and the thicker
envelope.

Finally, the red envelope and the embedded thin disk in our sample
appear to have a relative brightness comparable to the thick and thin
disks of the MW.  Measurements within the MW suggest a normalization
of the density of the thick disk between 4\% and 13\% of the thin disk
at the solar circle midplane.  (These normalizations are highly
anticorrelated with the derived thick disk scale height, with puffier
disks having lower local density; see error ellipses in Chen et
al.\ 2001).  Although detailed fitting of the thin disk and envelope
components will be deferred to a later paper, the surface brightness
level of the isophotes where the red population begins to dominate is
bright enough ($\mu_R\sim25-26\surfb$) that the surface brightness of
the envelope cannot be less than a few percent of the thin disk's.
Nor can the surface brightness be greater than 50\%, as the light from
the younger thin disk clearly dominates at the midplane.  This puts
the likely relative brightness of the thin and thick disks in the
range measured within the MW.

\subsubsection{Comparisons with Previous Detections of Thick Disks in Other Galaxies} \label{otherworksec}

In addition to the broad similarities between the red envelopes
discussed in this paper and the MW thick disk, there are close
connections between the thick disks reported in other external
galaxies and those we have detected in our sample.  This link is not
surprising, as reports of thick disks in other galaxies are also based
upon their similarity to the MW thick disk.  Since the first
detections of thick disks in external galaxies (Burstein 1979,
Tsikoudi 1979, van der Kruit \& Searle 1981) several studies have
attempted to identify thick disks in nearby galaxies.  Many of the
most recent galaxy decompositions of single band CCD images of edge-on
galaxies (e.g. Neeser et al.\ 2002, Abe et al.\ 1999, Rauscher et al
1998, Morrison et al 1997, N\"aslund \& J\"orsat\"er 1997, Morrison et
al.\ 1994, Shaw \& Gilmore 1990) find thick disks with scale heights
that are several times the height of the thin disk, and central
surface brightness normalizations of 5-15\% of the thin disk.  These
results are fully compatible with the red envelopes detected in our
sample.  Likewise, HST observations of resolved stars in the outer
disk of M31 tentatively suggest a detection of an old, slightly metal
poor ([Fe/H]$\sim$-0.2) thick disk (Sarajedini \& van Duyne 2001),
which is again in good agreement with the mean stellar population
properties estimated for our sample.

Even more striking is the agreement between our data and the few cases
where light has been traced to high latitudes in more than one
bandpass, allowing reliable detections of color gradients to large
scale heights.  Such detections are presented by Matthews et
al. (1999) although they do not specifically identify the color
changes with a thick disk; van Dokkum et al (1994) where they analyze
NGC 6504, a massive Sab galaxy, and find no significant color gradient
in agreement with our results for massive disks; Lequeux et al (1998)
for NGC 5907, although Zheng et al (1999) attribute part of the high
latitude light to a remnant ring from a disrupted satellite; Wainscoat
et al.\ (1989); and Jensen \& Thuan (1982) where the Sb galaxy NGC
4565 is traced to very faint levels and shows a clear color change
associated with the onset of a flattened ``corona''.  In all these
cases the color of the high latitude populations agrees remarkably
well with the typical high latitude colors of the galaxies in our
sample, generally reflecting an old, but not particularly metal poor,
stellar population.

Clearly, our results differ from claims of thick disk
{\emph{non-detections}} in several respects.  In particular, a recent
summary of the literature by Morrison (1999) argues that extended
thick disks have only been detected in galaxies with medium sized or
larger bulges, suggesting an evolutionary link between the two
components.  In contrast, we find clear evidence for a thick, red
stellar component in the majority of our sample of {\emph{bulgeless}}
galaxies.  Abe et al.\ (1999), and more recently Neeser et al.\
(2002), also show counterexamples to the Morrison (1999) claim, but
through surface brightness profile fitting.  The strongest claims of
non-detections in deep imaging of late-type galaxies (e.g. Fry et al
1999 analysis of NGC 4244, and Morrison et al 1994 analysis of NGC
5907) are based upon 2-dimensional image decomposition of a single
$R$-band image, and it is likely that without the structural contrast
provided by a bulge it is far more difficult to cleanly decompose a
thin disk, thick disk, and halo population.  Indeed, many of the
galaxies in Figure~\ref{profilefig} show no strong inflection points
in their surface brightness profiles, in spite of having strong color
gradients.  Moreover, in the late-type galaxy NGC 5907 there
{\emph{is}} a detection of high-latitude light, though at a lower
relative brightness that would be expected for a strict analog of the
MW thick disk, which lead the authors to attribute this light to a
stellar halo instead.  Morrison (1999) reports that on the basis of
deep $I$ band imaging, this extra component would be compatible with a
MW-like thick disk if the radial scale length were longer than the
thin disk, as we see in many of our sample galaxies with boxy outer
isophotes, and a lower overall surface brightness\footnote{We note,
however, a thick disk attribution for the high-latitude light in NGC
5907 may not be compatible with the very low metallicity implied by
the lack of detected giant stars in deep HST NICMOS imaging (Zepf et
al.\ 2000)}.  In a future paper we will attempt to characterize the
range of normalizations and radial scale lengths which are compatible
with our data, and better characterize the fraction of a galaxy's
light which typically makes up the thick disk.

We also note that, in many ways, detection of thick disks should be
easier in our sample than in previous studies.  All earlier
investigations have been of the very nearest, well-studied edge-on
galaxies, maximizing spatial resolution while making the observations
feasible on small Schmidt telescopes.  However, because of their
extremely large angular extent, accurate flat fielding over the
physical scales required to detect the thick disk is more difficult
and foreground star subtraction is far more critical.  In contrast,
our sample has a larger mean distance than previously studied galaxies
(by a factor of 2-5), leading to a small sacrifice in spatial
resolution, compensated by a tremendous gain in the area unaffected by
foreground stars.  In the majority of our sample, the spatial
resolution is more than sufficient for detecting thick disks, without
the complication of PSF subtraction.

\section{Implications for Thick Disk Formation}      \label{formationsec}

Assuming now that the red stellar envelopes are indeed analogs of the
MW thick disk, we can draw a number of conclusions about the general
properties of this apparently ubiquitous component.  

\begin{itemize}

\item {\sc Thick disks are common around galaxy disks of all masses.}
Previously, firm detections of thick disks were confined to the Milky
Way and other relatively massive, bulge-dominated galaxies
($V_c\gtrsim200\kms$).  This work extends the mass range of galaxies
with detected thick disks down to very low masses ($V_c\sim35\kms$).
Moreover, there is no reason to believe that the thick disk population
is not present in even lower mass galaxies.  Many observations of the
color magnitude diagrams (CMDs) of lower mass ``transitional'' dwarf
spheroidals and dwarf irregulars show that the younger stars of the
galaxies are embedded within an old, quiescent stellar component which
extends to much larger radii (e.g., Antlia (Aparicio et al 1997,
Sarajedini et al.\ 1997); Phoenix (Held et al.\ 2001, 1999,
Mart\'inez-Delgado et al.\ 1999); SagDIG (Held et al.\ 2001); DDO 187
(Aparicio et al.\ 2000); WLM (Minniti \& Zijlstra 1996); Harbeck et
al.\ 2001).  Recently, Held et al.\ (2001) have argued that this old
component around dwarf spheroidals is in fact flattened.  It therefore is
possible that these are face-on detections of stellar envelopes
similar to those we have detected here in edge-on, higher mass
galaxies. {\emph{The apparent universality of thick disks suggests
that the process that leads to formation of the thick disk must be a
generic feature of disk galaxy formation.}}

\item {\sc Thick disks are not a by-product of bulge formation.}
Previously, it was thought that the formation of the thick disk might
be closely tied to the formation of the bulge, due to the facts that
(i) thick disks were detected almost exclusively around galaxies with
medium to large bulges (Morrison 1999) and that (ii) there is
remarkable similarity in abundance patterns between thick disk and low
metallicity stars in the bulge (Prochaska et al.\ 2001).  In contrast,
our sample contains only bulgeless galaxies.  Thus, the dynamical
processes which lead to the formation of a thick disk do not
{\emph{necessarily}} lead to the formation of a bulge, although they
might in some cases.  Alternatively, the {\emph{epoch}} of thick disk
formation might be very similar to the epoch of bulge formation within
an individual galaxy, such that they both form from the same gas
resevoir, leading to similar enrichment patterns.  However, the
dynamical processes responsible for each component might quite
distinct, as might be expected given their drastically different
angular momenta (Wyse \& Gilmore 1992).

\item {\sc Thick disk formation takes place early, even for galaxies
with a very young mean stellar age.} Within our sample, the stellar
populations of the thick disks are much more similar from galaxy to
galaxy than are the stellar populations of the embedded thin
disks.  The thin disks span a very wide range in color, driven
primarily by differences in mean stellar age.  However, the thick
disks have colors compatible with being relatively old (at least
$\gtrsim6\Gyr$, if not older), even if the mean stellar age of
the thin disk is young.  This suggests that thick disk formation is a very
early process even in galaxies which will take many gigayears to
either accrete most of their gas and/or convert that gas into
stars\footnote{We note that we could potentially have a small bias
against very young thick disks in our sample, because we purposely
excluded galaxies which had obvious signs of on-going interactions.
If thick disks are formed in mergers, we may have biased ourselves
against the most recently formed endpoints of this process.  The same
bias could have been applied to the the Flat Galaxy Catalog from which
our sample was extracted. However, the number of galaxies which we
excluded due to warping was actually quite small (fewer than 5 out of
candidate lists of nearly 200 galaxies), and moreover we do have many
galaxies in the sample which show evidence for small past disturbances
(i.e.\ slight warping and asymmetries) at faint surface brightnesses
(see Paper I).  We therefore consider it unlikely that we are missing
a large population of much younger thick disks.}.

\item {\sc The formation of the thick and thin disks are largely
    decoupled.} As seen in our sample, the thick and thin disks have
  different ages and radial scale lengths within the same galaxy.
  Similar differences are seen between the thin and thick disks of the
  Milky Way, in addition to significant metallicity and relative
  abundance differences (Prochaska et al.\ 2001, Gratton et al.\ 2000,
  Furhmann 1998).  This evidence suggests that, as has been
  suggested for the Milky Way (Gilmore \& Wyse 1985, Wyse \& Gilmore
  1986, although see review by Norris 1999), the formation of the two
  components occured through two distinct processes.

\end{itemize}

The above conclusions can be compared with possible formation
mechanisms for the thick disk.  Almost all proposed models break into
two major classes.  In the first, the thick disk is very old and grew
monolithically during the initial dissipational collapse of the galaxy
(e.g.\ Sandage \& Fouts 1987, Wyse \& Gilmore 1988, Burkert et al.\
1992).  In the second, the thick disk is the by-product of a
significant merging event, through either (1) a single episode of energetic
vertical heating of an existing thinner disk via accretion of a massive
satellite galaxy (Quinn \& Goodman 1986), or (2) through direct accretion
of thick disk material from a cannibalized galaxy (Statler 1989).
Because the similar chemical enrichment patterns among stars in the
inner halo, metal-poor bulge, and thick disk of the MW argue
against the Statler (1989) direct accretion model for producing thick
disks, we will concentrate hereafter on the vertical heating mechanism
when considering the merger scenario.

On initial examination, the data do not rule
out either monolithic collapse or merger origins for the thick disk.
The ubiquity and old age of the detected
thick disks initially argues in favor of a monolithic formation
scenario, as the more stochastic, on-going process of merging seems 
less likely to produce such a pervasive thick disk
population as we observe.  However, upon deeper consideration, a
merger origin for the thick disk is equally well supported by our data,
and is additionally bolstered by other evidence.  First, if
galaxies assemble hierarchically, as expected in most favored
cosmogonies, then merging should be a generic feature of the evolution
of galaxies, and there would be many opportunities for a young disk to
be heated by a significant merging event, generating a thick disk.  In
such a scenario, thick disks should be pervasive, as we find in our
sample.  Second, the merging rate tends to decline with time,
suggesting that if merging produces thick disks, then these
thick disks should be old, as we also find.  Third, the lack of any
strong correlation between the sizes of the thick and thin disks
suggests a rather stochastic origin for the thick disk.  Collapse
models tend to lead to similar scale lengths for the thick and thin
disks (Ferrini et al.\ 1994), and thus the apparent lack of correlation is
more compatible with a merging scenario, where the exact timing of the
last merging event would set the relative masses and sizes of the
older thick disk and the thin disk which subsequently accreted.

The merger scenario for thick disk formation is additionally supported
by mounting evidence from the Milky Way.  A substantial ($\sim\!20$\%)
merging event is strongly suggested by the sharp
factor of 2 increase in velocity dispersion $\sim\!10\Gyr$ ago
(Quillen \& Garnett 2000), by the apparent floor in the metallicity of
thick disk stars (Chiba \& Beers 2000) suggesting substantial
pre-enrichment, by the lack of a vertical metallicity gradient in the
MW thick disk (Gilmore et al.\ 1995, Chiba \& Beers 2000), and by
numerical simulations (recently, Bekki \& Chiba 2001).  However, this
epoch of merging must have been extremely early, as argued recently by
Wyse (2000).  Detailed studies of stars and clusters in the Milky Way
suggest that the youngest stars in the MW thick disk are at least as
old as the thick disk globular cluster 47 Tuc (i.e.\ $12-13\Gyr$;
Carretta et al.\ 2000, Lui \& Chaboyer 2000, Gilmore et al.\ 1995),
and that star formation began in the thick disk concurrently with the
old inner halo (Chaboyer et al.\ 1999).

If merging is the origin of the thick disks seen in our sample as well
as in the Milky Way, then our data also confirms that the merging
epoch had to be early.  Based upon the optical-IR colors of the thick
disks in our sample, we estimate a {\emph{minimum}} formation age of
$\sim\!6-8\Gyr$ ago, with the true age likely to be even older.  In
addition, the merging epoch must have peaked sufficiently early in the
galaxies' mass accretion history to have left sufficient time for the
majority of the baryonic material to settle into a thin disk.

In spite of the above, the somewhat smaller possibility remains that
the thick disks did form during a monolithic collapse.  Our bias
towards the merger scenario is based in part upon observations within
the Milky Way, some of which are uncertain (see Norris 1999 for a
discussion).  Secure observations of a vertical metallicity and/or
kinematic gradient within the thick disk would favor more ``ELS''
flavored models for the dissipational formation of the thick disk.
Kinematic observations of the galaxies in our sample could also
provide discriminating constraints between the merging and collapse
formation models.

\section{Conclusions}                     \label{conclusionsec} 

In this paper we have presented several lines of evidence which lead
us to conclude that thick disks are a common product of disk galaxy
formation for galaxies of all masses.  Specifically, we have analyzed
the color maps, vertical color gradients, and faint isophote shapes
for a large sample of edge-on bulgeless disks.  Our observations
are consistent with the conclusion that that all galaxies in the
sample are embedded within somewhat
flattened ($\sim$4:1) red stellar envelopes whose properties vary
little from galaxy to galaxy although the galaxies themselves span
an enormous range in mass and color.  We have used stellar synthesis
models to argue that the stellar envelopes are old (at
least $6\Gyr$, and probably older), but not necessarily 
metal poor ([Fe/H]$<$-1).  We argue that the
properties of the red stellar envelopes are consistent with their
being close analogs of the MW thick disk.

We find that the evidence in hand, from our sample and the Milky
Way, is consistent with a picture for disk galaxy formation which
procedes as follows: (1) a thin stellar disk forms at high redshift
($z\gtrsim1-2$); (2) partial disruption occurs during a significant
merger capable of dramatically heating the thin disk, but not
necessarily leading to the formation of a bulge; (3) the
{\emph{majority}} of the galaxy's stars form from gas gradually
accreted after the merger and the creation of the thick disk; and (4)
no significant merger events follow.  This model will
only be consistent with cosmological scenarios where the merging rate
peaks early on (i.e.\ low $\Omega_m$ models).

The evidence suggests that this basic sequence of events is a generic
feature of the history of the majority of galaxies which appear as
thin disks today.  If so, then it places a number of strong
constraints on galaxy formation models:

\begin{itemize}

\item Many successful analytic models of disk formation treat the
formation of the disk as a monolithic collapse (i.e.\ Fall \&
Efstathiou 1980, Dalcanton et al.\ 1997, van den Bosch 1998, 2002).
These models tend to produce realistic disks, even
though their entire theoretical basis seems to be in conflict with a
hierarchical model for the assembly of galaxies.  However, our data
suggests that, for galaxies which are disk dominated today, major
hierarchical mass accretion probably ends early and involves only a
small fraction of the galaxies' mass.  This suggests that it is
probably legitimate to treat the formation of the disk as a monolithic
dissipative collapse.

\item The observation that significant mergers
are unlikely to have occured in the last $6-8\Gyr$ for very late-type disk
galaxies can place strong constraints on the input parameters for
semi-analytic models of galaxy formation, particularly once the
mass-accretion threshold for thick disk formation is better
constrained by realistic merging simulations of primordial thin disks.
Cosmological models which have merging rates that increase to the
present day would be less favored by these observations.

\item In most semi-analytic models of galaxy formation, it is assumed
that bulges form through merging of two galaxies with comparable
masses, and that any disk is accreted subsequent to the merger.
However, the observations of pervasive thick disks suggest that some
of the significant mergers early in a galaxy's history lead to the
formation of a thick disk, and do not neccessarily produce a bulge
(although they might, possibly depending on the gas mass
fraction of the merging progenitors).  Thus, it may be necessary to
revise the criteria for how bulges are produced in semi-analytic
models.

\item If merging leads to the formation of both a thick disk and a
bulge (e.g.\ Kauffmann et al.\ 1993), then the age constraints on
thick disks place indirect age limits on bulges that form via mergers.
Our data therefore suggests that bulges must form early, lest the
epoch of merging also create thick disks that are younger than is
observed.  However, bulge formation which takes place via secular
processes such as bar instabilities (e.g.\ Pfenniger 1993) is still
permitted at any epoch.

\item Assuming that the thick disks in our sample are produced by
merging, and that they persist down to the mass scale of transitional
dwarf spheroidals as we have argued above, then there must have been
merging sub-units on even smaller mass scales.  This decreasing mass
scale sets an upper limit on a possible smoothing scale for the
primordial power spectrum, and limits the masses of possible Warm Dark
Matter candidates.

\item Because the thick disk stars are older than those in the thin
disk, the thick disk isolates baryonic material from an earlier epoch.
Thus, it may be possible to use the relative dynamics and radial
distributions of the thick and thin disks to constrain how the
specific angular momentum distribution changes as a function of time.
This could potentially resolve subtle discrepancies between the
angular momentum distribution of thin disks and theoretical models
(van den Bosch 2001).

\item The apparent universality of thick disks down to very low mass
scales suggests that it may be difficult to measure truly
``primordial'' Helium abundances for constraining Big Bang
nucleosynthesis.  The existence of the red envelope suggests that even
the lowest mass (40-60$\kms$) galaxies with the youngest, bluest star
forming disks experienced an even earlier generation of star
formation.  This early star formation would be likely to
pollute the gas which is currently in the disk, and thus any
metallicity measurements made from HII regions would have been
enriched not just by the current generation of stars, but a previous
one as well.

\end{itemize}


\acknowledgements

JJD gratefully acknowledges discussions with Constance Rockosi, Beth
Willman, Vandana Desai, Andrew West, Suzanne Hawley, Craig Hogan, and
Christopher Stubbs on various aspects of this project.  She also thanks
Hazel Borden for frequent naps.

JJD was partially supported through NSF grant AST-990862 and the
Alfred P. Sloan Foundation.  Support for
RAB was provided by NASA through Hubble Fellowship grant HF-01088.01-97A
awarded by Space Telescope Science Institute.


\end{document}